\documentclass[11pt,a4paper]{article}
\usepackage[utf8x]{inputenc}
\usepackage{epsfig}
\usepackage[T1]{fontenc}    
\usepackage{graphics}
\usepackage{graphicx}
\usepackage{pstricks,pst-coil,pst-fill,pst-plot}
\usepackage[fleqn]{amsmath}    
\usepackage{amssymb}    
\usepackage{amsfonts}   
\usepackage{verbatim}   
\usepackage{mathrsfs}   
\usepackage{dsfont}
\usepackage{euscript}
\usepackage{yfonts}
\usepackage{enumerate}     
\usepackage{amsthm}         
\usepackage{txfonts}
\usepackage{marvosym}
\usepackage{vmargin}        
\usepackage{wasysym}		

\setmarginsrb{1.8cm}{2cm}{1.8cm}{2cm}{1cm}{1cm}{1cm}{1.6cm}
 \makeatletter
 \@addtoreset{equation}{section}
 \makeatother

\providecommand{\bysame}{\leavevmode\hbox to3em{\hrulefill}\thinspace}
\providecommand{\MR}{\relax\ifhmode\unskip\space\fi MR }

\providecommand{\href}[2]{#2}

\newcommand{\bra}[1]{\big< \,#1\,|}
\newcommand{\ket}[1]{|\,#1\, \big>}

\let\ua=\uparrow
\let\da=\downarrow
\let\tend=\rightarrow

\long\def\symbolfootnote[#1]#2{\begingroup%
\def\thefootnote{\fnsymbol{footnote}}\footnote[#1]{#2}\endgroup}

\newtheorem{theorem}{Theorem}[section]
\newtheorem{prop}[theorem]{Proposition}
\newtheorem*{theorem*}{Theorem}

\def\Proof{\medskip\noindent {\it Proof --- \ }}

\def\qed{\hfill\rule{2mm}{2mm}}

\newcommand\beq{\begin{equation}}
\newcommand\enq{\end{equation}}
\newcommand\bem{\begin{multline}}
\newcommand\enm{\end{multline}}

\def\beqa{\begin{eqnarray}}
\def\eeqa{\end{eqnarray}}
\def\ba{\begin{array}}
\def\ea{\end{array}}
\def\det{\operatorname{det}}

\newcommand{\f}[2]{{\ensuremath{%
    \mathchoice%
    {\dfrac{#1}{#2}}
    {\dfrac{#1}{#2}}
    {\frac{#1}{#2}}
    {\frac{#1}{#2}}
}}}
\newcommand{\tf}[2]{\ensuremath{#1/#2}}

\def\a{\alpha}

\def\ga{\gamma}
\def\Ga{\Gamma}

\def\de{\delta}

\def\De{\Delta}
\def\eps{\epsilon}
\def\veps{\varepsilon}
\def\la{\lambda}

\def\sg{\sigma}

\def\Ups{\Upsilon}
\def\ups{\upsilon}

\def\vth{\vartheta}

\def\Om{\Omega}
\def\om{\omega}
\def\vp{\varphi}

\newcommand{\mc}[1]{\ensuremath{\mathcal{#1}}}
\newcommand{\mf}[1]{\ensuremath{\mathfrak{#1}}}
\newcommand{\msc}[1]{\ensuremath{\mathscr{#1}}}

\newcommand{\bs}[1]{\ensuremath{\boldsymbol{#1}}}

\DeclareFontFamily{OT1}{pzc}{}
\DeclareFontShape{OT1}{pzc}{m}{it}{<-> s * [1.10] pzcmi7t}{}
\DeclareMathAlphabet{\mathpzc}{OT1}{pzc}{m}{it}

\def \i{ \mathrm i}

\newcommand{\ov}[1]{\ensuremath{\overline{#1}}}
\newcommand{\wt}[1]{\ensuremath{\widetilde{#1}}}
\newcommand{\wh}[1]{\ensuremath{\widehat{#1}}}

\newcommand{\Int}[2]{\ensuremath{\int\limits_{#1}^{#2}}}
\newcommand{\Oint}[2]{\ensuremath{\oint\limits_{#1}^{#2}}}

\newcommand{\sul}[2]{\ensuremath{\sum\limits_{#1}^{#2}}}
\newcommand{\pl}[2]{\ensuremath{\prod\limits_{#1}^{#2}}}

\newcommand{\R}{\ensuremath{\mathbb{R}}}
\newcommand{\Cx}{\ensuremath{\mathbb{C}}}

\newcommand{\Dp}[1]{\ensuremath{\partial_{#1}}}

\newcommand{\limit}[2]{\ensuremath{\underset{#1 \tend #2}{\longrightarrow} }}

\newcommand{\ex}[1]{\ensuremath{\e{e}^{#1}}}

\newcommand{\op}[1]{ \boldsymbol{ \texttt{#1} } }

\newcommand{\dd}{\mathrm{d}}
\newcommand{\e}[1]{\ensuremath{\mathrm{#1}}}

\newcommand{\intff}[2]{\ensuremath{ [  #1 \,; #2 ] }}
\newcommand{\intfo}[2]{\ensuremath{ [  #1 \,; #2 [ }}

\newcommand{\intoo}[2]{\ensuremath{ ]  #1 \,; #2 [ }}

\newcommand{\intn}[2]{\ensuremath{[\![ \, #1 \,;\, #2 \,]\!]}}

\begin{document}

\begin{center}
\begin{LARGE}
{\bf On the thermodynamic limit of form factor expansions of dynamical correlation functions in the massless regime of the XXZ spin $1/2$ chain.}
\end{LARGE}

\vspace{1cm}

\vspace{4mm}
{\large Karol K. Kozlowski \footnote{e-mail: karol.kozlowski@ens-lyon.fr}}%
\\[1ex]
Univ Lyon, Ens de Lyon, Univ Claude Bernard Lyon 1, CNRS, Laboratoire de Physique, F-69342 Lyon, France. \\[2.5ex]

\par 

\end{center}

\vspace{40pt}

\centerline{\bf Abstract} \vspace{1cm}
\parbox{12cm}{\small}

This work constructs a well-defined and operational form factor expansion in a model having a massless spectrum of excitations.
More precisely, the dynamic two-point functions in the massless regime of the XXZ spin-$\tf{1}{2}$ chain
are expressed in terms of properly regularised series of multiple integrals. These series are obtained by taking, in an appropriate way, the 
thermodynamic limit of the finite volume form factor expansions.  The series
are structured in way  allowing one to identify directly the contributions to the correlator stemming from the conformal-type excitations on the Fermi surface 
and those issuing from the massive excitations (deep holes, particles and bound states). 
The obtained form factor series opens up the possibility of a systematic and exact study of asymptotic regimes of dynamical correlation functions in 
the massless regime of the XXZ spin $1/2$ chain. Furthermore, the assumptions on the microscopic structure of the model's Hilbert space that are necessary so as to write down the series 
appear to be compatible with any model -not necessarily integrable- belonging to the Luttinger liquid universality class. 
Thus, the present analysis provides also the phenomenological structure of form factor expansions in massless models belonging to 
this universality class.

\vspace{20pt}

\begin{center}
 
 {\it Dedicated to the memory of Ludvig Faddeev }
 
\end{center}

\vspace{20pt}

\section*{Introduction}

It is usually impossible to compute explicitly, in a closed form the correlation functions, be it dynamic  or static ones, in a one-dimensional quantum model. 
The best one can hope for, and this is in fact enough for most of the applications, is to come up with an effective theory allowing one to grasp the salient features of the correlators. 
In doing so, the few cases of quantum integrable models where the calculations are amenable to an end may serve as a guiding principle, be it for building some aspects or testing the domain of applicability of such a theory. 
The computations of correlation functions in quantum integrable models has a rather long history which goes back to 
the late 40s when  Kaufman and Onsager   \cite{KafmanOnsagerFirstIntroDetRepCorrIsing2D} computed the row correlations of the 2D Ising model 
and to 
the early 60s when Lieb, Mattis and Schultz \cite{LiebMattisSchultzIsingAsFreeFermionModel}
computed, in closed form, the static correlators of the XY chain. Their approach was generalised or adapted to 
many other free fermion equivalent integrable models; the bottom line being  that the correlators in such models are representable in terms of Toepltz determinants
or Painlevé transcendent \cite{BarouchMcCoy2ptFctsInXYPlusLongDistAB,JimboMiwaMoriSatoSineKernelPVForBoseGaz,McCoyPerkShrockSpinTimeAutoCorrAsModSineKernel,BarouchTracyMcCOyWuScalingResultsForIsinfPainleveIII}. Such expressions can be considered as explicit enough in that 
they allow for a rather thorough analysis of the various physically interesting asymptotic regimes of the
correlators \cite{McCoySomeAsymptoticsForXYCorrelators,TracyVaidaTimeDependentTransverseIsingCorrelatorsAsymptotics,PerkAuYangTimeDpdtTransverseIsing}, this even including 
a rather extensive characterisation of the dynamic response functions \cite{BeckBonnerMullerThomasSpectralFctsXXXGeneralFeatures,MullerShrockDynamicCorrFnctsTIandXXAsymptTimeAndFourier}. 
Starting from the mid 90s new ideas led to explicit, multiple integral based,
representations for the correlators in the XXZ spin-$1/2$ chain, be it the static \cite{JimboMiwaElementaryBlocksXXZperiodicMassless,KitanineMailletTerrasElementaryBlocksPeriodicXXZ}, 
the dynamic \cite{KitanineMailletSlavnovTerrasDynamicalCorrelationFunctions}, the static at finite temperature \cite{GohmannKlumperSeelFinieTemperatureCorrelationFunctionsXXZ} 
or even the dynamic at finite temperature \cite{SakaiDynamicalAndTimeDependentCorrelatorsXXZ} ones. Some of the obtained representations for the static correlation functions could even
be analysed so as to extract, starting from first principles, the leading long-distance asymptotitc behaviour of the two point functions in the massless regime of the
chain \cite{KozKitMailSlaTerXXZsgZsgZAsymptotics}. 

Notwithstanding, the very existence of the above representations -just as the possibility to analyse these- is closely related to the integrable structure of the model
and hence, rather far away from the kind of representations usually dealt with in physics. Indeed, for a general model in finite volume, it appears convenient to characterise correlators by means of expansions into  
form factor series or closely related objects. Such series are obtained by introducing the closure relation
between each of the local operators building the correlator. Form factor expansions have the advantage of separating the dynamic (distance and space dependence)
part of a correlator from its operator content (specific dependence on a given operator). This property predestines form factor expansions to be powerful tools
for studying numerous properties of a model, this provided that one can make sense out of their thermodynamic, \textit{viz}. infinite volume, limit. 
It appears that such series are well-defined for models having a massive spectrum in the 
thermodynamic limit. In such case, the form factors -matrix elements of local operators taken between 
two excited states of the model- go, with an appropriate integer power in the volume, to a form factor density what allows one to replace, in the thermodynamic limit,
the summation over excited states by integrations over the continua of excited states. This property, 
along with a system of axioms \cite{KarowskiWeiszFormFactorsFromSymetryAndSMatrices,KirillovSmirnovFirstCompleteSetBootstrapAxiomsForQIFT} on the properties that ought to be satisfied by a form factor in
an integrable massive quantum field theory in 1+1 dimensions allowed for a construction of form factor expansions for numerous instances of such models, see \textit{e.g.} \cite{BabujianFoersterKarowskiSomeReviewOfBootstrap,SmirnovFormFactors}. 
Owing to the mentioned splitting between the dynamical and operator parts, the obtained series
were extremely effective for studying the long-distance behaviour of the correlators.

The situation becomes however problematic in the case of massless models. Indeed, it is expected \cite{CardyConformalDimensionsFromLowLSpectrum} -and these expectations have been confirmed for numerous massless quantum integrable
models \cite{ArikawaKabrachMullerWieleXXAsymptoticsofFFs,KozKitMailSlaTerEffectiveFormFactorsForXXZ,KozKitMailSlaTerThermoLimPartHoleFormFactorsForXXZ,KozRagoucyAsymptoticsHigherRankModels,SlavnovFormFactorsNLSE}-,
that the form factors have a non-uniform large-volume behaviour: the latter differs depending on whether one of the excited state contains massless excitations or not. 
In particular form factors involving massless excitations do exhibit a non-integer power-law behaviour in the volume what makes it impossible to take the thermodynamic limit of a finite
volume form factors series solely in terms of multiple integrals. Some attempts were made to obtain form factor expansions 
in certain massless quantum field theories by taking an appropriate large energy limit of the massive form factors \cite{DelfinoMussardoSimonettiMasslessFFFromMassiveQFTIntoMethod,MejanSmirnovFormFactorsinPincChiralModelMassless,PonsotFFDensityinMasslessSSG}. 
However, although systematic, the procedure was not applicable to numerous operators of interest; the reason being that, for such operators, it produces non-integrable form factor densities
what renders the associated form factor series divergent. This problem takes its origin in the lack of an appropriate regularising treatment of  the 
vicinity of the massless modes \cite{DelfinoMussardoSimonettiMasslessFFFromMassiveQFTIntoMethod}. 
The cluster property satisfied by the form factors of primary operators which allows for an effective exponentiation of the series in the infrared limit \cite{SmirnovReductionsAndClusterPropertyInSineGordonPlusSomeDiscussionsIRLimit}, 
indicates that it should be possible to make sense out of the mentioned massless limit, at least in the case of primary operators. However, obtaining explicit closed expressions through such an approach does not seem so evident and has not been achieved so far. 
Still, the mentioned clustering structure was used in \cite{LesageSaleurMasslessFFApproachtoFriedelOscillations,LesageSaleurSkorikMasslessFFApproachtoCurrentCorrelators}
to obtain the first two terms present in the massless form factor expansion of certain specific operators. 
Thus, at least in its present state of the art, this approach was not able to produce a clear, systematic and operational procedure allowing to re-absorb the various local divergences related to the presence of massless modes
and re-sum explicitly the original massive form factor series into an explicit massless form factor expansion.

A completely different approach for dealing with form factors series in massless integrable models was pioneered in \cite{KorepinSlavnovTimeDepCorrImpBoseGas} 
for the case of two-point functions in free fermion equivalent models. The idea was to start from a model in finite volume where there is no problem to define a form factor
expansion and then to re-sum the expansion over the excited states in terms of a finite size matrix. 
The latter, in the thermodynamic limit, goes to a Fredholm determinant of an operator $\e{id}+\op{V}$, with $\op{V}$ being an integrable integral operator \cite{ItsIzerginKorepinSlavnovDifferentialeqnsforCorrelationfunctions}. 
The technique appeared rather general and turned out to be applicable to dynamical correlation functions at finite temperature \cite{ColomoIzerginKorepinTognettiTempCorrFctXX},
and multi-point correlators \cite{SlavnovPDE4MultiPtsFreeNLSM} in free fermion equivalent models. The Fredholm determinant based representations allowed for an extensive and very effective characterisation of 
the asymptotic regimes of the free fermionic correlators by means of solving auxiliary Riemann--Hilbert problems. 
For the price of introducing certain auxiliary quantum fields \cite{KojimaKorepinSlavnovNLSEDeterminatFormFactorAndDualFieldTempeAndTime}, the so-called dual fields, 
it was argued that, at least formally, one can still perform the summation over all excited states, even in the case of interacting integrable models.
Then, the  correlators are expressed in terms of an average, over an auxiliary Hilbert space, of a dual field valued Fredholm determinant of an integrable integral operator. 
The presence of dual fields in the kernel of the integrable integral operator turned out to be an important issue for dealing with such representations. 
In particular, it posed serious problems when trying to extract physically interesting asymptotic regimes of the correlators out of such representations and only partial 
results could be obtained \cite{ItsSlavnovNLSTimeAndSpaceCorrDualFields,SlavnovComputationDualFieldVaccumExpLongTimeDistTempeRedDensNLSE}. 
The works \cite{KozReducedDensityMatrixAsymptNLSE,KozTerNatteSeriesNLSECurrentCurrent} proposed a more effective method of resumming a form factor series expansion for interacting integrable models; the latter was based on the concept 
of multi-dimensional deformation flows techniques. This led to explict representations of two-point functions in the non-linear Schrödinger model 
in terms of objects that can be interpreted as multidimensional generalisations of a Fredholm determinant. The latter construction allowed for a rather thorough and explicit analysis of the long-distance and large-time asymptotics of the 
correlators in this model \cite{KozReducedDensityMatrixAsymptNLSE,KozTerNatteSeriesNLSECurrentCurrent} by means of solving an auxiliary Riemann--Hilbert problem \cite{KozTimeDepGSKandNatteSeries}.

One can raise two criticisms. The resummation techniques were strongly relying on the integrable structure of the model and hence were model specific. 
Furthermore, the very use of resummations means that the approach circumvented the issue of defining directly an infinite volume form factor series of a massless model. Thus, 
despite some of its successes such as providing an exact confirmation of the universality based predictions for the asymptotics of correlation functions in massless integrable models, 
that analysis didn't allow to get a sufficient feeling of the very structure of a form factor expansion in a massless model. Yet developping the latter could in principle
allow for building up a phenomenological description of such series in the case of not exactly solvable, \textit{i.e.} generic, massless one dimensional quantum models.

Recently, the work \cite{KozKitMailSlaTerRestrictedSums} proposed a method allowing one to sum up directly, in the long-distance regime, the expansion of a static two-point function over the 
so-called  critical -\textit{i.e.} expectation values of local operators taken between the ground state 
and the low-lying excited states which exhibit a conformal structure-  form factors in massless integrable models such as the Bose gas or the spin-1/2 XXZ chain. 
The method was generalised to the cases of the large-distance behviour of form factor series representing static multi-point \cite{KozKitMailTerMultiRestrictedSums} correlation function of these models
or static two-point functions in higher rank (multi-species) integrable model \cite{KozRagoucyAsymptoticsHigherRankModels}. 
The approach could have  also been conformed to deal with dynamical correlation functions \cite{KozKitMailSlaTerRestrictedSumsEdgeAndLongTime} this in  large-time and distance regime and in the case of 
 the non-linear Schrödinger model. 
These summations allowed for a relatively easy check of the conformal field theory/non-linear Luttinger liquid based predictions for
various asymptotic behaviours of the correlators in these models. It also allowed to identify the microscopic mechanism, \textit{viz}. the structure of a model's Hilbert space, that is responsable for
the model to belong to the Luttinger liquid universality class \cite{KozMailletMicroOriginOfc=1CFTUniversality}. 
However, these summations could only have been done in the asymptotic regime, what allowed to argue and make use of a certain amout of simplifications in the 
form factor series. 

The purpose of the present work is to push further the observations of \cite{KozKitMailSlaTerRestrictedSums,KozKitMailSlaTerRestrictedSumsEdgeAndLongTime} 
so as to develop techniques allowing one to build explicit form factor expansions for the massless regime of the XXZ spin-$\tf{1}{2}$ chain, this in the thermodynamic limit of the model and for almost \textit{any} regime
of the distance and time. The analysis strongly relies on the results of \cite{KozProofOfAsymptoticsofFormFactorsXXZBoundStates} where, building 
on the determinant representations obtained in \cite{KitanineMailletTerrasFormfactorsperiodicXXZ}, the author obtained the large-volume behaviour of the form factors of local operators 
in the XXZ chain taken between the ground state and any excited state of the model. 
This work generalises the earlier analysis \cite{KozKitMailSlaTerEffectiveFormFactorsForXXZ,KozKitMailSlaTerThermoLimPartHoleFormFactorsForXXZ} of such form factor where only certain particle-hole excited states were considered. 

\subsection*{Main result of the paper}

The main result of this paper is the construction of a well-defined infinite volume form factor expansion for the dynamic two-point functions in the massless regime of the XXZ spin-$1/2$ chain.
This result has several important downfalls

\begin{itemize}

\item The derivation of the series stresses the great care that is needed so as to treat correctly the effects of the massless modes on the thermodynamic limit of the series. In particular, the summation over such modes
in finite volume cannot be grasped, when taking the thermodynamic limit, by an integration versus a Lebesgue continuous measure. The correct approach demands to recourse to more 
involved summation techniques. The treatment of the massless modes developed in this work, has strong analogies with an infrared renormalisation and can thus be seen
as an exact, non-perturbative, variant thereof. See the work where similar structures emerge within the constructive renormalisation approach to the XXZ chain.  

\item The otained  series is effective in that, as will be demonstrated in a forthcoming publication, it allows one to extract rather straightforwardly the large-time and long-distance asymptotic behaviour of the two-point functions. 
Since the series has a very close connection with the model's spectrum, it also allows one to show that,  
in the special case of static correlators, the bound states contribute to the large-distance asymptotics on the level of corrections exponentially small in the distance. 
Moreover, as will be shown in another subsequent work, the series also allows one to extract the so-called edge exponents characterising the power-law behaviour of the dynamic response functions
in the vicinity of the multi-particle or bound state excitation thresholds.

\item The assumptions on the microscopic structure of the model's Hilbert space that are necessary so as to write down the infinite volume form factor expansion
in the massless regime of the XXZ chain appear, in fact, to be satisfied by any model -not necessarily integrable- belonging to the Luttinger liquid universality class. 
Thus, the results of present paper also provides one with the phenomenological form taken by a form factor expansion in a massless model belonging to 
this universality class. 

\end{itemize}

\subsection*{Outline of the paper}
 
The paper is organised as follows. Section \ref{Secion resultats principaux et quelques notations} introduces the model of interest and some notations. 
This section also outlines the main result obtained in the paper. Section \ref{Section structure microscopique espace des etats} discusses the microscopic structure
of the Hilbert space associated with the massless regime of the XXZ spin-1/2 chain. Subsection \ref{SousSection parametrisation des etats} discusses the parametisation of the Eigenstates. 
Subsection \ref{SousSection spectre a grand L} discusses the structure of the spectum above the ground state in the large volume limit.  Subsection \ref{SousSection FF gd L} discusses the large-volume behaviour of 
form factors. Section \ref{Section Form factor expansion in thermo limit} is devoted to the computation of the thermodynamic limit of the form factor expansion of dynamical two-point functions
in the massless regime of the XXZ spin-1/2 chain. 
Various preliminary transformations are implemented in Subsections \ref{SousSection Serie auxiliaire C1}-\ref{Sousection serie auxiliare C2} while the 
\textit{per se} thermodynamic limit of the form factor expansion is obtained in Subsection \ref{Sousection limite thermo serie auxiliare C2}. 
This series is recast within the momentum representation in Subsection \ref{SousSectionFcts2PtsDansRepImpulsion} and 
the dynamic response functions are computed in Subsection \ref{SousSection fonctions de reponse dynamiques}. 
Finally, Section \ref{Section extension a autres modeles dans class de LL} discusses briefly how the present result 
provides one with the phenomenological form of form factor expansions in massless \textit{non-integrable} models belonging to the Luttinger liquid university class. 
The paper contains three appendices which gather some technical results of interest to the analysis. 
Appendix \ref{Appendix Observables dans XXZ} discusses the description of the observables in the XXZ spin-$1/2$ chain. Subappendix \ref{Appendix Lin Int Eqns Defs et al}
discusses the various solutions of linear integral equations which parmetrise the spectrum of the model. Subappendix \ref{Appendix Sous-section cordes}
recalls the existence conditions of the bound states. Appendix \ref{Appendix Restricted sums} evaluates the action of an operator that arises in the 
course of taking the thermodynamic limit of the form factor series. Finally, Appendix \ref{Appendix discreto continuous FT} evaluates an integral
which arises in the context of computing the dynamical response functions.

\section{Main results}
\label{Secion resultats principaux et quelques notations}

\subsection{The model}
The XXZ spin-$\tf{1}{2}$ chain refers to a system of interacting spins in one dimension described by the Hamiltonian  
\beq
\op{H} \, = \, J \sum_{a=1}^{L} \Big\{ \sigma^x_a \,\sigma^x_{a+1} +
  \sigma^y_a\,\sigma^y_{a+1} + \cos(\zeta)  \,\sigma^z_a\,\sigma^z_{a+1}\Big\} \, - \, \f{h}{2} \sul{a=1}{L} \sg_{a}^{z} \; . 
\label{ecriture hamiltonien XXZ}
\enq
$\op{H}$ is an operator on the Hilbert space $\mf{h}_{XXZ}=\otimes_{a=1}^{L}\mf{h}_a$ with $\mf{h}_a \simeq \Cx^2$. 
The matrices $\sg^{\ga}$, $\ga=x,y,z$ are the Pauli matrices and the operator $\sg_a^{\ga}$  acts as the Pauli matrix $\sg^{\ga}$
on $\mf{h}_a$ and as the identity on all the other spaces. 
The Hamiltonian depends on three coupling constants:
\begin{itemize}
 
 \item[$\bullet$] $J>0$ which represents the so-called exchange interaction ;
 
 \item[$\bullet$] $\cos(\zeta)$ which parametrises the anisotropy in the coupling between the spins in the longitudinal and transverse directions;
 
 \item[$\bullet$] $h>0$ which measures the intensity of the overall longitudinal magnetic field that is  applied to the chain.
 
 \end{itemize}

Throughout this work I shall focus on the following massless anti-ferromagnetic regime of the chain:  $-1< \cos(\zeta) <1$, \textit{i.e.} $\zeta \in \intoo{0}{\pi}$, 
and $h_{\e{c}}=8J \cos^2(\tf{\zeta}{2})>h >0$. I will assume periodic boundary conditions, \textit{viz.} $\sg_{a+L}^{\ga}=\sg_{a}^{\ga}$.

\noindent Given a spin operator $\op{O}_1$ acting on $\mf{h}_1$, the translation invariance of the chain ensures that the time and space evolved operator $\op{O}_{m+1}(t)$ takes the form 
\beq
\op{O}_{m+1}(t)\, = \, \ex{\i m \op{P} + \i \op{H}  t} \cdot \op{O}_1 \cdot \ex{ -\i t \op{H} -\i m \op{P}} \;, 
\label{ecriture evolution temporelle op local}
\enq
where $\op{P}$ is the momentum operator and hence, $\ex{\i  \op{P} } $ is  the translation operator by one-site.

At zero temperature, the finite volume two-point functions are given by the ground state $\ket{\Om}$ expectation values 
\beq
\bra{\Om}\sg_1^{\ga^{\prime}}\!(t)\,  \sg_{m+1}^{\ga} \ket{\Om} \;. 
\label{ecriture definition fct correlation zero temperature}
\enq
Note that this way of writing down the two-point function has been chosen so as to ensure that
the obtained representations for the correlators involve combinations of the dressed energies $\veps_{a}$ and momenta $p_a$ of the excitation in the form of a difference $m p_a -t \veps_a$. This plays only a cosmetic 
role in the analysis and the usually used two point function correlator $\bra{\Om}\sg_1^{\ga^{\prime}} \sg_{m+1}^{\ga}(t) \ket{\Om}$ can be recovered by the substitution $t \hookrightarrow -t$. 
Note also that the relative sign in the $t \& m$ dependent terms in \eqref{ecriture evolution temporelle op local} issues from the choice of the sign of the Hamiltonian and of the anisotropy $\cos(\zeta)$ in \eqref{ecriture hamiltonien XXZ}.
Furter, here and in the following, the operator $\sg_1^{\ga^{\prime}}$ is defined by  $\sg_1^{\ga^{\prime}}= (\sg_1^{\ga})^{\dagger}$ where $\dagger$ stands for the Hermitian conjugation. 
The presumably existing (this limit exists for $t=0$ as follows by putting together the results of \cite{KitanineMailletTerrasElementaryBlocksPeriodicXXZ,KozProofOfDensityOfBetheRoots})
infinite volume limit of the two-point function is the object of main interest to this work. It will be denoted as
\beq
\big< \sg_1^{\ga^{\prime}}\!(t)\, \sg_{m+1}^{\ga} \big> \; = \; \lim_{L \tend + \infty} \Big\{ \bra{\Om}\sg_1^{\ga^{\prime}}(t) \sg_{m+1}^{\ga} \ket{\Om}  \Big\} \;. 
\enq

\subsection{The main result}

The main result of this work consists in constructing a well-defined, \textit{viz}. free of any divergencies, form factor series
expansion for the thermodynamic limit of the dynamical two-point functions $\big< \sg_1^{\ga^{\prime}}(t) \sg_{m+1}^{\ga} \big>$
in the massless regime of the XXZ spin $1/2$ chain. The obtained form factor expansion is valid in the region where $\op{v}=\tf{m}{t} \not= \pm \op{v}_F$,
with $\op{v}_F$ the Fermi velocity of the model. It is given by the series of multiple integrals: 
\bem
\big< \sg_1^{\ga^{\prime}}\!(t)\,  \sg_{m+1}^{\ga} \big> \; = \;   (-1)^{m \op{s}_{\ga} }   \sul{  \bs{n}\in \mf{S}   }{}   
 \pl{ r \in  \mf{N} }{} \;\Bigg\{  \Int{ \big( \msc{C}_r^{(\de)} \big)^{n_r} }{}  \f{ \dd^{n_r}\nu^{(r)} }{ n_r! \cdot (2\pi)^{n_r}  }   \Bigg\} 
\cdot  \Int{ \big( \msc{C}_{h}^{(\de)} \big)^{n_h}   }{}  \f{ \dd^{n_h}\mu  }{ n_h! \cdot (2\pi)^{n_h}  } \cdot \mc{F}^{(\ga)}\big( \mf{Y} \big) \cdot 
\pl{\ups=\pm}{} \Bigg\{ \f{    \ex{\i \ups  m \ell_{\ups} p_F}    }{   \big[  -  \i m_{\ups}   \big]^{   \vth_{\ups}^2(\mf{Y}) }  }  \Bigg\}   \\
 \times
\pl{r \in \mf{N} }{}  \pl{a=1}{n_r} \Big\{ \ex{ \i m p_r(\nu_a^{(r)})-\i t \veps_r(\nu_a^{(r)}) } \Big\}  \cdot 
 \pl{a=1}{n_h} \Big\{ \ex{ \i t \veps_1(\mu_a)- \i m p_1(\mu_a) } \Big\}  \cdot  \bigg( 1+ \mf{r}_{\de,m, t}\big( \mf{Y} \big) \bigg)  \;. 
\label{ecriture serie FF limite thermo pour fct 2 pts}
\end{multline}
This formula demands some explanations.

\begin{itemize}

\item The series contains the relativistic combination of time and distance 
\beq
m_{\ups} \, = \, \ups m -\op{v}_F t \;. 
\label{definition mups}
\enq

\item $\op{s}_{\ga}$ corresponds to the pseudo-spin of the operator $\sg_{1}^{\ga}$:
\beq
\op{s}_{z}=0 \quad \e{and}  \quad \op{s}_{\pm}=\mp 1 \;. 
\label{definition spin operateur}
\enq
More precisely, $-\op{s}_{\ga}$ corresponds to the relative to the ground state longitudinal spin of the excited states that are connected to the ground state by the operator $\sg_{1}^{\ga}$.

\item The sum in \eqref{ecriture serie FF limite thermo pour fct 2 pts} runs through all possible choices of integers 
\beq
\bs{n}=\big( \{n_r\}_{r\in \mf{N}}, n_h , \ell_{\ups} \big) 
\enq
parametrising the various types of massive excitations. $n_r$  counts 
the various types, labelled by $r\in \mf{N} \subset \mathbb{Z}$, of massive excitations. Here $1 \in \mf{N}$ and more specifically $n_1$ counts the so-called particle excitations. 
$n_h$ counts the hole excitations. Finally, $\ell_{\pm}\in \mathbb{Z}$ are the Umklapp deficiencies which 
encode the difference between the numbers of massless particles and holes forming in the swarm of zero-energy excitations lying on the left ($+$) and right ($-$) Fermi boundary of the model. These integers are subject to the constraints
\beq
n_h \, +\, \op{s}_{\ga} \, = \, \sul{\ups =\pm }{} \ell_{\ups} \, + \, \sul{r \in \mf{N} }{} r n_{r}  \;. 
\enq
\item The massive excitations of type $r$ are parametrised by the rapidities $\nu_{a}^{(r)}$ and the massive hole excitations are parametrised by the rapidities $\mu_a$. 
The macroscopic variables (rapidities and Umklapp deficiencies) characterising a given excitation are all gathered in the set 
\beq
\mf{Y}\, = \, \Big\{ \{\mu_a\}_1^{n_h} ; \{ \{ \nu_a^{(r)} \}_{a=1}^{n_r} \}_{r \in \mf{N}} \, ; \, \{ \ell_{\ups} \} \Big\} \;. 
\enq
\item The rapidities of the various excitations evolve on curves $\msc{C}_{r}^{(\de)}$  and $\msc{C}^{(\de)}_{h}$. For $r \geq 2$ these curves are either given by $\R$
or by $\R+\i\tf{\pi}{2}$, this depending on the value of $r$, \textit{c.f.} Figure \ref{Figure Courbe finale integration modes cordes} and Section  \ref{Appendix Sous-section cordes}. These curves are $\de$ independent. 
The particle, resp. hole, rapidity curves   $\msc{C}_{1}^{(\de)}=\msc{C}_{p}^{(\de)}$, resp. 
$\msc{C}^{(\de)}_{h}$, are depicted in Figure \ref{Figure contour Gamma p et h reduits delta deformes a la limite thermo}. 
These curves coincide with $\Big\{ \R\setminus \intff{-q}{q}\big\} \cup \{-\R +\i\tf{\pi}{2} \}$, resp. $\intff{-q}{q}$, with the exception of 
a vicinity of $\pm q$. There, the curves stay at a distance of the order $\de$ from $\pm q$ and avoid this point, in such a way that the associated particle, resp. hole, oscillatory 
exponential factor in the second line of \eqref{ecriture serie FF limite thermo pour fct 2 pts} has modulus smaller than one. 

\item The particle, resp. hole, excitations carry a dressed momentum $p_1\big(\nu_a^{(1)}\big)$, resp. $p_1\big(\mu_a\big)$, and a dressed energy $\veps_{1}\big(\nu_a^{(1)}\big)$, resp. $\veps_1\big(\mu_a\big)$.
The $r$-bound state excitations, $r\in \mf{N}\setminus \{1\}$, carry a dressed momentum $p_r\big(\nu_a^{(r)}\big)$ and a dressed energy $\veps_{r}\big(\nu_a^{(r)}\big)$.
The definition of all these quantities can be found in \eqref{definition energie habille et energie nue}, \eqref{definition r energi habille} and \eqref{definition r moment habille}. 
 
 \item $\mc{F}^{(\ga)}\big( \mf{Y} \big)$ corresponds to the form factor density squared of the operator $\sg_1^{\ga}$ taken between the ground state and an excited state 
whose massive excitations are parametrised by $\mf{Y}$. More details on the form factor density can be found in Sub-section\ref{SousSection FF gd L}. 
 
 \item $\vth_{\ups}^2(\mf{Y}) $ corresponds to  the critical exponent governing the large-distance long-time decay associated with the excitations characterised by the collection of rapidities
$\mf{Y}$.  Its explicit expression can be found in \eqref{definition phase habilee totale excitation} and \eqref{definition exponsant critique ell shifte}. 

\item $ \mf{r}_{\de,m, t}\big( \mf{Y} \big) $ is a remainder that is controlled as
\beq
 \mf{r}_{\de,m, t}\big( \mf{Y} \big)  \, = \, \e{O}\bigg( \de |\ln \de |\, +\,    \sum_{ \ups = \pm }  \Big( \de^2  | m_{\ups} | \, + \,  \de   \ln |  m_{\ups} | \, + \, \ex{-\de |m_{\ups}|} \Big)  \bigg)  
\label{ecriture estimee sur reste de la serie FF finale}
\enq
with $m_{\ups}$ given in \eqref{definition mups}. Although the remainder is not provided explicitly in this work, it can, in principle, be computed, order-by-order up to the desired precision 
in powers of $\tf{1}{m_{\ups} }$ and $\tf{ \ln |m_{\ups}| }{m_{\ups} }$.

\item The form factor series  depends on an auxiliary control parameter $\de>0$ which is arbitrary and can be taken as small as necessary. This dependence
manifests itself on the level of the remainder $ \mf{r}_{\de,m, t}\big( \mf{Y} \big)$  and in the way the integration curves  $\msc{C}_{1}^{(\de)}$ and 
$\msc{C}^{(\de)}_{h}$ are deformed in the vicinity of the endpoints $ \pm q$ of the Fermi zone.

\end{itemize}

\subsection{Comments}

The parameter $\de$ emerges in the course of the analysis as a means to separate between the massive and massless constituents of the spectrum. 
Indeed, due to a qualitatively different structure of the form factors associated with these two kinds of excitations, each of these demands a separate and 
completely different treatment. The magnitude of this separation is arbitrary what means that, on the level of \eqref{ecriture serie FF limite thermo pour fct 2 pts}, the parameter $\de$ can be taken as small as necessary. 
However, it cannot be set directly to zero. Indeed, the form factor densities $\mc{F}^{(\ga)}\big( \mf{Y} \big) $
exhibit non-integrable singularities when a particle's rapidity $\nu^{(1)}_a$ or a hole's rapidity $\mu_a$ approaches one of the endpoints $\pm q$ of the Fermi zone $\intff{-q}{q}$. 
Furthermore, the integration curve $\msc{C}^{(\de)}_{p}$, resp. $\msc{C}^{(\de)}_{h}$, approaches $\Big\{ \R\setminus \intff{-q}{q} \Big\}\cup\Big\{ -\R+\i\tf{\pi}{2}\Big\}$, resp. $\intff{-q}{q}$,
when $\de \tend 0^+$. Then one can argue that any given multiple particle-hole integral produces, in the $\de\tend 0^+$ limit, logarithmically diverging contributions of the form $P(\ln \de )$ with $P$
a polynomial depending on the integral.
These divergences are reminiscent of the problems with dealing correctly with the infrared divergencies of a theory.
It is important to stress nonetheless that the correlator $\big< \sg_1^{\ga^{\prime}}\!(t)\,  \sg_{m+1}^{\ga} \big> $ does not depend on $\de$ and thus, obviously,  
has a well-defined $\de\tend 0^+$ limit. However, obtaining the latter on the level of the series would demand non-trivial resummations that would end up in 
destroying the very structure of the form factor expansion which is at the root of its usefulness for applications. 
Thus the presence of a regularising parameter $\de$ appears necessary if one wants to preserve the salient features of a form factor expansion.

I refer to the core of the paper so as to obtain a deeper understanding of the expansion and its constituents, Section \ref{SousSection spectre a grand L} in what concerns the spectrum,
Section \ref{SousSection FF gd L} in what concerns the large volume behaviour of the form factors of local operators and, finally, Section \ref{Sousection limite thermo serie auxiliare C2}
where the above representation is obtained.

Finally, one should mention that the above series can be recast in the momentum representation, \textit{c.f.} Section \ref{SousSectionFcts2PtsDansRepImpulsion}, in particular equation \eqref{ecriture limit thermo fct 2 pts forme finale}. 
Also, the form factor expansion \eqref{ecriture serie FF limite thermo pour fct 2 pts} allows one to compute the dynamical response function subordinate to the two-point function 
\beq
\msc{S}^{(\ga)}(k,\om) \, = \, \sul{ m \in \mathbb{Z} }{} \Int{ \R }{} \big< \sg_1^{\ga^{\prime}}\!(t)\, \sg_{m+1}^{\ga} \big> \cdot \ex{ \i(\om t - k m) } \dd t  \;. 
\label{definition dynamic response function}
\enq
The explicit expression of $\msc{S}^{(\ga)}(k,\om)$ can be found in  \eqref{ecriture dynamic response factor}-\eqref{ecriture contrib excitation donnee facteur structure}. 

\vspace{3mm}

One could get the wrong impression that the presence of a regularisation parameter $\de$ and of a non-explicit remainder $ \mf{r}_{\de,m, t}\big( \mf{Y} \big)$
limits the usefulness of the obtained form factor expansion. First of all, I stress that within the setting developed in the core of the work, 
$\mf{r}_{\de,m, t}\big( \mf{Y} \big)$ can be computed, at least in principle, to an arbitrary precision. However, the  estimates \eqref{ecriture estimee sur reste de la serie FF finale}
are already enough for practical applications. As a matter of fact, in its present form,  the series is fully operational on a technical level. 
It will be show in forthcoming publications that, by choosing suitable values for $\de$, the series allows one, by solely relying on saddle-point techniques, to access the 
asymptotic regimes of the two-point functions: their long-time and large-distance asymptotics on the one hand and, on the other hand, 
the power law-behaviour in the frequency that characterises the non-smooth behaviour of the dynamical response functions in the vicinity of the edges
of the excitations thresholds -in particular to access to the so-called edge exponents-.

\section{The microscopic structure of the XXZ chain's Hilbert space}
\label{Section structure microscopique espace des etats}

\subsection{Parametrisation of the Eigenstates}
\label{SousSection parametrisation des etats}

The Eigenstates and energies of the XXZ spin-$\tf{1}{2}$ chain can be obtained within the Bethe Ansatz. In this approach, the Eigenstates $\ket{\Ups}$ and associated energies $\wh{\mc{E}}_{\Ups}$ 
are parametrised by a set $\Ups$ consisting of auxiliary parameters, the so-called Bethe roots \cite{BetheSolutionToXXX,OrbachXXZCBASolution}. In the following, 
I will reserve the notation $\ket{\Om}$ for the ground state of $\op{H}$ and denote by $\Om$ the collection of Bethe roots characterising the ground state.  

One expects that, in the thermodynamic limit $L \tend +\infty$, only Eigenstates having a finite excitation energy relatively to the ground state, \textit{i.e.} such that
$\lim_{L\tend +\infty} \wh{\mc{E}}_{\Ups}- \wh{\mc{E}}_{\Om}< +\infty$, should contribute 
to the zero-temperature  correlation functions of the model (see \cite{KozReducedDensityMatrixAsymptNLSE} for a more precise discussion of this phenomenon.
Notably, in this work, the property has been proven to hold for free fermion equivalent models.). 
I will thus only discuss the structure of these Eigenstates.

According to the general Bethe Ansatz reasonings, the ground state $\ket{\Om}$ is described by a collection of $N$ real parameters,
the ground state Bethe roots. The value of $N$ is fixed by the magnetic field $h_{\e{c}}>h>0$. When $L\tend + \infty$, one has that $N\tend + \infty$
in such a way that $\tf{N}{L}\tend D \in \intfo{0}{1/2}$ and the value of $D$ is uniquely determined by $h$. 
The associated ground state Bethe roots then form a dense distribution on a symmetric segment $\intff{-q}{q}$ whose
endpoint $q$ is fixed by $h$  \cite{KozDugaveGohmannThermoFunctionsZeroTXXZMassless}. The interval $\intff{-q}{q}$ is called the Fermi zone of the model.

Eigenstates of the XXZ spin-$\tf{1}{2}$ chain having a \textit{finite} relative excitation energy above the ground state can be 
seen as a collection of dressed excitations above the Fermi zone $\intff{-q}{q}$ of the model 
\cite{BogoliubiovIzerginKorepinBookCorrFctAndABA,DescloizeauxGaudinExcitationsXXZ+Gap,DescloizeauxPearsonExcitationsXXX,FaddeevTakhtadzhanSpinOfExcitationsInXXX,IshimuraShibaLinIntEqnsForFiniteHForXXZ}
which take place in a spin $\op{s}$ sector relatively to the ground state. 
These excitations can be gathered into three classes: the particles, the holes and the bound states. Any bound state excitation is necessarily massive. However, the particle or hole excitations can be either massless or massive. 

There exist several branches of bound state excitations in the XXZ chain. They are labelled by an integer $r$. 
In the Bethe Ansatz language, an "$r$-bound state" corresponds to a collection of Bethe roots forming a string of length $r\geq 2$ and, hence, is called
an $r$-string \cite{BetheSolutionToXXX}. For given $\zeta \in \intoo{0}{\pi}$ there may or may not exist an upper bound 
$r_{\e{max}}$ on the allowed length of a string. This depends on whether $\tf{\zeta}{\pi}$ is rational or not \cite{TakahashiThermodynamics1DSolvModels}. 
In the following, if there is no upper bound, one should simply take $r_{\e{max}}=+\infty$. 
Furthermore, for fixed $\zeta$, an $r$-string may only arise when $r$ takes values in a subset $\mf{N}_{\e{st}}$ of $\intn{ 2 }{ r_{\e{max}} }$. 
The fact that an $r$ string exists or not depends on the continued fraction decomposition of $\tf{\zeta}{\pi}$. 
The conditions fixing the allowed lengths of strings were first argued by Suzuki and Takahashi in \cite{TakahashiSuzukiFiniteTXXZandStrings} and, subsequently,
Korepin \cite{KorepinAnalysisofBoundStateConditionMassiveThirring} argued the form of these conditions on the basis of the normalisability of the wave function for the Thirring model.
Korepin's arguments were applied to the XXZ chain simultaneously by Hida \cite{HidaConditionExistenceStringsXXZ} and by Fowler, Zotos in \cite{FowlerZotosConditionsOfStringExistenceXYZAndSineGordon}. 
However, \textit{per se}, these arguments only hold in the ferromagnetic sector $D=0$, \textit{i.e.} for the model at $h$ large enough. Other fine effects come into play 
when $\tf{N}{L}$ has non-zero limit. The author \cite{KozProofOfStringSolutionsBetheeqnsXXZ} corrected the form of these conditions by taking rigorously the large-$L$ limit. 
This correct form of the constraints for the existence of an $r$ bound state is recalled in Appendix \ref{Appendix Sous-section cordes}.

When $L$ is large, a given excited state $\ket{\Ups}$ having a finite excitation energy relatively to the ground state can thus be parametrised by its spin $\op{s}_{\Ups}$ 
relatively to the ground state and by the collection of rapidities of the various excitations:
\beq
\mf{R}_{\Ups}\, = \, \Big\{ \big\{ \la_{ m_a }^{(p)} \big\}_1^{ n_p^{(\e{tot})} }\cup  \big\{ \la_{m_a^{\prime}}^{(h)} \big\}_1^{n_h^{(\e{tot})} }  \Big\} \cup 
\Big\{ \big\{ \nu_{ d_a^{(r)} }^{(r)} \big\}_{a=1}^{ n_r } \Big\}_{ r \in \mf{N}_{\e{st}}  } \;.
\label{ecriture rapidites macr decrivant etat}
\enq
For finite $L$, the rapidities in \eqref{ecriture rapidites macr decrivant etat} satisfy to the higher level Bethe Ansatz equations
\beq
\wh{\xi}_{1}\big( \la_{ m_a }^{(p)}  \mid \mf{R}_{\Ups} \big)  \; = \; \tfrac{  2\pi  }{ L }  m_a  \; , \quad 
\wh{\xi}_{1}\big( \la_{ m_a^{\prime} }^{(h)}  \mid \mf{R}_{\Ups} \big)  \; = \; \tfrac{  2\pi  }{ L }  m_a^{\prime} 
\quad \e{and} \quad 
\wh{\xi}_{r}\Big( \nu_{ d_a^{ (r)} }^{(r)}   \mid \mf{R}_{\Ups} \Big)  \; = \;   \tfrac{ 2\pi }{ L }  d_a^{(r)} \; . 
\label{equations pour rapidities part trous et cordes}
\enq
$\wh{\xi}_{a}$ are the so-called counting functions \cite{DeVegaWoynarowichFiniteSizeCorrections6VertexNLIEmethod,DestriLowensteinFirstIntroHKBAEAndArgumentForStringIsWrong,KlumperBatchelorNLIEApproachFiniteSizeCorSpin1XXZIntroMethod}.
When $a=1$, they are associated with the particle-holes rapidities and, more generally, when $a=r\geq 2$ with the $r$-strings rapidities. The equations \eqref{equations pour rapidities part trous et cordes} involve three types of integers 
 $m_a$, $m^{\prime}_{a}$ and $d_a^{(r)}$ which  belong to the sets 
\beq
m_a \in  \intn{-M_{-} }{ M_{+} } \setminus \intn{ 1 }{ N+\op{s}_{\Ups} } \quad, \quad 
m_a^{\prime} \in  \intn{ 1 }{ N + \op{s}_{\Ups} }  \quad , \quad 
d_a^{(r)} \in \intn{-M^{(r)}_{-} }{ M^{(r)}_{+} } \;. 
\label{definition domaine appartenance entiers des modes excites}
\enq
The integers $M_{\pm}$, $M^{(r)}_{\pm}$ and $N$ all go linearly with $L$ to infinity.  

A given excited state $\ket{\Ups}$ in the spin $\op{s}_{\Ups}$ sector above the ground state corresponds to the choice of the integers $ n_p^{(\e{tot})}, n_h^{(\e{tot})} $ and $ n_r $,   $r \in \mf{N}_{\e{st}}  $,
satisfying to the constraint 
\beq
 n_h^{(\e{tot})}  \, + \, \op{s}_{\Ups} \, = \, n_p^{(\e{tot})}  \, + \, \sul{ r \in \mf{N}_{\e{st}}   }{} r n_r  \qquad n_h^{(\e{tot})} \in \intn{0}{\varkappa_L}
\label{definition contrainte sur entiers etats non collapse en mode masse nulle}
\enq
and then to the choice of increasing sequences of integers
\beq
m_1<\dots < m_{ n_p^{(\e{tot})} } \quad,  \quad  m_1^{\prime}<\dots < m_{ n_h^{(\e{tot})} }^{\prime} \qquad  \e{and}  \qquad d_1^{(r)} < \dots < d_{n_r}^{(r)} \; , 
\label{definition contrainte sur entiers labellant rapidities des modes excites}
\enq
such that each of the involved integers belongs to its respective domain as given in \eqref{definition domaine appartenance entiers des modes excites}. Here, $\varkappa_{L}$ is some arbitrary sequence
such that $\tf{ \varkappa_L }{ \sqrt{L} }\limit{L}{+\infty} +\infty$ and $\tf{ \varkappa_L }{ L }\limit{L}{+\infty} 0$. It is believed that only excited states such that $0\leq n_h^{(\e{tot})} \leq  \varkappa_{L}$
may have a finite thermodynamic limit of their excitation energy above the ground state. 
Note that since $n_h^{(\e{tot})}$ is finite, \eqref{definition contrainte sur entiers etats non collapse en mode masse nulle} necessarily implies that, for the given excited state under consideration,
$\big\{ r \in \mf{N}_{\e{st}} \, : \, n_r\not=0 \big\}$ is a finite set.  This is
a non-trivial constraint when $\tf{\zeta}{\pi}$ is irrational as the set of allowed string lengths $\mf{N}_{\e{st}}$ is unbounded.

\noindent The rapidities arising in \eqref{ecriture rapidites macr decrivant etat} have different origins:

\vspace{2mm}

\begin{itemize}

\item[$\bullet$]  $\la_{ a }^{(h)}$ are the rapidities of the hole excitations. In the $L\tend + \infty$ limit and when $a$ runs through  $\intn{1 }{ N +\op{s}_{\Ups} } $, $\la_{ a }^{(h)}$ runs through $\intff{-q}{q}$.

\vspace{2mm}

\item[$\bullet$] $\la_{a }^{(p)}$ are the rapidities of the particle excitations. When $L\tend + \infty$ and $a$ runs through the subset of integers  $\intn{-M_{-} }{ M_{+} } \setminus \intn{1}{N+\op{s}_{\Ups}} $, the rapidities vary on 
$\big\{ -\R + \i\tf{\pi}{2}  \big\} \cup \big\{ \R \setminus \intff{-q}{q}\big\}$. Here $-\R$ indicates that the set is skimmed through along the opposite orientation.   
 
\vspace{2mm}
\item[$\bullet$] $\nu_{ a }^{(r)}$ are the rapidities of the $r$-strings. An $r$-string is characterised by a definite parity $\de_{r}$ which is either $0$ or $1$.  
If the $r$-string has zero parity, then its rapidity is real valued, whereas, if it has parity one, its rapidity belongs to $\R+\i\tf{\pi}{2}$. 
When $L\tend + \infty$ and $d_a^{(r)}$ runs through  $ \intn{-M^{(r)}_- }{ M^{(r)}_+ } $, the $r$-string rapidity varies on 
$s_r\R+\de_{r}\i\tf{\pi}{2} $. Here $s_r=1$ or $-1$ and encodes the orientation along which the set is skimmed through. 
 
\end{itemize}

\vspace{1mm}

Given any configuration $\mf{R}$ of parameters -not necessarily solving \eqref{equations pour rapidities part trous et cordes}- 
\beq
\mf{R} \, = \, \Big\{ \big\{ \la_{ a }^{(p)} \big\}_1^{ n_p^{(\e{tot})} }\cup  \big\{ \la_{a}^{(h)} \big\}_1^{n_h^{(\e{tot})} }  \Big\} \cup 
\Big\{ \big\{ \nu_{a }^{(r)} \big\}_{a=1}^{ n_r } \Big\}_{ r \in \mf{N}_{\e{st}}  }  
\label{ecriture rapidites macr cas generique sous integrale}
\enq
such that the cardinalities of the respective sets satisfy \eqref{definition contrainte sur entiers etats non collapse en mode masse nulle}, the counting functions takes the form 
\beq
\wh{\xi}_{r}\big( \om  \mid \mf{R} \big)  \; = \; p_{r}\big( \om \big) \, - \,     \f{1}{L}  \wh{F}_{r}\big( \om  \mid \mf{R} \big) \, + \, \de_{r,1} \f{N+1}{2 L } \;. 
\label{equations donnant fct cptge r}
\enq
The functions $p_{r}$ have the interpretation of the dressed momenta of the excitations associated with $\wh{\xi}_{r}$. They are defined as solutions to linear integral equations, 
\textit{c.f.} Appendix \ref{Appendix Lin Int Eqns Defs et al} for more details. 
The functions $\wh{F}_{r}$ appearing in \eqref{equations donnant fct cptge r} depend on $\mf{R}$ and are analytic  functions of their arguments provided that these stay in a neighbourhood of the domains
 of condensation of the associated rapidities. They are such that 
\beq
|\wh{F}_r\big( \om \mid \mf{R} \big)| \leq C \cdot  \Big( n_p^{(\e{tot})}+ n_h^{(\e{tot})} + |\op{s}_{\Ups}| + \sul{r \in \mf{N}_{\e{st}} }{} r n_r \Big)\, ,
\enq
uniformly in the rapidities contained in $\mf{R}$ and in $\om$ belonging to a neighbourhood of $\R \cup \big\{\R+\i\tf{\pi}{2} \big\}$. Furthermore, they obey reduction properties when some of the rapidities of the particles or holes 
become close to the Fermi boundaries $\pm q$. More precisely, assume that the particle and hole rapidities present in $\mf{R}$ partition as 
\beq
\Big\{ \la_{a}^{(p)} \Big\}_1^{n_p^{(\e{tot})} } \; = \;   \Ups^{(p)}_{+} \cup \big\{ \nu_{a }^{(1)} \big\}_{1}^{n_1}  \cup   \Ups^{(p)}_{-}  
\quad \e{and} \quad 
\Big\{ \la_{a}^{(h)} \Big\}_1^{n_h^{(\e{tot})} } \; = \;    \Ups^{(h)}_{+} \cup \big\{ \mu_{a} \big\}_1^{n_h} \cup  \Ups^{(h)}_{-}  \;. 
\label{ecriture patritionnement rapidites particules et trous en massive et massless}
\enq
The rapidities building up the sets  $\Ups^{(p)}_{\pm} \, = \, \big\{ \la_a^{\pm}  \big\}_1^{ n_p^{\pm} } $, resp.  $\Ups^{(h)}_{\pm} \, = \, \big\{ \mu_a^{\pm}  \big\}_1^{ n_h^{\pm} } $, 
are at most within a distance $\de$ of the endpoints $\pm q$ of the Fermi zone:
\beq
|\la_a^{\ups} - \ups q|  \leq \de \qquad \e{and} \qquad  |\mu_a^{\ups} - \ups q|  \leq \de \;. 
\label{ecriture borne delta sur position part trous type massless}
\enq
Here, $\de>0$ is some fixed, sufficiently small parameter that is $L$-independent. In their turn, the 
"bulk" rapidities $\mu_{a}$ and $\nu_{a}^{(1)} $ are all uniformly away from the endpoints of the Fermi zone 
\beq
\big| \mu_{a} - \ups q  \big| \, > \,  \de \qquad \e{and} \qquad  \big| \nu_{a}^{(1)} -  \ups q  \big|  \, > \,  \de  \qquad \e{for} \quad \ups \in \{ \pm \}\;.
\label{ecriture borne delta sur position part trous type massive}
\enq

Up to precision $\de$, the rapidities in $ \Ups^{(p/h)}_{\pm} $ can be thought of as collapsing on the right $(+)$, resp. left $(-)$, endpoint of the Fermi zone. 
The integers 
\beq
\ell_{\ups}=n_p^{\ups} - n_h^{\ups}
\label{definition entier ell ups}
\enq
encode the discrepancy between the numbers of particles and holes collapsing on either of the endpoints of the Fermi zone.

If  the decomposition \eqref{ecriture patritionnement rapidites particules et trous en massive et massless} holds, the rapidities gathered in $\mf{R}$ split
\beq
\mf{R} \; = \;\mf{Y} \bigcup\limits_{\ups = \pm }^{} \Big\{ \Ups^{(p)}_{\ups} \cup \Ups^{(h)}_{\ups}\Big\}
\label{decomposition de R mode massifs et massless}
\enq
into a collection of macroscopic variables of the massive modes
\beq
\mf{Y} \; = \; \Big\{  \big\{ \mu_{a} \big\}_{1}^{ n_{h} } \, ; \, \big\{ \big\{ \nu_{a}^{(r)} \big\}_{a=1}^{n_r} \big\}_{r\in \mf{N} }  \, ; \,  \big\{ \ell_{\ups}\big\}  \Big\}
\qquad \e{with} \quad \mf{N} \, = \, \mf{N}_{\e{st}}\cup \{ 1 \}
\label{definition rapidites massives reduites}
\enq
and a collection of rapidities $ \Ups^{(p/h)}_{\pm} $ of the particles and holes belonging to a neighbourhood of the Fermi endpoints, \textit{viz}. those giving rise to the massless 
part of the spectrum. 

Then, given the decomposition \eqref{decomposition de R mode massifs et massless} subordinate to $\de>0$ as in \eqref{ecriture borne delta sur position part trous type massless}, the building blocks of the counting functions satisfy
\beq
\wh{F}_r\big( \om \mid \mf{R} \big) \; = \; \wh{F}_r\big( \om \mid \mf{Y} \big) \, + \, \e{O}\big(\de \big)
\enq
what entails that 
\beq
\wh{\xi}_{r}\big( \om  \mid \mf{R} \big) \; = \; \wh{\xi}_{r}\big( \om  \mid \mf{Y} \big) \, + \, \e{O}\Big( \tfrac{ \de }{ L } \Big) \;. 
\label{ecriture propriete reduction fct comptage}
\enq
Finally, $\wh{F}_r\big( \om \mid \mf{R} \big) $ admits the large $\om$ asymptotic expansion
\beq
\wh{F}_r\big( \om \mid \mf{R} \big)  \, = \, \check{c}^{\pm}_r \, + \, \check{C}^{\pm}_r \ex{\mp 2 \om} \; + \; \e{O}\Big( \ex{\mp 4 \om}\Big)
\label{asymptotiques fct shift}
\enq
when $\Re(\om)\tend \pm \infty$ and $\om$ belongs to an open neighbourhood of $\R+\i \de_r \tf{\pi}{2}$ and for some constants $\check{c}^{\pm}_r, \check{C}^{\pm}_r$.

\subsection{The spectrum at large-$L$}
\label{SousSection spectre a grand L}

\noindent The relative to the ground state excitation energy and momentum of an excited state $\ket{\Ups}$  parametrised 
by the rapidities \eqref{ecriture rapidites macr decrivant etat} and located in the  -$\op{s}_{\Ups}$ spin sector takes the form 
\beq
 \wh{\mc{E}}_{\Ups\setminus \Om }  \;  =  \; \mc{E}\big( \mf{R}_{\Ups} \big)   \; + \; \e{O}\Big( L^{-1} \Big)
\qquad \e{and} \qquad 
 \wh{\mc{P}}_{\Ups\setminus \Om }  \;  =  \; \mc{P}\big( \mf{R}_{\Ups} \big)  \, + \, \pi \op{s}_{\Ups} \; + \; \e{O}\Big( L^{-1} \Big)
\label{ecriture forme energie relative excitation grand L}
\enq
where, given $\mf{R}$ as in \eqref{ecriture rapidites macr cas generique sous integrale}, one has  
\beqa
 \mc{E}\big( \mf{R}  \big) & = & \sul{ a=1 }{ n_p^{(\e{tot})} } \veps_1\big( \la_{a}^{(p)} \big)  \, - \, \sul{ a=1 }{ n_h^{(\e{tot})} } \veps_1( \la_{a}^{(h)}  )   
					   \;  +\hspace{-2mm}  \sul{ r\in \mf{N}_{\e{st}} }{ } \sul{a=1}{n_r } \veps_{r}\big( \nu_{a}^{(r)} \big)    
\label{ecriture forme leading energie ex}\\
 \mc{P}\big( \mf{R}  \big)  & = &  \sul{ a=1 }{ n_p^{(\e{tot})} } p_1\big( \la_{a}^{(p)} \big)  \, - \, \sul{ a=1 }{ n_h^{(\e{tot})} } p_1( \la_{a}^{(h)} )    
					     \;  +\hspace{-2mm}   \sul{ r\in \mf{N}_{\e{st}} }{ } \sul{a=1}{n_r } p_{r}\big( \nu_{a}^{(r)} \big)     \;. 
\label{ecriture forme leading impulsion ex}
\eeqa

The functions $\veps_{a}$, resp. $p_{a}$, are the dressed energies and momenta of the individual excitations. They are defined as solutions to linear integral equation, 
\textit{c.f.} Appendix \ref{Appendix Lin Int Eqns Defs et al}. Finally, $p_F=p_1(q)$ is the Fermi momentum. 

The form taken by the excitation energy $ \wh{\mc{E}}_{\Ups\setminus \Om }$ and the properties of the dressed energies $\veps_{r}$, \textit{c.f.} Appendix \ref{Appendix Lin Int Eqns Defs et al}, ensure that  
\vspace{2mm}
\begin{itemize}
\item[$\bullet$] the bound state excitations are massive, \textit{i.e.} $\veps_{r}(\la) > c_r >0$ for some constant $c_r$ on $\R+\de_{r}\i\tf{\pi}{2} $, for $r \in \mf{N}_{\e{st}}$;
\item[$\bullet$] the particle-hole excitation have a massless and a massive branch: one has 
$$ \veps_1 >0  \quad \e{on} \quad \Big\{ \R + \i\tfrac{ \pi }{ 2 } \Big\} \cup \Big\{ \R \setminus \intff{ -q }{ q } \Big\} \qquad  
\e{and} \quad  \veps_1<0 \quad \e{ on } \quad  \intoo{ - q }{ q } \; .$$
In particular $\veps_1(\pm q) =0$ so that the massless excitations will correspond to particles, resp. holes, whose rapidities $\la^{(p)}_a$, resp. $\la^{(h)}_a$, collapse, in the thermodynamic limit,  
on the endpoints of the Fermi zone. Such a collapse is achieved if the integer $m_a$, resp. $m_a^{\prime}$, in \eqref{equations pour rapidities part trous et cordes}
is of the type $N+\e{o}(L)$ -meaning that the rapidity collapses on $q$- or $1+\e{o}(L)$ -meaning that the rapidity collapses on $-q$-. Hence, the distinction made in 
\eqref{ecriture borne delta sur position part trous type massless}-\eqref{ecriture borne delta sur position part trous type massive}
\end{itemize}

\vspace{2mm} 

For further purposes, it is convenient to introduce a special notation for the ratio 
\beq
\op{v}=\f{m}{t}
\enq
of the distance to time as well as for the combination 
\beq
\mc{U}\big( \mf{R}, \op{v} \big)\; = \;  \mc{P}\big( \mf{R}  \big) \, - \, \f{1}{\op{v}} \cdot \mc{E}\big( \mf{R}  \big)  
\enq
of the excitation momenta and energies which can be recast as 
\beq
\mc{U}\big( \mf{R}, \op{v} \big)\; = \; \sul{ a=1 }{ n_p^{(\e{tot})} } u_1\big( \la_{a}^{(p)}, \op{v} \big)  \, - \, \sul{ a=1 }{ n_h^{(\e{tot})} } u_1( \la_{a}^{(h)}, \op{v} )    
					     \;  +\hspace{-2mm}   \sul{ r\in \mf{N}_{\e{st}} }{ } \sul{a=1}{n_r } u_{r}\big( \nu_{a}^{(r)}, \op{v} \big)    \;. 
\label{definition U des rapidites}
\enq
There, I agree upon 
\beq
u_{r}\big( \la, \op{v} \big)     \; = \; p_r(\la) - \f{ \veps_{r}(\la) }{ \op{v} } \;. 
\label{definition fct ur et vitesse v}
\enq
If the configuration of rapidities in $\mf{R}$ decomposes as \eqref{decomposition de R mode massifs et massless}, then the function $\mc{U}\big( \mf{R}, \op{v} \big)$
simplifies to  
\beq 
\mc{U}\big( \mf{R}, \op{v} \big)\; = \; \msc{U}\big( \mf{Y}, \op{v} \big)\; + \; \e{O}\big( \de \big)
\enq
where 
\beq
\msc{U}\big( \mf{Y}, \op{v} \big)\; = \; \sul{ r\in \mf{N} }{ } \sul{a=1}{ n_r } u_{r}\big( \nu_{a}^{(r)}, \op{v} \big)   \, - \, \sul{ a=1 }{ n_h  } u_1( \mu_{a}, \op{v} )    \, + \, \sul{\ups=\pm }{} \ups \ell_{\ups} p_{F}\;. 
\label{definition de energi impuslion reduite condensation massless}
\enq
Above, I have introduced the Fermi momentum $p_F=p_1(q)$. 

I refer to \cite{KozProofOfAsymptoticsofFormFactorsXXZBoundStates} where all this is discussed at length.

\subsection{The form factors at large-$L$}
\label{SousSection FF gd L}

The analysis to come will build on the explicit form taken by  the large-volume behaviour of the form factors of the local operators $\sg^{\ga}_a$, $\ga \in \{\pm , z\}$, namely the expectation values 
$\big| \bra{\Ups} \sg_a^{\ga} \ket{\Om} \big|^2$. 
The large volume behaviour of $\big| \bra{\Ups} \sg_a^{\ga} \ket{\Om} \big|^2$  when $\ket{\Ups}$ is an excited state as described in Section \ref{SousSection parametrisation des etats} and $\ket{\Om}$ the ground state  
has been determined in \cite{KozProofOfAsymptoticsofFormFactorsXXZBoundStates} on the basis of a rigorous analysis. One concludes that there exists an $L$-dependent
function $ \wh{\msc{F}}^{(\ga)}$ of the rapidities $\mf{R}_{\Ups}$ associated with the excited state $\ket{\Ups}$, \textit{c.f.}  \eqref{ecriture rapidites macr decrivant etat}, such that 
\beq
\big| \bra{\Ups} \sg_a^{\ga} \ket{\Om} \big|^2  \; = \;   \wh{\msc{F}}^{(\ga)}\big( \mf{R}_{\Ups} \big)   \cdot  \Big( 1\, + \, \e{O}\Big( \tfrac{\ln L}{L}  \Big) \Big) \;. 
\label{ecriture DA gd L du FF GS vers Ups}
\enq
The explicit expression of $ \wh{\msc{F}}^{(\ga)}\big( \mf{R} \big)  $ is rather cumbersome and will be of no use  in the following. It can be found in \cite{KozProofOfAsymptoticsofFormFactorsXXZBoundStates}. 
For further purposes, it is enough to know that it is an analytic function on some small neighbourhood of the 
region where the various rapidities evolve. Furthermore, for all $r$-string rapidities and particle or hole rapidities uniformly away  from $\pm q$, this neighbourhood may be taken $L$-independent.

The large-$L$ expansion given above is not uniform in respect to the position of the particle-hole rapidities $\mf{R}_{\Ups}$ given in \eqref{ecriture rapidites macr decrivant etat}, 
or equivalently in respect to the integers arising in \eqref{equations pour rapidities part trous et cordes}.
It can be further simplified if one provides additional information  on the proximity of the particle-hole rapidities to the endpoints $\pm q$ of the Fermi zone. 
This non-uniformness has to do with the fact that the particle or hole excitations on the Fermi boundaries $\pm q$ correspond to the massless excitations of the model.

In order to state the form of the simplified expression, consider a collection of rapidities $\mf{R}$, not necessarily solving \eqref{equations pour rapidities part trous et cordes},  
which partitions as in \eqref{decomposition de R mode massifs et massless}. It is convenient to re-parametrise the rapidities in $\Ups^{(p/h)}_{\pm}$ as 
\beq
  \mf{Z}^{\ups}_{\mf{Y}}   \; = \; \bigg\{ \Big\{  \ups \big[ \tfrac{L}{2\pi} \, \wh{\xi}_1(\mu \mid \mf{Y} )-N_{\ups} \big] \Big\}_{ \mu\in\Ups_{\ups}^{(p)} }  \, ; \,  
  \Big\{\ups \big[N_{\ups} -  \tfrac{L}{2\pi} \, \wh{\xi}_1(\mu\mid \mf{Y} ) \big] -1 \Big\}_{ \mu\in \Ups_{\ups}^{(h)} }   \bigg\} \;. 
\label{definition des variables Z ups}
\enq
Then, given the partitioning \eqref{decomposition de R mode massifs et massless}, one gets 
\beq
 \wh{\msc{F}}^{(\ga)}\big( \mf{R} \big) \; = \; {\bf : } \pl{ \ups \in \{ \pm \} }{} \Big\{ \mf{F}_{\ups}\big( \mf{Y} \mid \mf{Z}^{\ups}_{\mf{Y}} \big)  \Big\}  {\bf :}\,
 \f{ \mc{F}^{(\ga)}\big( \mf{Y} \big)   \cdot  \Big( 1\, + \, \e{O}\Big( \tfrac{\ln L}{L}  \, + \, \de \ln \de \Big) \Big) }
{ \prod_{ a=1 }^{ n_h} \Big\{  L \, \wh{\xi}^{\prime}_1( \mu_{a} \mid \mf{Y} )   \Big\} 
\cdot \prod_{ r\in \mf{N}   }^{ } \prod_{ a=1 }^{ n_r }  \Big\{  L \, \wh{\xi}^{\prime}_r\big( \nu_{a}^{(r)} \mid \mf{Y} \big)   \Big\}   }  
\label{ecriture DA FF apres partition modes massifs et masse nulle}
\enq
where the counting functions are as given in \eqref{equations pour rapidities part trous et cordes}.  The ${ \bf : } * {\bf :}$ symbol appearing in the first factor 
corresponds to an operator normal-like ordering; its precise meaning will be discussed below. 
The form factor $ \wh{\msc{F}}^{(\ga)}\big( \mf{R} \big)$ factorises into three parts:
\begin{itemize}

\item[$\bullet$] $\mc{F}^{(\ga)}\big( \mf{Y} \big) $ which should be thought of as the form factor density squared associated with the massive modes of the model; 

\item[$\bullet$] $\mf{F}_{\ups}\big( \mf{Y} \! \mid \! \mf{Z}^{\ups}_{\mf{Y}} \big) $  which should be thought of as the form factor density squared that is associated with the massless excitations of the model;

\item[$\bullet$] the denominator containing the counting functions which corresponds to the densities of the massive modes parametrised by $\mf{Y}$. 
 
\end{itemize}

\vspace{2mm}
The form factor density of the massive modes $ \mc{F}^{(\ga)}\big( \mf{Y} \big) $ is a smooth function of $\mf{Y}$.  Its explicit expression can be found in \cite{KozProofOfAsymptoticsofFormFactorsXXZBoundStates}. 
The only properties that will be needed is that 
\begin{itemize}
  
  \item[$i)$] it has at least a double zero when some rapidities of the same type (\textit{i.e.} particle, hole or $r$-string) coincide;

  \item[$ii)$] it decays as $\ex{ \mp 2 r \nu_a^{(r)}}$ when the real part of a rapidity $\nu^{(r)}_a$ goes to $\pm \infty$;
  
    \item[$iii)$] it has power-law singularities  at $\pm q$ in respect to the massive particle-hole variables $\nu_{a}^{(1)}$ and $\mu_a$;
  
  \item[$iv)$] it extends to an analytic function of any of the variables present in $\mf{Y}$, where each rapidity belongs to an $L$-independent, sufficiently small, neighbourhood of the domain where the corresponding excitation rapidities live. 
The size of this neighbourhood depends on $\de$ in what concerns the particle-hole rapidities.   
The form factor density is also analytic in a neighbourhood of $\infty$, with the exception of the lines 
\beq
\nu_a^{(1)} \in \intoo{-\infty}{-q} + \i \mf{f}_{ \zeta } \quad \e{and} \quad 
\nu_a^{(r)} \in \intoo{-\infty}{-q} + \i \mf{f}_{ \f{r \pm 1}{2}\zeta } \quad \e{for} \quad r \in \mf{N}_{\e{st}} \quad \e{and} \; modulo \; \i\pi
\label{ecriture des domaines de saut pour FF density}
\enq
this upon agreeing that $\mf{f}_{ \eta }=\e{min}\big( \eta- \pi \lfloor \tf{\eta}{\pi} \rfloor, \pi - \eta +  \pi \lfloor \tf{\eta}{\pi} \rfloor \big)$. On the above rays, the form factor has a jump discontinuity  
which takes the form 
\beq
\mc{F}^{(\ga)}\big( \mf{Y}_{\pm,s}^{\ua} \big)  \;  = \; \mc{F}^{(\ga)}\big( \mf{Y}_{\pm,s}^{\da} \big) \;. 
\label{ecriture conditions saut FF}
\enq
This boundary value problem involves two sets of variables:
\beqa
\mf{Y}_{\sg,s}^{\ua} & = & \Big\{  \big\{ \mu_{a} \big\}_{1}^{ n_{h} } \, ; \, 
\big\{ \big\{ \nu_{a}^{(r)} \big\}_{a=1}^{n_r} \big\}_{r\in \mf{N} }  \, ; \,  \big\{ \ell_{\ups}\big\}  \Big\}_{ \nu^{(s)}_b=x+\i\mf{f}_{ \f{s +\sg }{2}\zeta } + \i 0^+ } \\
\mf{Y}_{\sg,s}^{\da} & = & \Big\{  \big\{ \mu_{a} \big\}_{1}^{ n_{h} } \, ; \, \big\{ \big\{ \nu_{a}^{(r)} \big\}_{a=1}^{n_r} \big\}_{r\in \mf{N} }  \, ; \, 
\big\{ \ell_{\ups}+\ups u_s^{\sg} \big\}  \Big\}_{ \nu^{(s)}_b=x+\i\mf{f}_{ \f{s +\sg }{2}\zeta } - \i 0^+ } 
\label{ecriture variables pour saut FF}
\eeqa
with $x<-q$.  All the rapidities in $\mf{Y}_{\sg}^{\ua/\da}$ are taken generic with the exception of $\nu^{(s)}_b$ which ought to be specialised as stated. 
Above, $\sg \in \{ \pm 1\}$ if $s\geq 2$ and $\sg =1$ if $s=1$. Furthermore, one has 
\beq 
u_r^{\sg} \, = \,  -\e{sgn}\Big( \pi + 2\pi \lfloor \tfrac{r+ \sg}{2\pi}\zeta \rfloor - (r+ \sg)\zeta \Big)\cdot \big(1-\de_{\sg,-}\de_{r,1}\big) \;.
\label{definition parametre u r sigma}
\enq
\end{itemize}
The first three properties, put all together, can be summarised within the representation 
\beq
\mc{F}^{(\ga)}\big( \mf{Y} \big) \; = \;  \pl{r \in \mf{N}_{\e{st}} }{}   \pl{a\not= b}{n_r} \Big(\nu_{a}^{(r)} - \nu_{b}^{(r)} \Big)
\cdot  D_{n_1;n_h}\Big( \big\{  \nu_{a}^{(1)} \big\}_1^{n_1} \, ; \,  \big\{  \mu_{a} \big\}_1^{n_h}  \Big) \cdot  \mc{F}^{(\ga)}_{\e{reg}}\big( \mf{Y} \big)
\label{ecriture comportement local FF smooth a rapidite coincidantes}
\enq
where one has 
\bem
 D_{n_1;n_h}\Big( \big\{  \nu_{a}^{(1)} \big\}_1^{n_1} \, ; \,  \big\{  \mu_{a} \big\}_1^{n_h}  \Big)  \; = \; 
 \pl{a=1}{n_1} \bigg( \f{ \nu^{(1)}_{a} + q }{   \nu_{a}^{(1)} - q   }  \bigg)^{2 \vth (\nu_a^{(1)} \mid \, \mf{Y} )  }  \cdot 
\pl{a=1}{n_h} \bigg( \f{ \mu_{a} - q }{   \mu_{a} + q   }  \bigg)^{2 \vth (\mu_a  \mid \,  \mf{Y})  }  \\
\times \pl{\ups = \pm }{} \Bigg\{  \f{ \pl{a=1}{n_1}(\nu_a^{(1)}-\ups q )  }{   \pl{a=1}{n_h}(\mu_a-\ups q ) }  \Bigg\}^{ 2 \ell_{\ups} } \cdot   \f{   \pl{a \not= b}{n_h} \Big( \mu_{a} - \mu_{b} \Big) \cdot \pl{a \not= b}{n_1} \Big(  \nu^{(1)}_{a} -  \nu^{(1)}_{b} \Big)   }{ \pl{a=1}{n_1}\pl{b=1}{n_h} \big(  \nu^{(1)}_a-\mu_b \big)^2 } \;. 
\label{ecriture fonction singuliere D}
\end{multline}
Above, the exponents $\vth ( \om \mid \mf{Y} )$ are related to the opposite of the thermodynamic limit of the shift function associated with the massive excitation $\mf{Y}$.  
They take the explicit form 
\beq
\vth ( \om \mid \, \mf{Y} ) \, = \,  \sul{ a=1 }{ n_h } \phi_1( \om, \mu_{a} )   
\, + \, \tfrac{ 1 }{ 2 } \op{s}_{\ga}  Z( \om ) \; - \; \sul{  r \in \mf{N}  }{  } \sul{ a=1 }{ n_r } \phi_{r}(\om,\nu_a^{(r)} )
\; - \hspace{-1mm} \sul{ \ups^{\prime} \in \{ \pm \} }{}\ell_{ \ups^{\prime} }  \phi_1(\om, \ups^{\prime} q \, )  \;. 
\label{definition phase habilee totale excitation}
\enq
Here $Z$ is the dressed charge \eqref{definition dressed charge} and $\phi_r$ the dressed phase \eqref{definition dressed phase} of an $r$-string. 
Note that, for generic parameters, the function $ \mc{F}^{(\ga)}_{\e{reg}}\big( \mf{Y} \big)$ appearing in the decomposition  \eqref{ecriture comportement local FF smooth a rapidite coincidantes}
does not vanish when some rapidities coincide or approach the endpoints $\pm q$ of the Fermi zone. Note also that the singularities at $\pm q$ in \eqref{ecriture comportement local FF smooth a rapidite coincidantes} that are present in 
the $D$-factor are reminiscent of the fact that the form factor $ \wh{\msc{F}}^{(\ga)}\big( \mf{R} \big)$ does not admit a uniform in the hole and particle rapidity,
large-$L$ asymptotics. Therefore, these singularities do not correspond to singularities of  $ \wh{\msc{F}}^{(\ga)}\big( \mf{R} \big)$ since, by construction, 
the rapidities $\mu_a$ and $\nu_a^{(1)}$ are at least at distance $\de$ from $\ups q$.

 The discrete form factor density  $\mf{F}_{\ups}\big( \mf{Y}  \mid \mf{Z}^{\ups} \big)$ associated with the massless modes of the model solely depends on the rapidities of the massive modes 
through $\vth_{\ups}\big( \mf{Y} \big)$ which takes the form 
\beq
\vth_{\ups}(\mf{Y}) \, = \, \vth( \ups q\mid \mf{Y}) \, - \, \ups \ell_{\ups}  \;. 
\label{definition exponsant critique ell shifte}
\enq
This discrete form factor reads 
\bem
 \mf{F}_{\ups}\big( \mf{Y}   \mid \mf{Z}^{\ups}_{\mf{Y}} \big) \; = \; 
  \pl{ \mu \in \Ups^{(p)}_{\ups} }{}\ex{  \bs{\mf{g}}_{\ups}(\mu) }       \pl{ \mu \in \Ups^{(h)}_{\ups} }{}\ex{ - \bs{\mf{g}}_{\ups}(\mu) }   \cdot 
\f{ G^2\big(1- \ups   \vth_{\ups}(\mf{Y}) - \ups \nu_{\ups} - \ell_{\ups} \big)     }{ G^2\big(1- \ups   \vth_{\ups}(\mf{Y}) - \ups \nu_{\ups} \big) } \\
\times  \mc{R} \Big(   \mf{Z}^{\ups}_{\mf{Y}} \mid - \ups   \vth_{\ups}(\mf{Y}) - \ups \nu_{\ups} - \ell_{\ups}    \Big)  \cdot \bigg( \f{ 2 \pi }{ L } \bigg)^{  ( \vth_{\ups}(\mf{Y})  + \nu_{\ups} )^2 } \;. 
\label{ecriture explicite FF discret}
\end{multline}
 $ \mf{Z}^{\ups}_{ \mf{Y} }$ has been defined in \eqref{definition des variables Z ups}, while $\bs{\mf{g}}_{\ups}(\mu)$ is the differential operator
\beq
\bs{\mf{g}}_{\ups}(\mu) \; = \; \sul{ \ups^{\prime} = \pm }{} \Big\{ \phi_{1}\big( \, \ups^{\prime} q ,\mu\big)  \, - \,   \phi_{1}\big( \, \ups^{\prime} q  , \ups q \big)   \Big\} \cdot 
\f{ \Dp{}   }{ \Dp{} \nu_{ \ups^{\prime} }  }  \mid_{ \nu_{\ups^{\prime}} = 0  } \;. 
\label{definition explicite operateur diff g}
\enq
Note that the only explicit, \textit{i.e.} non-rescaled as in the definition of $\mf{Z}^{\ups}_{\mf{Y}}$, dependence in \eqref{ecriture explicite FF discret} on the rapidities contained in $\Ups^{(p/h)}_{\ups}$ is through the action of the operator 
$\bs{\mf{g}}_{\ups}$. The ${\bf :} * {\bf :}$ symbol in \eqref{ecriture DA FF apres partition modes massifs et masse nulle} is relatively to the ordering that has to be imposed on the action of this operator: 
all the derivatives should appear to the 
left of the expression in \eqref{ecriture DA FF apres partition modes massifs et masse nulle} prior to be computed and the $\nu_{\pm}\tend 0$ limit should be taken at the very end 
of the calculations. 

Finally, the ratio of Barnes functions $G$ is a normalisation pre-factor while 
$\mc{R} \big( \mf{Z}_{\ups} \mid-\ups   \vth_{\ups}(\mf{Y}) - \ell_{\ups}   \big)$ contains the non-trivial part of the discrete form factor. 
Given a parameter $\nu$ and for $\mf{Z} \, = \, \big\{  \{p_a\}_1^{n_p} ;\{h_a \}_1^{n_h } \big\}$, it takes the form:
\beq
\mc{R}\big( \mf{Z} \mid \nu  \big)  \; = \; 
\bigg( \f{\sin(\pi \nu) }{  \pi } \bigg)^{2 n_h} \f{ \pl{a<b}{n_h} (h_a-h_b)^2 \cdot  \pl{a<b}{n_p} (p_a-p_b)^2 }{  \pl{a=1}{n_h} \pl{b=1}{n_p} (h_a+p_b+1)^2   } 
\cdot  \pl{ a=1 }{n_p} \f{ \Ga^2 ( 1+p_a+\nu ) }{\Ga^2(1+p_a) } \cdot \pl{ a=1 }{n_h} \f{ \Ga^2 ( 1+h_a-\nu ) }{\Ga^2(1+h_a) } \;. 
\label{definition densite R discrete}
\enq

It is important for further purposes to state that the exponent $\vth_{\ups}\big(\mf{Y} \big)$ defined in \eqref{definition exponsant critique ell shifte} and the reduced momentum-energy combination $\msc{U}\big( \mf{Y}, \op{v} \big)$
given in \eqref{definition de energi impuslion reduite condensation massless} satisfy jump discontinuities that are identical to those satisfied by the from factor density $\mc{F}^{(\ga)}\big(\mf{Y}\big)$ \eqref{ecriture conditions saut FF}, namely:
\beq
\vth_{\ups}\big(\mf{Y}^{\ua}_{\sg,s} \big) \; = \; \vth_{\ups}\big(\mf{Y}^{\da}_{\sg,s} \big) \qquad \e{and} \qquad 
\msc{U}\big( \mf{Y}^{\ua}_{\sg,s}, \op{v} \big) \; = \;  \msc{U}\big( \mf{Y}^{\da}_{\sg,s}, \op{v} \big)
\label{ecriture conditions saut pour exposant et impulsion energie}
\enq
with $s \in \mf{N}$, $\sg \in \{\pm 1\}$ and  
where I have used the parametrisation \eqref{ecriture variables pour saut FF}. Note also that although it follows from \eqref{definition r moment habille} that
$\msc{U}(\mf{Y};\op{v})$ has cuts along $\nu_a^{(r)}\in \R^+\pm \tf{ \i r \zeta }{ 2}$  \textit{modulo} $\i\pi$, the function $\exp\big\{ \i \msc{U}(\mf{Y},\op{v}) \big\}$ is continuous 
across these lines. 




\section{The form factor expansion in the thermodynamic limit}
\label{Section Form factor expansion in thermo limit}

\subsection{The auxiliary series $\wh{C}_1^{\, (\ga)}(m,t)$}
\label{SousSection Serie auxiliaire C1}

The two-point functions admit the form factor series expansion over all the excited states $\Ups$ of the model 
\beq
\bra{\Om}\sg_1^{\ga^{\prime}}(t) \sg_{m+1}^{\ga} \ket{\Om} \; = \; \sul{ \Ups }{} \;  \ex{ \i m \wh{\mc{P}}_{\Ups\setminus \Om} - \i t \wh{\mc{E}}_{\Ups\setminus \Om} } 
  \cdot \big| \bra{\Ups} \sg_a^{\ga} \ket{\Om} \big|^2 \;. 
\label{ecriture expression fct 2 pts cmme serie FF}
\enq
Since the Hilbert space of the model is finite dimensional, the form factor expansion is well defined when $L$ is finite. However, convergence related subtleties can arise when sending $L\tend +\infty$
on the level of the form factor series. To avoid some of these complications and regularise the calculation of the thermodynamic limit of the two-point function, 
one should consider the two-point function in the distributional sense with, furthermore, $t$ being continued to the upper half-plane. This improves significantly the convergence properties of the series. 
In particular, the expressions that follow should be understood in the sense of a $\Im(t)\tend 0^+$ limit. 

The presumably existing  infinite volume limit of the two-point function will be denoted as
\beq
\big< \sg_1^{\ga^{\prime}}(t) \sg_{m+1}^{\ga}  \big> \; = \; \lim_{L \tend + \infty} \Big\{ \bra{\Om}\sg_1^{\ga^{\prime}}(t) \sg_{m+1}^{\ga}  \ket{\Om}  \Big\} \;. 
\enq
As already discussed, one expects that, in the thermodynamic limit $L \tend +\infty$, 
\begin{itemize}

\item[$i)$] only those Eigenstates having a finite thermodynamic limit of the 
relative excitation energy above the ground state $\wh{\mc{E}}_{\Ups\setminus \Om}$ should contribute 
to the sum over the excited states defining the form factor expansion of the correlator. This means that, effectively speaking, the sum over the excited states boils down to a summation 
over all integers $n_p^{(\e{tot})},  n_h^{(\e{tot})}$ and $n_r$ subject to the constraint \eqref{definition contrainte sur entiers etats non collapse en mode masse nulle}  
and then to a summation over all the possible choices of integers \eqref{definition contrainte sur entiers labellant rapidities des modes excites}
belonging to the sets \eqref{definition domaine appartenance entiers des modes excites}.

\item[$ii)$] Only the leading large-$L$ behaviour of the summands in 
\eqref{ecriture expression fct 2 pts cmme serie FF} will contribute to the thermodynamic limit of the two-point function.

\end{itemize}

The two above statements were shown to hold true for a generalised free fermion model 
in \cite{KozReducedDensityMatrixAsymptNLSE} and various supporting arguments and checks, in the case of the interacting non-linear Schrödinger model, were given in \cite{KozReducedDensityMatrixAsymptNLSE,KozTerNatteSeriesNLSECurrentCurrent}.

Upon dropping the terms in \eqref{ecriture expression fct 2 pts cmme serie FF} that are irrelevant for the thermodynamic limit, one gets
\beq
\big< \sg_1^{\ga^{\prime}} \sg_{m+1}^{\ga}(t) \big> \; = \; \lim_{L \tend + \infty} \Big\{  \wh{\mc{C}}^{\,(\ga)}_1(m,t)   \Big\}
\enq
where $ \wh{\mc{C}}^{\,(\ga)}_1(m,t) $ contains the summation over the excited states whose excitation integers  $ n_p^{(\e{tot})}, n_h^{(\e{tot})}, \{n_r\}_{r \in \mf{N}_{\e{st}}} $, 
collected in $\bs{n}= \big( n_p^{(\e{tot})}, n_h^{(\e{tot})}, \{n_r\}_{r \in \mf{N}_{\e{st}}} \big)$, belong to 
\beq
\mf{S}_{\e{tot}} \; = \; \bigg\{ \big( n_p^{(\e{tot})}, n_h^{(\e{tot})}, \{n_r\}_{r \in \mf{N}_{\e{st}}} \big) \; : \; 
n_h^{(\e{tot})} \, + \, \op{s}_{\ga} \; = \; n_p^{(\e{tot})} \, + \, \sul{ r\in \mf{N}_{\e{st}} }{} r n_r \;\;, \; n_h^{(\e{tot})} \in \intn{ 0 }{ \varkappa_L } \bigg\}  
\label{ecriture domaine de sommation S tot}
\enq
and with rapidities $\mf{R}_{\Ups}$ as in \eqref{ecriture rapidites macr decrivant etat} and solving the system of equations \eqref{equations pour rapidities part trous et cordes}:
\beq
 \wh{\mc{C}}^{\,(\ga)}_1(m,t) \; = \;  (-1)^{m \op{s}_{\ga}} \sul{   \bs{n}\in \mf{S}_{\e{tot}}   }{}   
\sul{ \mf{R}_{\Ups} }{} \;  \wh{\msc{F}}^{(\ga)}\big( \mf{R}_{\Ups} \big) \cdot \ex{\i m \mc{U}(\mf{R}_{\Ups}, \op{v}) } \;. 
\label{definition premiere du correlateur effectif C1}
\enq
Note that $\op{s}_{\ga}$ appearing in \eqref{ecriture domaine de sommation S tot}-\eqref{definition premiere du correlateur effectif C1} corresponds to the pseudo-spin of the operator $\sg_{1}^{\ga}$
which has been defined in \eqref{definition spin operateur}.  Finally, the function $\mc{U}(\mf{R}_{\Ups}, \op{v})$ appearing in \eqref{definition premiere du correlateur effectif C1} has been defined in \eqref{definition U des rapidites}.

By using the analiticity of the summand, the summation over the excited states in \eqref{definition premiere du correlateur effectif C1} can be recast in terms of contour integrals 
by means of the multidimensional residue theorem \cite{AizenbergYuzhakovIntRepAndMultidimensionalResidues}: 
\beq
 \wh{\mc{C}}^{\,(\ga)}_1(m,t) \; = \;   (-1)^{m \op{s}_{\ga}}  \sul{   \bs{n}\in \mf{S}_{\e{tot}}  }{}   
 \pl{ r \in  \mf{N}_{st} }{} \;\Bigg\{  \Oint{ \Ga^{(r)}_{\e{tot}} }{}  \mc{D}^{n_r} \nu^{(r)} \Bigg\} 
 \cdot \Oint{ \Ga^{(p)}_{\e{tot}} *  \Ga^{(h)}_{\e{tot}} }{}  \mc{D}^{n_p^{(\e{tot})} } \hspace{-1mm} \la^{(p)}  \cdot     \mc{D}^{n_h^{(\e{tot})}} \hspace{-1mm} \la^{(h)} 
 \cdot \wh{\mc{H}}\big( \mf{R} \big) \cdot \wh{\msc{F}}^{(\ga)}\big( \mf{R} \big) \cdot \ex{\i m \mc{U}(\mf{R}, \op{v}) } \;. 
\label{ecriture C1 en serie int mult}
\enq
Above, $\mf{R}$ contains all the integration variables as in \eqref{ecriture rapidites macr cas generique sous integrale}, while the integration measures read
\beq
\mc{D}^{n_r} \nu^{(r)} \; = \; \pl{a=1}{n_r} \bigg\{  \f{ L \, \wh{\xi}_{r}^{\prime}\big(\nu_a^{(r)} \mid \mf{R} \big) }{  \ex{\i L \wh{\xi}_{r}(\nu_a^{(r)} \mid \mf{R} )} -1  }   \bigg\} \cdot \f{ \dd^{n_r} \nu^{(r)} }{ n_r! \, (2\pi)^{n_r} }  
\quad , \quad
\mc{D}^{ n_p^{(\e{tot})} } \hspace{-1mm} \la^{(p)} \; = \; \pl{a=1}{ n_p^{(\e{tot})} } \bigg\{  \f{ L \, \wh{\xi}_{1}^{\prime}\big(\la_a^{(p)} \mid \mf{R} \big) }{  \ex{\i L \wh{\xi}_{1}(\la_a^{(p)} \mid \mf{R} )}  -1 }   \bigg\} 
\cdot \f{ \dd^{ n_p^{(\e{tot})} } \la^{(p)} }{ n_p^{(\e{tot})} ! \, (2\pi)^{ n_p^{(\e{tot})} } }  
\enq
and $\mc{D}^{n_h^{(\e{tot})} } \hspace{-1mm}\la^{(h)} $ is obtained from $\mc{D}^{n_p^{(\e{tot})} } \hspace{-1mm} \la^{(p)} $ by the exchange of superscripts $p \leftrightarrow h$. The counting functions
arising in the definition of the measures are as defined in \eqref{equations pour rapidities part trous et cordes}.   $\wh{\mc{H}}\big( \mf{R} \big) $ is there so as to compensate for the Jacobian
arising from the computation of the multidimensional residues
\beq
\wh{\mc{H}}\big( \mf{R} \big) \; = \; \pl{ a=1 }{ n_h^{(\e{tot})} } \Bigg\{ \f{  \ex{-\i L \, \wh{\xi}_{1}(\la_a^{(h)} \mid \, \mf{R} )}   }{ \i L \, \wh{\xi}^{\, \prime}_1( \la_{a}^{(h)} \mid \mf{R} )  } \Bigg\} 
\cdot \pl{ a=1 }{ n_p^{(\e{tot})} } \Bigg\{ \f{ \ex{-\i L \, \wh{\xi}_{1}(\la_a^{(p)} \mid \, \mf{R} )}    }{ \i L \, \wh{\xi}^{\, \prime}_1( \la_{a}^{(p)} \mid  \mf{R}  )  } \Bigg\} \cdot 
 \prod_{ r\in \mf{N}_{\e{st}}   }^{ } \pl{ a=1 }{ n_r }  \Bigg\{ \f{  \ex{-\i L \, \wh{\xi}_{r}(\nu_a^{(r)} \mid \, \mf{R} )}  }{ \i L \, \wh{\xi}^{\, \prime}_r\big( \nu_{a}^{(r)} \mid \mf{R}  \big) }  \Bigg\}  
 \cdot \det \big[ \op{D}{\bs \vp }\big]  \;,
\enq
where $ \op{D}{\bs \vp }$ is the differential of the map: 
\beq
{\bs \vp } \; :  \; \bs{r} \;  \mapsto  \;
\Bigg( \Big(  \ex{\i L \, \wh{\xi}_{1}(\la_a^{(p)} \mid \, \mf{R} )}  -1   \Big)_{a=1}^{ n_p^{(\e{tot})} } \, , \,   \Big(  \ex{\i L \, \wh{\xi}_{1}(\la_a^{(h)} \mid \, \mf{R} )}  -1  \Big)_{a=1}^{ n_h^{(\e{tot})} }  \Bigg) \, , \, 
\Bigg( \Big(  \ex{\i L \, \wh{\xi}_{r}(\nu_a^{(r)} \mid \, \mf{R} )} -1  \Big)_{a=1}^{ n_r } \Bigg)_{ r \in \mf{N}_{\e{st}}  }  \Bigg) \;. 
\enq
Here, $\bs{r}=(\bs{\la}^{(p)},\bs{\la}^{(h)}, \bs{\nu}^{(r_1)}, \dots, \bs{\nu}^{(r_{\e{max}})}) $ with $\mf{N}_{\e{st}}=\big\{ r_1, \dots, r_{\e{max}} \big\}$ is the vector whose 
entries are the $n_{p/h}^{(\e{tot})}$-dimensional vectors 
\beq
\bs{\la}^{(p/h)}=\big( \la^{(p/h)}_1, \dots , \la^{(p/h)}_{n_{p/h}^{(\e{tot})} } \big) 
\label{definition des vecteurs part trous}
\enq
and the $n_r$-dimensional vectors $\bs{\nu}^{(r)}=\big(\nu_{1}^{(r)},\cdots, \nu_{ n_r }^{(r)} \big)$  whose corrdinates issue from the elements of the respective sets in $\mf{R}$. 
Note that the order of choosing the coordinates is irrelevant in that all the functions of interest are symmetric in the coordinates of $\bs{\la}^{(p)}$, $\bs{\la}^{(h)}$, or any of the $\bs{\nu}^{(r)}$
taken singly. In the following, such an identification between vectors and a collection of sets $\mf{R}$ will be made sometimes, this without further notice. Also, one should note that since 
$|\bs{n}| = n_p^{ (\e{tot}) } + n_h^{ (\e{tot}) } + \sul{ r \in \mf{N}_{\e{st}} }{} n_r$ is finite for each $\bs{n}\in \mf{S}_{\e{tot}}$, 
$\bs{r}=(\bs{\la}^{(p)},\bs{\la}^{(h)}, \bs{\nu}^{(r_1)}, \dots, \bs{\nu}^{(r_{\e{max}})}) $
has, effectively speaking, a finite number of coordinates.

The integration \eqref{ecriture C1 en serie int mult} runs along two kinds of contours:
\begin{itemize}
 \item[$\bullet$] The contour $\Ga^{\,(r)}_{\e{tot}}=\big( \Ga^{(r)} \big)^{n_r}$ is associated with a summation over the possible choices of rapidities for the $r$-bound states. 
 The contour $\Ga^{(r)}$ is depicted in Figure \ref{Figure contour Gamma r string et sa decomposition} and consists of a small counterclockwise loop around $ \R \, + \, \i \tfrac{\pi}{2} \de_{r}$.
 \item[$\bullet$] The integration domain $\Ga^{(h)}_{\e{tot}} * \Ga^{(p)}_{\e{tot}}$ is associated with a summation over all the possible choices of the hole, resp. particle, rapidities. 
It corresponds to an integration over an $n_p^{(\e{tot})}+n_h^{(\e{tot})}$ real dimensional sub-manifold of $\Cx^{ n_p^{(\e{tot})}+n_h^{(\e{tot})} }$. To define it, 
one first fixes a given value of the $r$-string rapidities $ \big\{ \{ \nu_a^{(r)} \}_{a=1}^{n_r} \big\}_{r \in \mf{N}_{\e{st}} } $. 
Then one introduces the map 
\beq
\bs{\psi}\; : \; \big(\bs{\la}^{(p)}, \bs{\la}^{(h)} \big) \; \mapsto  \; 
\Big( \, \wh{\xi}_{1}\big(\la_1^{(p)}\mid \mf{R} \big) , \dots,  \wh{\xi}_{1}\big(\la^{(p)}_{n_p^{(\e{tot})} }\mid \mf{R} \big)  , 
 \, \wh{\xi}_{1}\big(\la_1^{(h)}\mid \mf{R} \big) , \dots,  \wh{\xi}_{1}\big(\la^{(h)}_{n_h^{(\e{tot})} }\mid \mf{R} \big) \Big)
\label{definition map psi small}
\enq
where the $r$-string rapidities entering in the definition of $\mf{R}$ are given by the above choice and where 
the vectors $\bs{\la}^{(p/h)}$ are as in \eqref{definition des vecteurs part trous} and built from the respective variables present in $\mf{R}$. 
Then,  $\Ga^{(p)}_{\e{tot}} * \Ga^{(h)}_{\e{tot}}$ is defined as the pre-image
\beq
\Ga^{(p)}_{\e{tot}} *  \Ga^{(h)}_{\e{tot}} \; =  \; \bs{\psi}^{-1}\Big( \big(\mf{C}^{(p)}\big)^{ n_p^{(\e{tot})} } \times \big(\mf{C}^{(h)}\big)^{ n_h^{(\e{tot})} }  \Big)
\enq
where the one-real dimensional contours $\mf{C}^{(p/h)}$ are as defined on Figure~\ref{Figure contour Gamma p h dans espace cible}.
Although the submanifold $\Ga^{(p)}_{\e{tot}} * \Ga^{(h)}_{\e{tot}}$ is \textit{not} a Carthesian product, effectively speaking and owing to the Carthesian product structure in the target space of $\bs{\psi}$, 
one can think of each particle, resp. hole, rapidity as being integrated on the one-dimensional contour depicted in Figure~\ref{Figure contour Gamma p et h et leurs decompositions}. 
In fact, such a reduction occurs in the large-$L$ limit upon replacing $\wh{\xi}_1$ by $p_1$. 
\end{itemize}

\begin{figure}[ht]
\begin{center}

\begin{pspicture}(4.5,2.5)

\psline[linestyle=dashed, dash=3pt 2pt](0,1)(4,1)
\psline[linewidth=2pt]{->}(4,1)(4.1,1)

 \rput(5,1){$\R + \i \de_{r} \tf{\pi}{2}$}

 \psline[]{<->}(0.5,1)(0.5,1.5)
\rput(0.2,1.2){$\a$}

 \psline[]{<->}(3,1)(3,0.5) 
\rput(2.8,0.7){$\a$} 
 
\psline(0,1.5)(4,1.5)
\psline(0,0.5)(4,0.5)
\psline[linewidth=2pt]{->}(2,1.5)(1.9,1.5)
\psline[linewidth=2pt]{->}(2,0.5)(2.1,0.5)

\rput(3.5,0.2){$\Ga^{(r)}_-$}
\rput(3.5,1.8){$\Ga^{(r)}_+$}

\end{pspicture}
\caption{ Contour $\Ga^{(r)} = \Ga_+^{(r)}\cup \Ga_-^{(r)}$, $r \in \mf{N}_{\e{st}}$. These contours are taken as $\Ga^{(r)}_{\eps}=-\eps \R \, + \, \i \de_r \tfrac{\pi}{2}+\i\eps \a $ for some $\a>0$ and small enough. 
\label{Figure contour Gamma r string et sa decomposition} }
\end{center}

\end{figure}
\begin{figure}[ht]
\begin{center}

\begin{pspicture}(12,2.5)

\psline[linestyle=dashed, dash=3pt 2pt](0,1.25)(11.8,1.25)
\psline[linewidth=2pt]{->}(11.8,1.25)(11.9,1.25)

\psline[linestyle=dashed, dash=3pt 2pt](11.3,2)(11.5,2)
\psline[linestyle=dashed, dash=3pt 2pt](11.3,0.5)(11.5,0.5)

 \rput(12.1,1.25){$\R $}

 \rput(12,2){$\R + \i\de  $}
 \rput(12,0.5){$\R  - \i\de  $}
 
\psdots(0.75,1.25)(3,1.25)(3.25,1.25)(5.75,1.25)(6.25,1.25)(8.25,1.25) (8.75,1.25)(11,1.25)

 \psline(0.75,0.5)(0.75,2)
 \psline(0.75,2)(3,2)
 \psline(3,2)(3,0.5)
 \psline(3,0.5)(0.75,0.5)
 \psline[linewidth=2pt]{->}(2.2,2)(2.1,2)

 \psline(5.75,0.5)(5.75,2)
 \psline(5.75,2)(3.25,2)
 \psline(3.25,2)(3.25,0.5)
 \psline(3.25,0.5)(5.75,0.5)
\psline[linewidth=2pt]{->}(4,2)(3.9,2)

 \psline(5.75,0.5)(5.75,2)
 \psline(5.75,2)(3.25,2)
 \psline(3.25,2)(3.25,0.5)
 \psline(3.25,0.5)(5.75,0.5)
\psline[linewidth=2pt]{->}(4,2)(3.9,2)

 \psline(6.25,0.5)(6.25,2)
 \psline(6.25,2)(8.25,2)
 \psline(8.25,2)(8.25,0.5)
 \psline(8.25,0.5)(6.25,0.5) 
\psline[linewidth=2pt]{->}(6.75,0.5)(6.85,0.5)
 
 \psline(8.75,0.5)(8.75,2)
 \psline(8.75,2)(11,2)
 \psline(11,2)(11,0.5)
 \psline(11,0.5)(8.75,0.5) 
\psline[linewidth=2pt]{->}(9.5,2)(9.4,2)

\rput(2,2.3){$\mf{C}^{(p)}_{1;+}$}
\rput(2,0.1){$\mf{C}^{(p)}_{1;-}$}

\rput(4.5,2.3){$\mf{C}^{(p)}_{0;+}$}
\rput(4.5,0.1){$\mf{C}^{(p)}_{0;-}$}

\rput(10,2.3){$\mf{C}^{(p)}_{0;+}$}
\rput(10,0.1){$\mf{C}^{(p)}_{0;-}$}

\rput(7.5,2.3){$\mf{C}^{(h)}_+$}
\rput(7.5,0.1){$\mf{C}^{(h)}_-$}

\rput(-0.2,1.6){$ \tfrac{2\pi }{ L} \big( M_- \!-\!  \tfrac{1}{2}\big) $}
\rput(1.9,0.9){$\tfrac{2\pi }{ L} \big( M_-^{(\mf{r})}  \!+\! \tfrac{1+ 0^+}{2}\!\big)$}

\rput(4.4,1.6){$ \tfrac{2\pi }{ L} \big( M_-^{(\mf{r})} \!+\!  \tfrac{1-0^+}{2}\!\big) $}
\rput(10.1,1.6){$\tfrac{2\pi }{ L} \big( M_+  \!+\!  \tfrac{1}{2}\big)$}

\rput(5.45,0.9){$\tfrac{2\pi }{4 \cdot L}  $}
\rput(6.55,0.9){$\tfrac{2\pi \cdot 3}{4 \cdot L}  $}

\rput(7.4,1.6){$\tfrac{2\pi }{ L} \big( N_+ \!-\!  \tfrac{3}{4}\big)  $}
\rput(9.6,0.9){$\tfrac{2\pi }{ L} \big( N_+ \!-\!  \tfrac{1}{4}\big)  $}

\end{pspicture}
\caption{ Contours $\mf{C}^{(p)} \, = \, \bigcup_{\ups=\pm} \Big\{ \mf{C}^{(p)}_{1;\ups} \cup \mf{C}^{(p)}_{0;\ups} \Big\}$ and 
$\mf{C}^{(h)} \, = \, \bigcup_{\ups=\pm} \mf{C}^{(h)}_{\ups} $ arising in the definition of $\Ga^{(p)}_{\e{tot}}$ and $\Ga^{(h)}_{\e{tot}}$. Here $N_+=N+1+\op{s}_{\ga} $ and $\de>0$ is some fixed
but sufficiently small parameter. The contour $ \bigcup_{\ups=\pm}  \mf{C}^{(p)}_{0;\ups}$ corresponds to integrating around solutions to $ \ex{\i L \, \wh{\xi}_1( \om \mid \, \mf{R} ) }-1=0 $
that are very close to $\R$ while the contour $ \bigcup_{\ups=\pm}  \mf{C}^{(p)}_{1;\ups}$ corresponds to integrating around solutions that are close to 
$\R+\i\tfrac{\pi}{2}$. These contours are infinitesimally close to each other as indicate the $\pm 0^+$ shifts.  
\label{Figure contour Gamma p h dans espace cible} }
\end{center}

\end{figure}
\begin{figure}[ht]
\begin{center}

\begin{pspicture}(8.5,3.5)

\psline[linestyle=dashed, dash=3pt 2pt](0,1)(7.5,1)
\psline[linewidth=2pt]{->}(7.5,1)(7.6,1)

 \rput(7.8,1){$\R$}

 \psdots(2.2,1)(2.6,1)(5.2,1)(5.6,1)

\rput(3,0.75){$\wh{q}^{\,-}_{h}$}
\rput(4.9,0.75){$\wh{q}^{\, +}_{h}$}

\rput(1.9,1.25){$\wh{q}^{\, -}_{p}$}
\rput(5.9,1.25){$\wh{q}^{\, + }_{p}$}

\psline[linestyle=dashed, dash=3pt 2pt](0,3)(7.5,3)
\psline[linewidth=2pt]{->}(7.5,3)(7.6,3)

 \rput(8.2,3){$\R + \i \tfrac{\pi}{2}$}

\psline(0,3.25)(7.5,3.25)
\psline(0,2.75)(7.5,2.75) 
\psline[linewidth=2pt]{->}(4,3.25)(3.9,3.25)
\psline[linewidth=2pt]{->}(4,2.75)(4.1,2.75)

\rput(2,3.6){$\Ga^{(p)}_{+;1}$}
\rput(5,2.4){$\Ga^{(p)}_{-;1}$}

\psline(2.2,0.5)(2.2,1.5)
\psline(0,0.5)(2.2,0.5)
\psline(0,1.5)(2.2,1.5)
\psline[linewidth=2pt]{->}(1,1.5)(0.9,1.5)
\psline[linewidth=2pt]{->}(1,0.5)(1.1,0.5)

\rput(-0.2,0.5){$\Ga^{(p)}_{-;0}$}
\rput(-0.25,1.5){$\Ga^{(p)}_{+;0}$}

\psline(5.6,0.5)(5.6,1.5)
\psline(5.6,0.5)(7.5,0.5)
\psline(5.6,1.5)(7.5,1.5)
\psline[linewidth=2pt]{->}(6.5,1.5)(6.4,1.5)
\psline[linewidth=2pt]{->}(6.5,0.5)(6.6,0.5)

\rput(7.9,0.5){$\Ga^{(p)}_{-;0}$}
\rput(7.9,1.5){$\Ga^{(p)}_{+;0}$}

\psline(2.6,0.5)(2.6,1.5)
\psline(5.2,0.5)(2.6,0.5)
\psline(2.6,1.5)(5.2,1.5)
\psline(5.2,0.5)(5.2,1.5)
\psline[linewidth=2pt]{->}(3.5,1.5)(3.4,1.5)
\psline[linewidth=2pt]{->}(3.5,0.5)(3.6,0.5)

\rput(4.3,0.1){$\Ga^{(h)}_{-}$}
\rput(4.3,1.8){$\Ga^{(h)}_{+}$}

\end{pspicture}
\caption{ Contours $\Ga^{(p)} = \Ga_{+;0}^{(p)}\cup \Ga_{-;0}^{(p)}\cup \Ga_{+;1}^{(p)}\cup \Ga_{-;1}^{(p)}$ and $\Ga^{(h)} = \Ga_{+}^{(h)}\cup \Ga_{-}^{(h)}$.  Here, the endpoints of integration are defined as
$\wh{q}^{\, -}_p=p_1^{-1}\Big( \tfrac{2\pi }{4 L} \Big)$,  $\wh{q}^{\, -}_h=p_1^{-1}\Big( \tfrac{2\cdot 3 \pi }{4 L} \Big)$, 
$\wh{q}^{\, +}_h=p_1^{-1}\Big( \tfrac{2\pi }{ L} \big( N_{+}-\tfrac{3}{4}\big) \Big)$,  $\wh{q}^{\, +}_p=p_1^{-1}\Big( \tfrac{2\pi }{ L} \big( N_{+}-\tfrac{1}{4}\big) \Big)$. 
\label{Figure contour Gamma p et h et leurs decompositions} }
\end{center}

\end{figure}

\subsection{The auxiliary series $\wh{C}_2^{\, (\ga)}(m,t)$}
\label{Sousection serie auxiliare C2}

The series of multiple contour integrals for $ \wh{\mc{C}}^{\,(\ga)}_1(m,t) $ has still to be transformed so as to be able to compute its thermodynamic limit. 
To start with, I shall argue that one can drop several terms that should not contribute to its thermodynamic limit. 

First of all, for fixed $\bs{n}\in \mf{S}_{\e{tot}}$, the total number of variables in the associated multiple integral appearing in \eqref{ecriture C1 en serie int mult}
is 
\beq
|\bs{n}| \, = \, n_p^{(\e{tot})} \, + \, n_h^{(\e{tot})} \, +\,  \sul{ r \in \mf{N}_{\e{st}} }{} n_r \;.  
\enq
Hence $\det \big[ \op{D}{\bs \vp }\big]$ is the determinant of a finite matrix. Upon using the explicit expression for the counting functions 
given in \eqref{equations donnant fct cptge r}, and computing the matrix representing the differential, one gets that the matrix has the block structure
\beq
\op{D}{\bs \vp } \, = \, \left( \ba{ccccc}         \op{M}_{pp}	& 	\op{M}_{ph} & \dots & \op{M}_{ p r^{\prime} }	& \dots  \\  
						    \op{M}_{hp}	& 	\op{M}_{hh} & \dots & \op{M}_{ h r^{\prime} }	& \dots  \\
							\vdots &   \vdots         &     \ddots & \dots 	& \dots    \\
							\op{M}_{rp}	& 	\op{M}_{rh} & \vdots & \op{M}_{ r r^{\prime}}	& \dots  \\
						      \vdots &   \vdots         &     \vdots & \vdots 	& \vdots      \ea  \right)
\label{ecriture matrice par blocks jacobien vp}
\enq
where $\op{M}_{pp}$ is a $n_p^{(\e{tot})} \times n_p^{(\e{tot})}$ matrix, $\op{M}_{ph}$ is a $n_p^{(\e{tot})} \times n_h^{(\e{tot})}$ matrix, 
 $\op{M}_{ p r^{\prime} }$ is a $n_p^{(\e{tot})} \times n_{r^{\prime}}$ matrix \textit{etc}. Further, the indices $r, r^{\prime}$ in \eqref{ecriture matrice par blocks jacobien vp} 
 run through $\{ r \in \mf{N}_{\e{st}} \, : \, n_r\not= 0 \}$,  while the blocks building up the first block line of the matrix read:
\beq
 \Big( \op{M}_{pp} \Big)_{ab} \, = \, \Big\{ \i L \de_{ab} \wh{\xi}^{\, \prime}_1\big(\la_a^{(p)} \mid \mf{R} \big) -\i \Dp{ \la_a^{(p)} }   \wh{F}_1(\om \mid \mf{R})_{\mid \om=\la_b^{(p)} } \Big\}
\cdot  \ex{ \i L \wh{\xi}_1 (\la_a^{(p)} \mid \, \mf{R}  ) } \; , 
\enq
and 
\beq
 \Big( \op{M}_{ph} \Big)_{ab} \, = \,  -\i \Dp{ \la_a^{(p)} }   \wh{F}_1(\om \mid \mf{R})_{\mid \om=\la_b^{(h)} } 
\cdot  \ex{ \i L \wh{\xi}_1 (\la_a^{(h)} \mid \, \mf{R} ) } \quad ,  \qquad 
 \Big( \op{M}_{pr} \Big)_{ab} \, = \,  -\i \Dp{ \la_a^{(p)} }   \wh{F}_r(\om \mid \mf{R})_{\mid \om=\nu_b^{(r)} } 
\cdot  \ex{ \i L \wh{\xi}_r(\nu_b^{(r)} \mid \, \mf{R} ) }  \;. 
\enq
Analogours expressions hold for the other block matrices building up $\op{D}{\bs \vp }$. 
Thus, upon factorising the entries appearing on the diagonal, one gets that 
\beq
\wh{\mc{H}}\big( \mf{R} \big) \, = \, \det\Big[ \op{I}_{|\bs{n}|}+ \tfrac{1}{L}\op{M}\big( \mf{R} \big) \Big]
\enq
where $ \op{I}_{|\bs{n}|}$ is the identity matrix in dimension $|\bs{n}|$ and $\op{M}\big( \mf{R} \big)$
is an $|\bs{n}|\times |\bs{n}|$ is a matrix with bounded in $L$ coefficients for bounded rapidities $\mf{R}$. 
Thus, the below pointwise limit holds
\beq
\lim_{L\tend + \infty} \Big\{ \wh{\mc{H}}\big( \mf{R} \big)  \Big\} \; = \; 1 \;. 
\label{ecriture limite determinant H R}
\enq
It is thus reasonable to expect that one can replace $\wh{\mc{H}}\big( \mf{R} \big)$ by $1$ in the series expansion for $ \wh{\mc{C}}^{\,(\ga)}_{1}(m,t) $ 
without altering the value of its thermodynamic limit. This assumption will be made in the following. We stress that this assumption is basically
equivalent to point $ii)$ mentioned in Sub-Section \ref{SousSection Serie auxiliaire C1}, since $\wh{\mc{H}}\big( \mf{R} \big)$ corresponds  to  $1+\e{O}\big(1/L\big)$
corrections.

\vspace{3mm}

Second, observe that the one-dimensional $r$-string contour integral can be decomposed as 
\beq
\Oint{ \Ga^{(r)} }{}   \f{   \wh{\xi}_{r}^{\, \prime}\big(\nu \mid \mf{R} \big)   f(\nu)}{  \ex{\i L \, \wh{\xi}_{r}(\nu \, \mid \,  \mf{R} )} -1  } \cdot  \dd \nu   \; = \; 
-\Int{ \Ga^{\, (r)}_{s_r}  }{}   \wh{\xi}_{r}^{\, \prime}\big(\nu \mid \mf{R} \big) f(\nu)  \cdot \dd \nu   \; + \; 
\sul{\eps = \pm }{} \Int{ \Ga^{\, (r)}_{ \eps s_r }  }{}     \f{ - \eps  \,  \wh{\xi}_{r}^{\, \prime}\big(\nu \mid \mf{R} \big)  f(\nu)  }{  \ex{ -\i \eps L \wh{\xi}_{r}(\nu \, \mid \,  \mf{R} ) } -1  } \cdot \dd \nu 
\label{decomposition integrale sur Ga r string}
\enq
where the contours $ \Ga^{\, (r)}_{ \eps s_r }$ are defined in Figure~\ref{Figure contour Gamma r string et sa decomposition}, 
\beq
s_r = \e{sgn}\big[ p^{\prime}_r\big]_{\mid \R + \i \de_{r} \tf{\pi}{2} }  
\enq
and $f$ is some holomorphic function inside of $\Ga^{(r)}$ such that both $f$ and $f^{\prime}$ decay at least as $\ex{-2|\Re(\nu)| }$ at $\infty$. 
Here, $\nu$ corresponds to any of the $r$-string variables contained in $\mf{R}$. All the other variables present in $\mf{R}$
are assumed to belong to their respective domains of integration, as appearing in \eqref{ecriture C1 en serie int mult}. 
It is then straightforward to estimate the large-$L$ behaviour of the sum over $\eps=\pm$ in \eqref{decomposition integrale sur Ga r string}
by using the properties of the dressed momentum $p_r$ and the form of the counting function. An integration by parts yields 
\beq
\sul{\eps = \pm }{} \Int{ \Ga^{\, (r)}_{ \eps s_r }  }{}     \f{ - \eps  \,  \wh{\xi}_{r}^{\, \prime}\big(\nu \mid \mf{R} \big)  f(\nu)  }{  \ex{ -\i \eps L \wh{\xi}_{r}(\nu \, \mid \,  \mf{R} ) } -1  } \cdot \dd \nu 
\; = \; 
\sul{\eps = \pm }{} \f{\i}{L}  \Int{ \Ga^{\, (r)}_{ \eps s_r }  }{}   f^{\prime}(\nu) \cdot   \ln \Big[ 1-  \ex{ \i \eps L \wh{\xi}_{r}(\nu \, \mid \,  \mf{R} ) }  \Big]    \cdot \dd \nu \;. 
\enq
It is then readily seen from the explicit expressions for the building blocks for $\wh{\xi}_r$ \eqref{equations donnant fct cptge r}, \eqref{asymptotiques fct shift} and \eqref{definition r moment habille}, 
that for $\Re(\om) \tend  \pm \infty$ with $\om$ belonging to an open neighbourhood of $\R+\i \tf{ \pi \de_r }{ 2 }$, it holds
\beq
\wh{\xi}_{r}(\om \, \mid \,  \mf{R} ) \; = \; c_{r}^{\pm} \, + \, \wt{C}_{r}^{\pm}\cdot \ex{\mp 2 \om}  \; + \, \e{O}\Big( \ex{\mp 4 \om}  \Big) \qquad \e{with} \qquad 
 c_{r}^{\pm} \, , \,  \wt{C}_{r}^{\pm} \in \R   \;. 
\enq
Furthermore, \eqref{equations donnant fct cptge r} and the defintion of $s_r$ allow one to conclude that 
\beq
s_r= - \e{sgn}\Big( 2 (-1)^{\de_r} \wt{C}_{r}^{\pm} \Big)  \;. 
\enq

According to Fig.~\ref{Figure contour Gamma r string et sa decomposition}, $\om \in \Ga^{\, (r)}_{ \eps s_r }$ may be parameterised as 
\beq
\om=\pm x + \i \tfrac{ \pi \de_r }{ 2 } +\i \eps s_r \a \; , 
\label{parametrisation omega pour courbe voisinage r corde}
\enq
for some $\a>0$  and small enough and with $x \geq 0$. Then, for such $\om$'s, one has
\beq
\i \eps \, \wh{\xi}_{r}(\om \, \mid \,  \mf{R} ) \; = \; \i\eps c_{r}^{\pm} \, - \i \eps s_r \, |\wt{C}_{r}^{\pm}| \cdot \ex{- 2 x  - 2\i \eps s_r \a}  \; + \, \e{O}\Big( \ex{- 4 x}  \Big) \;. 
\enq
This entails that there exists $M_0>0$ large enough such that it holds
\beq
\Re\Big( \i \eps \, \wh{\xi}_{r}(\om \, \mid \,  \mf{R} ) \Big) \; \leq  \;  - \f{ |\wt{C}_{r}^{\pm}| }{ 2} \sin(2\a) \cdot \ex{- 2 x }  \quad \e{for} \qquad x \geq \ln M_0 \;. 
\enq
Thus, for such $x$'s, one has 
\beq
\Big|   \ln \Big[ 1-  \ex{ \i \eps L \wh{\xi}_{r}(\om \, \mid \,  \mf{R} ) }  \Big]  \, \Big| \; \leq \; 
-  \ln \Big[ 1-  \exp\big[  - \kappa_{\pm} \cdot \ex{- 2 x } \big]   \Big]  \qquad \e{with} \qquad \kappa_{\ups}\;= \; \f{ |\wt{C}_{r}^{\ups}| }{ 2} \sin(2\a)  \;, 
\label{ecriture borne sur logarithme avec fct cptge}
\enq
and $\om$ as in \eqref{parametrisation omega pour courbe voisinage r corde}. 
Therefore, using that $\Ga^{\, (r)}_{ \eps s_r } \, = \, \R + \i\tfrac{\pi}{2} \de_r +  \i \eps s_R \a$ for some $\a>0$ and small enough, one has the decomposition 
\bem
  \Int{ \Ga^{\, (r)}_{ \eps s_r }  }{}   f^{\prime}(\nu) \cdot   \ln \Big[ 1-  \ex{ \i \eps L \wh{\xi}_{r}(\nu \, \mid \,  \mf{R} ) }  \Big]    \cdot \dd \nu \; = \; 
\Int{ -\ln M_0 + \i\eps s_r \a  }{ \ln M_0 + \i\eps s_r \a } \hspace{-3mm}  f^{\prime}(\nu) \cdot   \ln \Big[ 1-  \ex{ \i \eps L \wh{\xi}_{r}(\nu \, \mid \,  \mf{R} ) }  \Big]    \cdot \dd \nu \\
\; + \; \sul{\ups = \pm}{}  \; \ups \hspace{-3mm} \Int{ \ups \ln M_0 + \i\eps s_r \a  }{ \ups \infty+ \i\eps s_r \a } \hspace{-3mm}
      f^{\prime}(\nu) \cdot   \ln \Big[ 1-  \ex{ \i \eps L \wh{\xi}_{r}(\nu \, \mid \,  \mf{R} ) }  \Big]    \cdot \dd \nu \;. 
\end{multline}
The first term in the \textit{rhs} produces exponentially small corrections in $L$ since the bound \eqref{ecriture equation positivite pr prime} implies
that, for $L$ large enough,  
\beq
\Re\Big( \i \eps \, \wh{\xi}_{r}(\nu \, \mid \,  \mf{R} ) \Big) \; < \; 0 \qquad \e{uniformly} \; \e{in} \quad \nu \in \Big\{ \intff{ -\ln M_0   }{ \ln M_0  }   \, + \i\eps s_r \a \Big\}  \;. 
\enq
Furthermore, the bound \eqref{ecriture borne sur logarithme avec fct cptge} and the fact that for the concerned $\nu$'s  it holds 
$|f^{\prime}(\nu) | \leq C_f \ex{-2x}$  for some $f$-dependent constant $C_f>0$, entail that 
\bem
\Bigg| \Int{ \ups \ln M_0 + \i\eps s_r \a  }{ \ups \infty+ \i\eps s_r \a } \hspace{-3mm}  f^{\prime}(\nu) \cdot   \ln \Big[ 1-  \ex{ \i \eps L \wh{\xi}_{r}(\nu \, \mid \,  \mf{R} ) }  \Big]    \cdot \dd \nu  \; \Bigg|
\; \leq \; 
- C_f \! \Int{ \ln M_0}{ + \infty} \! \ex{-2x} \ln \Big[ 1\, - \, \ex{ -\kappa_{\ups} L \ex{-2 x} } \Big] \cdot \dd x = \e{O}\Big(L^{-1} \Big)\;.  
\end{multline}
All-in-all, this allows one to conclude that 
\beq
\Oint{ \Ga^{(r)} }{}   \f{   \wh{\xi}_{r}^{\, \prime}\big(\nu \mid \mf{R} \big)   f(\nu)}{  \ex{\i L \, \wh{\xi}_{r}(\nu \, \mid \,  \mf{R} )} -1  } \cdot  \dd \nu   \; = \; 
-\Int{ \Ga^{\, (r)}_{s_r}  }{}   \wh{\xi}_{r}^{\, \prime}\big(\nu \mid \mf{R} \big) f(\nu)  \cdot \dd \nu   \; + \; \e{O}\Big(L^{-1} \Big) \;. 
\enq

 \vspace{3mm}

It is readily seen that the same property does hold for a $n_r$-dimensional analogue of the integral. More precisely, given a symmetric function $f$ and  
understanding that  $\bs{\nu} =(\nu_1,\dots,\nu_{n_r} )$ corresponds to the $r$-string rapidities present in $\mf{R}$ while all the other variables present in $\mf{R}$
belong to their respective domains of integration as given in \eqref{ecriture C1 en serie int mult},  one has 
\bem
\Oint{    \Ga^{(r)}_{\e{tot}}    }{}  \pl{a=1}{n_r} \bigg\{ \f{   \wh{\xi}_{r}^{\, \prime}\big(\nu_a \mid \mf{R} \big)  }{  \ex{\i L \, \wh{\xi}_{r}(\nu_a \, \mid \,  \mf{R} )} -1  } \bigg\}
 f( \bs{\nu} ) \cdot  \dd^{n_r} \nu   \; = \; 
(-1)^{n_r}\Int{   \big( \Ga^{\, (r)}_{s_r} \big)^{n_r}  }{}    \pl{a=1}{n_r} \bigg\{    \wh{\xi}_{r}^{\, \prime}\big(\nu_a \mid \mf{R} \big)   \bigg\}  f( \bs{\nu} )  \cdot  \dd^{n_r} \nu       \\
+ \, \sul{p=1}{n_r} C^{p}_{n_r} \cdot (-1)^{n_r-p}  
\pl{a=1}{p} \Bigg\{   \sul{\eps_a=\pm}{}  \Int{ \Ga^{\, (r)}_{ \eps_a s_r }  }{} \!\! \dd \nu_a  \,  \f{ - \eps_a  \,  \wh{\xi}_{r}^{\, \prime}\big(\nu_a \mid \mf{R} \big)   }{  \ex{ -\i \eps_a L \wh{\xi}_{r}(\nu_a \, \mid \,  \mf{R} ) } -1  }  \Bigg\}
  \pl{ a = p+1 }{ n_r } \Bigg\{   \Int{     \Ga^{\, (r)}_{s_r}     }{}  \!\! \dd \nu_a  \,    \wh{\xi}_{r}^{\, \prime}\big(\nu_a \mid \mf{R} \big)   \Bigg\}  f( \bs{\nu} )   \;. 
\label{decomposition integrale multiple sur Ga r string}
\end{multline}
By following the same strategy as in the analysis of the one dimensional integral, one concludes that each summand appearing in the second line of 
\eqref{decomposition integrale multiple sur Ga r string}
contains at least one integration giving rise to $\e{O}\big( L^{-1} \big)$ corrections in \eqref{decomposition integrale sur Ga r string}. Hence, 
one concludes that 
\bem
\Oint{    \Ga^{(r)}_{\e{tot}}    }{}  \pl{a=1}{n_r} \bigg\{ \f{   \wh{\xi}_{r}^{\, \prime}\big(\nu_a \mid \mf{R} \big)  }{  \ex{\i L \, \wh{\xi}_{r}(\nu_a \, \mid \,  \mf{R} )} -1  } \bigg\}
 f( \bs{\nu} ) \cdot  \dd^{n_r} \nu   \; = \; 
(-1)^{n_r}\Int{   \big( \Ga^{\, (r)}_{s_r} \big)^{n_r}  }{}    \pl{a=1}{n_r} \bigg\{    \wh{\xi}_{r}^{\, \prime}\big(\nu_a \mid \mf{R} \big)   \bigg\}  f( \bs{\nu} )  \cdot  \dd^{n_r} \nu   
 \; + \; \e{O}\Big( L^{-1} \Big) \;. 
\label{decomposition integrale multiple sur Ga r string}
\end{multline}

Quite similarly, by using the Cartesian product structure of  $\Ga^{(p)}_{\e{tot}} *  \Ga^{(h)}_{\e{tot}} $ in the target space of $\bs{\psi}$, 
the lower bounds \eqref{ecriture equation positivite pr prime} on the dressed momentum and elementary bounds see \textit{e.g.}\cite{KozProofOfDensityOfBetheRoots} for some details, one can decompose 
the particle-hole integrals as 
\bem
\Oint{  \Ga^{(p)}_{\e{tot}} * \Ga^{(h)}_{\e{tot}} }{}  
\pl{a=1}{ n_p^{(\e{tot})} } \Bigg\{ \f{   \wh{\xi}_{1}^{\, \prime}\big(\la_a^{(p)} \mid \mf{R} \big)  }{  \ex{\i L \, \wh{\xi}_{1}(\la_a^{(p)} \, \mid \,  \mf{R} )} -1  } \Bigg\}
\cdot \pl{a=1}{n_h^{(\e{tot})} } \Bigg\{ \f{   \wh{\xi}_{1}^{\, \prime}\big(\la_a^{(h)} \mid \mf{R} \big)  }{  \ex{\i L \, \wh{\xi}_{1}(\la_a^{(h)} \, \mid \,  \mf{R} )} -1  } \Bigg\}
\cdot f\big( \bs{\la}^{(p)}, \bs{\la}^{(h)} \big) \cdot \dd^{n_p^{(\e{tot})} } \hspace{-1mm}\la^{(p)} \cdot \dd^{n_h^{(\e{tot})} } \hspace{-1mm}\la^{(h)} \\
\; = \; \Oint{  \big(\Ga^{(p)}_{\e{tot}} * \Ga^{(h)}_{\e{tot}} \big)_{+} }{}  \hspace{-3mm}
\pl{a=1}{ n_p^{(\e{tot})} } \Big\{ -  \wh{\xi}_{1}^{\, \prime}\big(\la_a^{(p)} \mid \mf{R} \big)   \Big\}
\cdot \pl{a=1}{n_h^{(\e{tot})} } \Big\{  - \wh{\xi}_{1}^{\, \prime}\big(\la_a^{(h)} \mid \mf{R} \big)   \Big\}
\cdot f\big( \bs{\la}^{(p)}, \bs{\la}^{(h)} \big) \cdot \dd^{n_p^{(\e{tot})} } \hspace{-1mm}\la^{(p)} \cdot \dd^{n_h^{(\e{tot})} } \hspace{-1mm}\la^{(h)} 
\; + \; \e{O}\Big( L^{-1} \Big)
\label{decomposition integrale sur Gatot p et h}
\end{multline}
where the submanifold $\big(\Ga^{(p)}_{\e{tot}} * \Ga^{(h)}_{\e{tot}} \big)_{+}$ giving rise to the leading contribution is defined as being the pre-image 
\beq
\big(\Ga^{(p)}_{\e{tot}} * \Ga^{(h)}_{\e{tot}} \big)_{+} \; = \;
\bs{\psi}^{-1}\bigg(    \Big( \mf{C}_{+;1}^{(p)}\cup \mf{C}_{+;0}^{(p)} \Big)^{ n_p^{(\e{tot})} } \times \Big( \mf{C}_{+;0}^{(h)} \Big)^{ n_h^{(\e{tot})} }  \bigg)
\enq
by the map $\bs{\psi}$ introduced in \eqref{definition map psi small}. In terms of the "model" contour depicted in Figure~\ref{Figure contour Gamma p et h et leurs decompositions},
one can think, in a leading order in $L$ approximation, of the contour $\big(\Ga^{(p)}_{\e{tot}} * \Ga^{(h)}_{\e{tot}} \big)_{+} $ as the Cartesian product of the one-dimensional contours
$ \Ga^{(p)}_{+;0}\cup \Ga^{(p)}_{-;1} $ relatively 
to the particle rapidities and of the  one-dimensional contours $\Ga^{(h)}_{+}$ relatively to the hole rapidities as  depicted in Figure~\ref{Figure contour Gamma p et h et leurs decompositions}. 
Note that, in the large-$L$ limit, $\mf{C}_{+;1}^{(p)}$ is mapped onto $\Ga_{-;1}^{(p)}$ due to $p^{\prime}_1(\la+\i \tf{\pi}{2})<0$.

Clearly, the precise bounds of both remainders in \eqref{decomposition integrale multiple sur Ga r string} and \eqref{decomposition integrale sur Gatot p et h} 
do depend on the function $f$ as well as on the number of integrations involved. It seems reasonable to assume that the class
of functions involved in the effective form factor series $ \wh{\mc{C}}^{\,(\ga)}_{1}(m,t)$ makes all such remainder
uniform and thus allows one to drop these when computing the thermodynamic limit of 
$ \wh{\mc{C}}^{\,(\ga)}_{1}(m,t) $. In other words, I shall assume that it is licit to make the below replacements in $ \wh{\mc{C}}^{\,(\ga)}_{1}(m,t) $ without altering the value of its thermodynamic limit:
\vspace{2mm}
\begin{itemize}
 \item[$\bullet$] $  \Ga^{(r)} \; \hookrightarrow  \;  \Ga^{\, (r)}_{s_r} \quad   $ and  $\quad   \Ga^{(p)}_{\e{tot}}*\Ga^{(h)}_{\e{tot}}   \; \hookrightarrow \;   \big(\Ga^{(p)}_{\e{tot}} * \Ga^{(h)}_{\e{tot}} \big)_{+}\;   $ ; \vspace{2mm}
 \item[$\bullet$] $ \mc{D}^{n_r} \nu^{(r)} \; \hookrightarrow  \;
 \mc{D}^{n_r}_{\e{red}} \nu^{(r)}  \; = \; \prod_{a=1}^{n_r} \Big\{   - L \wh{\xi}_{r}^{\prime}\big(\nu_a^{(r)} \mid \mf{R} \big)  \Big\} \cdot \tfrac{ \dd^{n_r} \nu^{(r)} }{ n_r!\,  (2\pi)^{n_r} }  $; \vspace{2mm}
 \item[$\bullet$]$   \mc{D}^{ n_{p}^{(\e{tot})}  }  \hspace{-1mm} \la^{(p)} \; \hookrightarrow  \;  
	\mc{D}^{ n_{p}^{(\e{tot})} }_{\e{red}} \hspace{-1mm}\la^{(p)}   \; = \;\prod_{a=1}^{ n_{p}^{(\e{tot})} } \Big\{  - L \wh{\xi}_{1}^{\prime}\big(\la_a^{(p)} \mid \mf{R} \big)    \Big\} 
	\cdot \tfrac{ \dd^{ n_{p}^{(\e{tot})} } \la^{(p)} }{ n_{p}^{(\e{tot})} ! \,  (2\pi)^{ n_{p}^{(\e{tot})} } }  \; \;  $ 
and likewise with $p\leftrightarrow h$. 
\end{itemize}
On the basis of the above discussion, one expects that 
\beq
\lim_{L\tend +\infty} \Big\{  \wh{\mc{C}}^{\,(\ga)}_1(m,t)   \Big\} \; = \; \lim_{L\tend +\infty} \Big\{  \wh{\mc{C}}^{\,(\ga)}_2(m,t)   \Big\}
\enq
where 
\beq
 \wh{\mc{C}}^{\,(\ga)}_2(m,t) =  (-1)^{m \op{s}_{\ga}}  \sul{ \bs{n} \in \mf{S}_{\e{tot}}  }{}   
 \pl{ r \in  \mf{N}_{st} }{} \;\Bigg\{  \Int{  \big( \Ga^{\, (r)}_{s_r} \big)^{n_r}  }{}  \mc{D}^{n_r}_{\e{red}} \nu^{(r)} \Bigg\} 
 \cdot \hspace{-3mm} \Int{  \big(\Ga^{(p)}_{\e{tot}} * \Ga^{(h)}_{\e{tot}} \big)_{+} }{} \hspace{-4mm}  \mc{D}^{ n_{p}^{(\e{tot})} }_{\e{red}} \hspace{-1mm} \la^{(p)} \cdot \mc{D}^{ n_{h}^{(\e{tot})} }_{\e{red}} \hspace{-1mm} \la^{(h)}
\cdot \wh{\msc{F}}^{(\ga)}\big( \mf{R} \big) \cdot \ex{ \i m \mc{U}(\mf{R},\op{v})  } \;. 
\label{definition series eff C hat 2 de gamma}
\enq

It is convenient to slightly re-organise the contours of integration in \eqref{definition series eff C hat 2 de gamma} by deforming them back 
to the contours where the physical rapidities, solving the higher-level Bethe Ansatz equations, condense. 
Some care should however be taken in the treatment of the part associated to the particle-hole rapidities. There, one should treat slightly differently the vicinity 
of the endpoints of the Fermi zone: the curves linking the endpoint of the Fermi zone to some points at distance $C \de$, for some $C>0$,
should be deformed  into a path along which  $\ex{ \i m u_{1}(\la,\op{v})}$ -in what concerns the particles- and $\ex{ -\i m u_{1}(\la,\op{v})}$ -in what concerns the holes- is decaying.  
Observe that for 
\beq
| \op{v} | >\op{v}_F  \quad \e{one} \, \e{has} \quad   u_{1}^{\prime}(\pm q,\op{v})=p^{\prime}_1(q)\cdot \Big( 1 \mp \f{ \op{v}_F }{ \op{v} } \Big) > 0 \;,
\quad \e{where} \quad   \op{v}_F= \f{\veps_1^{\prime}(q) }{ p^{\prime}_1(q) }\;, 
\enq
and hence $\ex{ \i m u_{1}(\la,\op{v})}$, resp. $\ex{ -\i m u_{1}(\la,\op{v})}$, is decaying when $\la$ moves from $\pm q$ in the direction of positive, resp. negative,
imaginary parts.  Likewise, for 
\beq
\op{v}_F > \op{v}>0 \quad \e{one} \, \e{has} \quad  \mp u_{1}^{\prime}(\pm q,\op{v}) > 0
\enq
and hence the situation remains unchanged in what concerns $-q$ but the directions of decay are inverted around $q$. 
Finally, for 
\beq
0>\op{v}> -\op{v}_F  \quad \e{one} \, \e{has} \quad  \pm u_{1}^{\prime}(\pm q,\op{v}) > 0
\enq
the situation remains as in the first case in what concerns $q$ but the directions of decay are inverted around $-q$.

The analiticity of the integrand allows one to deform the contours in \eqref{definition series eff C hat 2 de gamma}  as \vspace{2mm}
\begin{itemize}

 \item[$\bullet$] $ \Ga^{\, (r)}_{s_r}  \hookrightarrow  -\msc{C}_{r}$ where $\msc{C}_r\, = \, s_r \R \, + \, \i \de_r \tfrac{\pi}{2}$ is depicted in Figure \ref{Figure Courbe finale integration modes cordes}; \vspace{2mm}
 
 \item[$\bullet$]  $\big(\Ga^{(p)}_{\e{tot}} * \Ga^{(h)}_{\e{tot}} \big)_{+} \hookrightarrow  \msc{C}_{p;\e{tot}} * \msc{C}_{h;\e{tot}} $ where 
\beq
 \msc{C}_{p;\e{tot}} * \msc{C}_{h;\e{tot}}\; = \; \bs{\psi}^{-1}\Big( \big(\mf{C}_{p;\infty}\big)^{n_p^{(\e{tot})}} \times  \big( \mf{C}_{h;\infty} \big)^{n_h^{(\e{tot})}} \Big)
\enq
with $\bs{\psi}$ as in \eqref{definition map psi small} and the contours  $\mf{C}_{p/h;\infty}$ are as defined in Figure~\ref{Figure contour Gamma p et h reduits et deforme quasi vers R}. 
 
\end{itemize}

\begin{figure}[ht]
\begin{center}

\begin{pspicture}(12,9)

\psline[linestyle=dashed, dash=3pt 2pt](0,7.25)(11.8,7.25)
\psline[linewidth=2pt]{->}(11.8,7.25)(11.9,7.25)

 \rput(11.8,6.9){$\R $}
 
 \rput(11.5,8.1){$\R + \i \de$}
 \rput(11.5,6.4){$\R - \i \de$}

\psline[linestyle=dashed, dash=3pt 2pt](10.5,8.1)(11,8.1)
\psline[linestyle=dashed, dash=3pt 2pt](10.5,6.4)(11,6.4) 

\psdots(0.75,7.25)(3,7.25)(4,7.25)(8,7.25) (9,7.25)(11,7.25)
 
 \psline[linecolor=blue,linewidth=2pt]{->}(1.5,7.25)(1.4,7.25)
 \psline[linecolor=blue,linewidth=2pt]{->}(10.3,7.25)(10.2,7.25)
 \psline[linecolor=orange,linewidth=2pt]{->}(6,7.25)(5.9,7.25)

 \psline[linecolor=blue](0.75,7.25)(2.25,7.25)
 \pscurve[linecolor=blue](2.25,7.25)(2.5,6.9)(3,6.4)
 \psline[linestyle=dotted,linecolor=blue](3,6.4)(3,7.25)
 
 \psline{<->}(2.25,7.5)(3,7.5)
 \rput(2.6,7.7){$\de$}

 \psline[linestyle=dotted,linecolor=orange](4,7.25)(4,8.1)
 \pscurve[linecolor=orange](4,8.1)(4.5,7.6)(4.75,7.25)
 \psline[linecolor=orange](4.75,7.25)(7.25,7.25)
 \pscurve[linecolor=orange](7.25,7.25)(7.5,6.9)(8,6.4)
 \psline[linestyle=dotted,linecolor=orange](8,6.4)(8,7.25)

 \psline[linestyle=dotted,linecolor=blue](9,7.25)(9,8.1)
 \pscurve[linecolor=blue](9,8.1)(9.5,7.6)(9.75,7.25)
 \psline[linecolor=blue](9.75,7.25)(11,7.25)
 

\rput(-0.2,7.6){$ \tfrac{2\pi }{ L} \big( M_- \!-\!  \tfrac{1}{2}\big) $}

\rput(11.2,7.6){$\tfrac{2\pi }{ L} \big( M_+  \!+\!  \tfrac{1}{2}\big)$}

\rput(3.3,7.6){$\tfrac{2\pi }{4 \cdot L}  $}
\rput(4.2,6.9){$\tfrac{2\pi \cdot 3}{4 \cdot L}  $}

\rput(7.6,7.6){$\tfrac{2\pi }{ L} \big( N_+ \!-\!  \tfrac{3}{4}\big)  $}
\rput(9.3,6.9){$\tfrac{2\pi }{ L} \big( N_+ \!-\!  \tfrac{1}{4}\big)  $}


\psline(0,5.7)(12,5.7)


\psline[linestyle=dashed, dash=3pt 2pt](0,4.25)(11.8,4.25)
\psline[linewidth=2pt]{->}(11.8,4.25)(11.9,4.25)

 \rput(11.8,3.9){$\R $}
 
 \rput(11.5,5.1){$\R + \i \de$}
 \rput(11.5,3.4){$\R - \i \de$}

\psline[linestyle=dashed, dash=3pt 2pt](10.5,5.1)(11,5.1)
\psline[linestyle=dashed, dash=3pt 2pt](10.5,3.4)(11,3.4) 

\psdots(0.75,4.25)(3,4.25)(4,4.25)(8,4.25) (9,4.25)(11,4.25)

 \psline[linecolor=blue,linewidth=2pt]{->}(1.5,4.25)(1.4,4.25)
 \psline[linecolor=blue,linewidth=2pt]{->}(10.3,4.25)(10.2,4.25)
 \psline[linecolor=orange,linewidth=2pt]{->}(6,4.25)(5.9,4.25)

 \psline[linecolor=blue](0.75,4.25)(2.25,4.25)
 \pscurve[linecolor=blue](2.25,4.25)(2.5,4.6)(3,5.1)
 \psline[linestyle=dotted,linecolor=blue](3,5.1)(3,4.25)
 
 \psline{<->}(2.25,4)(3,4)
 \rput(2.6,3.8){$\de$}

 \psline[linestyle=dotted,linecolor=orange](4,4.25)(4,3.4)
 \pscurve[linecolor=orange](4,3.4)(4.5,3.9)(4.75,4.25)
 \psline[linecolor=orange](4.75,4.25)(7.25,4.25)
 \pscurve[linecolor=orange](7.25,4.25)(7.5,4.6)(8,5.1)
 \psline[linestyle=dotted,linecolor=orange](8,5.1)(8,4.25)

 \psline[linestyle=dotted,linecolor=blue](9,4.25)(9,3.4)
 \pscurve[linecolor=blue](9,3.4)(9.5,3.9)(9.75,4.25)
 \psline[linecolor=blue](9.75,4.25)(11,4.25)
 

\rput(-0.2,4.6){$ \tfrac{2\pi }{ L} \big( M_- \!-\!  \tfrac{1}{2}\big) $}

\rput(11.2,4.6){$\tfrac{2\pi }{ L} \big( M_+  \!+\!  \tfrac{1}{2}\big)$}

\rput(3.3,3.9){$\tfrac{2\pi }{4 \cdot L}  $}
\rput(4.2,4.6){$\tfrac{2\pi \cdot 3}{4 \cdot L}  $}

\rput(7.6,3.9){$\tfrac{2\pi }{ L} \big( N_+ \!-\!  \tfrac{3}{4}\big)  $}
\rput(9.3,4.6){$\tfrac{2\pi }{ L} \big( N_+ \!-\!  \tfrac{1}{4}\big)  $}


\psline(0,2.7)(12,2.7)


\psline[linestyle=dashed, dash=3pt 2pt](0,1.25)(11.8,1.25)
\psline[linewidth=2pt]{->}(11.8,1.25)(11.9,1.25)

 \rput(11.8,0.9){$\R $}
 
 \rput(11.5,2.1){$\R + \i \de$}
 \rput(11.5,0.4){$\R - \i \de$}

\psline[linestyle=dashed, dash=3pt 2pt](10.5,2.1)(11,2.1)
\psline[linestyle=dashed, dash=3pt 2pt](10.5,0.4)(11,0.4) 

\psdots(0.75,1.25)(3,1.25)(4,1.25)(8,1.25) (9,1.25)(11,1.25)

 \psline[linecolor=blue,linewidth=2pt]{->}(1.5,1.25)(1.4,1.25)
 \psline[linecolor=blue,linewidth=2pt]{->}(10.3,1.25)(10.2,1.25)
 \psline[linecolor=orange,linewidth=2pt]{->}(6,1.25)(5.9,1.25)

 \psline[linecolor=blue](0.75,1.25)(2.25,1.25)
 \pscurve[linecolor=blue](2.25,1.25)(2.5,1.6)(3,2.1)
 \psline[linestyle=dotted,linecolor=blue](3,2.1)(3,1.25)
 
 \psline{<->}(2.25,1)(3,1)
 \rput(2.6,0.8){$\de$}

 \psline[linestyle=dotted,linecolor=orange](4,1.25)(4,0.4)
 \pscurve[linecolor=orange](4,0.4)(4.5,0.9)(4.75,1.25)
 \psline[linecolor=orange](4.75,1.25)(7.25,1.25)
 \pscurve[linecolor=orange](7.25,1.25)(7.5,0.9)(8,0.4)
 \psline[linestyle=dotted,linecolor=orange](8,0.4)(8,1.25)

 \psline[linestyle=dotted,linecolor=blue](9,1.25)(9,2.1)
 \pscurve[linecolor=blue](9,2.1)(9.5,1.6)(9.75,1.25)
 \psline[linecolor=blue](9.75,1.25)(11,1.25)
 

\rput(-0.2,1.6){$ \tfrac{2\pi }{ L} \big( M_- \!-\!  \tfrac{1}{2}\big) $}

\rput(11.2,1.6){$\tfrac{2\pi }{ L} \big( M_+  \!+\!  \tfrac{1}{2}\big)$}

\rput(3.3,0.9){$\tfrac{2\pi }{4 \cdot L}  $}
\rput(4.2,1.6){$\tfrac{2\pi \cdot 3}{4 \cdot L}  $}

\rput(7.6,1.6){$\tfrac{2\pi }{ L} \big( N_+ \!-\!  \tfrac{3}{4}\big)  $}
\rput(9.3,0.9){$\tfrac{2\pi }{ L} \big( N_+ \!-\!  \tfrac{1}{4}\big)  $}

\end{pspicture}
\caption{ Particle $\mf{C}_{p;\infty}$ -in blue- and hole $\mf{C}_{h;\infty}$ -in orange- contours in the target space and plotted for the three regimes of the velocity 
$\op{v}=\tf{m}{t}$ appearing from bottom to top $|\op{v}|>\op{v}_F$, then $\op{v}_F>\op{v}>0$ and finally $0 > \op{v} > -\op{v}_F$. 
Here $N_+=N+1+\op{s}_{\ga}$. 
The contour $\mf{C}_{p;\infty}$, resp. $\mf{C}_{h;\infty}$,
partitions into three parts $\mf{C}_{p;\infty}=\mf{C}_{p;\infty}^{(-)}\cup\mf{C}_{p;\infty}^{(0)}\cup\mf{C}_{p;\infty}^{(+)}$, resp. 
$\mf{C}_{h;\infty} \, = \, \mf{C}_{h;\infty}^{(-)}\cup\mf{C}_{h;\infty}^{(0)}\cup\mf{C}_{h;\infty}^{(+)}$. The contours 
$ \mf{C}_{p/h;\infty}^{(\ups)} $ are depicted in dotted lines while the contours $ \mf{C}_{p/h;\infty}^{(0)} $ are depicted in solid lines. 
$ \mf{C}_{p/h;\infty}^{(+)} $ appears to the right while $ \mf{C}_{p/h;\infty}^{(-)} $ to the left. 
\label{Figure contour Gamma p et h reduits et deforme quasi vers R} }
\end{center}

\end{figure}

\noindent Doing so, recasts $ \wh{\mc{C}}^{\,(\ga)}_2(m,t) $  as
\beq
 \wh{\mc{C}}^{\,(\ga)}_2(m,t) =   (-1)^{m \op{s}_{\ga}}  \sul{ \bs{n}\in \mf{S}_{\e{tot}}  }{}   
 \pl{ r \in  \mf{N}_{st} }{} \;\Bigg\{  \Int{  \big( - \msc{C}_r \big)^{n_r}  }{}  \mc{D}^{n_r}_{\e{red}} \nu^{(r)} \Bigg\} \hspace{2mm}
 \cdot \hspace{-4mm} \Int{  \msc{C}_{p;\e{tot}} *  \msc{C}_{h;\e{tot}} }{} \hspace{-4mm}  \mc{D}^{ n_{p}^{(\e{tot})} }_{\e{red}} \hspace{-1mm} \la^{(p)} \cdot \mc{D}^{ n_{h}^{(\e{tot})} }_{\e{red}} \hspace{-1mm} \la^{(h)}
\cdot \wh{\msc{F}}^{(\ga)}\big( \mf{R} \big) \cdot \ex{ \i m \mc{U}(\mf{R},\op{v})  } \;. 
\enq
\begin{figure}[ht]
\begin{center}

\begin{pspicture}(4.5,2.5)

\psline[linestyle=dashed, dash=3pt 2pt](0,1)(4,1)
\psline[linewidth=2pt]{->}(4,1)(4.1,1)

 \rput(5,1){$\R + \i \de_{r} \tfrac{\pi}{2}$}

\end{pspicture}
\caption{ Contour $\msc{C}_{r}= s_r \R+\i\de_r \tfrac{\pi}{2}$ where $s_r=\e{sgn}\big[p_r^{\prime} \big]_{ \mid \R+\i\de_r \tf{\pi}{2}  }$ and $\de_r\in \{0,1\}$ depending on the value of $r$.  \label{Figure Courbe finale integration modes cordes} }
\end{center}

\end{figure}

 \subsection{The thermodynamic limit of $\wh{C}_2^{\, (\ga)}(m,t)$}
\label{Sousection limite thermo serie auxiliare C2}

One more step is necessary so as to be able to take the thermodynamic limit of $ \wh{\mc{C}}^{\,(\ga)}_2(m,t) $, namely decompose the 
contours of integration for the particle-hole rapidities into their parts that are "close", with precision $\de$, to $\pm q$ and the parts that are at least at 
distance $C \de$, for some $C>0$. This will allow to make a sharp distinction between the "massive" and "massless" parts of the particle-hole spectrum
and hence make use of the more refined large-$L$ asymptotics of the form factor $\wh{\msc{F}}^{(\ga)}\big( \mf{R} \big)$
given in \eqref{ecriture DA FF apres partition modes massifs et masse nulle}. The decomposition of the particle and hole contours can be done by breaking up the contours 
$\mf{C}_{p/h;\infty}$ in the target space as 
\beq
\mf{C}_{p/h;\infty} \, = \, \mf{C}_{p/h;\infty}^{(-)} \cup \mf{C}_{p/h;\infty}^{(0)} \cup \mf{C}_{p/h;\infty}^{(+)} \;. 
\label{ecriture partition ctr espace cible et parties loc et loin de Fermi bd}
\enq
The contours appearing on the \textit{rhs} are given in  Figure~\ref{Figure contour Gamma p et h reduits et deforme quasi vers R}:
$ \mf{C}_{p/h;\infty}^{(0)} $ are depicted in full lines whereas $ \mf{C}_{p/h;\infty}^{(\pm)} $ are depicted in dotted lines.


Let $f$ be a symmetric function of $\bs{\la}^{(p)}$ and $\bs{\la}^{(h)}$ taken singly and consider the integral 
\beq
\mc{I}_{ n_{p}^{(\e{tot})},  n_{h}^{(\e{tot})}}[f] \; = \; \Int{  \msc{C}_{p;\e{tot}} *  \msc{C}_{h;\e{tot}} }{} \hspace{-4mm}  \mc{D}^{ n_{p}^{(\e{tot})} }_{\e{red}} \hspace{-1mm} \la^{(p)} 
\cdot \mc{D}^{ n_{h}^{(\e{tot})} }_{\e{red}} \hspace{-1mm} \la^{(h)}
\; f\big( \bs{\la}^{(p)} , \bs{\la}^{(h)} \big) \;. 
\enq
When  decomposing each contour arising in the Cartesian product decomposition as in \eqref{ecriture partition ctr espace cible et parties loc et loin de Fermi bd} 
and summing up over all possible decompositions,  owing to the symmetry of the function $f$, it is enough to sum up over all the possible decompositions
of the original set of integration variables into
\beq
\big\{ \la^{(p)}_a  \big\}_1^{ n_{p}^{(\e{tot})} } \; = \; \big\{ \la_a^{+} \big\}_1^{ n_{p}^{+} } \cup  \big\{ \nu_a^{(1)} \big\}_1^{ n_{p} } \cup  \big\{ \la_a^- \big\}_1^{ n_{p}^{-} }
\quad \e{and} \quad
\big\{ \la^{(h)}_a  \big\}_1^{ n_{h}^{(\e{tot})} } \; = \; \big\{ \mu_a^{+} \big\}_1^{ n_{h}^{+} }  \cup \big\{ \mu_a \big\}_1^{ n_{h} } \cup  \big\{ \mu_a^{-} \big\}_1^{ n_{h}^{-} } 
\enq
where $\la^{\ups}_a$, resp. $\mu^{\ups}_a$, correspond to integrations, in the target space, over $\mf{C}_{p;\infty}^{(\ups)}$, resp.  $\mf{C}_{h;\infty}^{(\ups)}$, and 
$\nu^{(1)}_a$, resp. $\mu_a$, to an integration over $\mf{C}_{p;\infty}^{(0)}$, resp. $\mf{C}_{h;\infty}^{(0)}$. Such a sum is weighted by a multi-nomial coefficient 
and the integration runs through 
\beq
\wh{\msc{C}}_{ n_p^{*} ; n_h^{*} } \; = \; \bs{\psi}^{-1}\bigg( \Big( \mf{C}_{p;\infty}^{(+)} \Big)^{ n_p^+ } \times \Big( \mf{C}_{p;\infty}^{(0)} \Big)^{ n_p } \times  \Big( \mf{C}_{p;\infty}^{(-)} \Big)^{ n_p^- }  
\times  \Big( \mf{C}_{h;\infty}^{(+)} \Big)^{ n_h^+ } \times \Big( \mf{C}_{h;\infty}^{(0)} \Big)^{ n_h } \times  \Big( \mf{C}_{h;\infty}^{(-)} \Big)^{ n_h^- }  \bigg)  \;, 
\enq
with $\bs{\psi}$ as in \eqref{definition map psi small}. One eventually gets 
\beq
\mc{I}_{ n_{p}^{(\e{tot})},  n_{h}^{(\e{tot})}}[f] \; = \; \sul{   \substack{ n_{p/h}^{(\e{tot})} = n_{p/h}^{+} \\ +n_{p/h}+n_{p/h}^{-}   }  }{} 
\Int{  \wh{\msc{C}}_{ n_p^{*} ; n_h^{*} }  }{} \hspace{-2mm}  \mc{D}^{ n_{p} }_{\e{red}}  \nu^{(1)} 
\cdot \mc{D}^{ n_{h} }_{\e{red}}  \,  \mu \cdot  \pl{\ups=\pm}{} \Bigg\{  \mc{D}^{ n_{p}^{\ups} }_{\e{red}}  \la^{\ups} \cdot \mc{D}^{ n_{h}^{\ups} }_{\e{red}} \,  \mu^{\ups} \Bigg\} 
\cdot f\big( \bs{\la}^{(p)} , \bs{\la}^{(h)} \big) \;. 
\enq
Here, the integration measure  $ \mc{D}^{n_{p}^{\ups} }_{\e{red}} \la^{\ups}$, resp. $ \mc{D}^{n_{h}^{\ups} }_{\e{red}} \, \mu^{\ups}$ \textit{etc}., 
is obtained from $ \mc{D}^{n_{p} }_{\e{red}} \la^{(p)}$ upon the substitution $\la^{(p)} \hookrightarrow \la^{\ups}$, resp. $\la^{(h)} \hookrightarrow \mu^{\ups}$  \textit{etc}. 
Also, above and in the following, one should understand   that $ \bs{\la}^{(p)} =  \big( \bs{\la}^{+}, \bs{\nu}^{(1)} , \bs{\la}^{-} \big)$, $ \bs{\la}^{(h)} =  \big( \bs{\mu}^{+},\bs{\mu} , \bs{\mu}^{-} \big)$.

The variables $\big\{ \la_a^{ \ups } \big\}_1^{ n_{p}^{\ups} } $ and $\big\{ \mu_a^{\ups} \big\}_1^{ n_{h}^{\ups} } $ are at a distance $\e{O}(\de)$
from $\ups q$. This means that, up to $1 \, + \,  \e{O}(\de/L)$ corrections, one can replace the variables $\mf{R}$ in the functions $ \wh{\xi}^{\, \prime}_{1}\big( * \mid \mf{R}\big) $ occurring in each of the $\mc{D}$ measures 
by the variables $\mf{Y}$ given in \eqref{definition rapidites massives reduites}, \textit{i.e.} by the functions $ \wh{\xi}^{\, \prime}_{1}\big( * \mid \mf{Y}\big) $. Again, up to such corrections, one can make the same replacement in the function $\bs{\psi}$
defining the integration contour $\wh{\msc{C}}_{ n_p^{*} ; n_h^{*} }$. Thus, introduce 
\beq
\bs{\Psi}\; : \; \big( \nu_1^{(1)}, \cdots , \nu_{n_p}^{(1)} , \mu_1, \cdots , \mu_{n_h }; \ell_{\ups} \big) \; \mapsto  \; 
\Big( \,  \wh{\xi}_{1}\big( \nu_1^{(1)} \mid \mf{Y} \big) \, , \cdots, \,  \wh{\xi}_{1}\big( \nu_{n_p}^{(1)} \mid \mf{Y} \big)  \, , 
 \wh{\xi}_{1}\big(\mu_1 \mid \mf{Y} \big) \, , \cdots,  \, \wh{\xi}_{1}\big(\mu_{n_h } \mid \mf{Y} \big) \Big)
\label{definition map psi}
\enq
and define
\beq
\wh{\msc{C}}_{ n_p ; n_h }^{\; (\de)}  \;=\;  \bs{\Psi}^{-1}\bigg(  \Big( -\mf{C}_{p;\infty}^{(0)} \Big)^{ n_p } \times  \Big( -\mf{C}_{h;\infty}^{(0)} \Big)^{ n_h }  \bigg)  \;. 
\enq
Here, the $-$ sign means that the contours ought to be endowed in the opposite orientation. 
Also, for fixed $\mf{Y}$, define the \textit{one}-dimensional contours 
%
%
\beq
\wh{\msc{C}}_{p}^{\ups} \; = \; \wh{\xi}_1^{-1}\Big(\ups \intff{ \tfrac{ 2\pi }{ L } \big( N_{\ups} -\tfrac{\ups}{4} \big) }{   \tfrac{ 2\pi }{ L } \big( N_{\ups} -\tfrac{\ups}{4} \big)  + \i \de \mf{s}_{u}^{\ups} } \mid \mf{Y} \Big)
\enq
and
\beq
\wh{\msc{C}}_{h}^{ \ups} \; = \; \wh{\xi}_1^{-1}\Big( -\ups \intff{ \tfrac{ 2\pi }{ L } \big( N_{\ups} -\tfrac{3\ups}{4} \big) }{   \tfrac{ 2\pi }{ L } \big( N_{\ups} -\tfrac{3 \ups}{4} \big)  - \i \de \mf{s}_{u}^{\ups}  }  \mid \mf{Y} \Big) \;, 
\enq
where the $\pm \ups $ prefactor in the interval indicates the orientation, 
\beq
\mf{s}_{u}^{\ups} \, = \, \e{sgn}\Big(u_1^{\prime}(\ups q ; \op{v}) \Big) \quad, \quad  N_+=N+1+\op{s}_{\ga} \quad , \quad N_{-}=0 \;. 
\enq
All of this allows one to recast the original integral as 
\beq
\mc{I}_{ n_{p}^{(\e{tot})},  n_{h}^{(\e{tot})}}[f] \; =  \hspace{-3mm} \sul{   \substack{ n_{p/h}^{(\e{tot})} = n_{p/h}^{+} \\ +n_{p/h}+n_{p/h}^{-}   }  }{} 
\Int{  \wh{\msc{C}}_{ n_p ; n_h }^{\; (\de)}  }{} \hspace{-2mm}  \mf{D}^{ n_{p} }_{\e{red}}  \nu^{(1)} \cdot \mf{D}^{ n_{h} }_{\e{red}}  \,  \mu \cdot  
\pl{\ups=\pm}{} \Bigg\{ \Int{ \big(\wh{\msc{C}}_{p}^{ \ups}\big)^{ n_p^{\ups} }   }{}   \mf{D}^{ n_{p}^{\ups} }_{\e{red}}  \la^{\ups} 
\cdot \Int{ \big( \wh{\msc{C}}_{h}^{\ups} \big)^{n_h^{\ups} }   }{}   \mf{D}^{ n_{h}^{\ups} }_{\e{red}} \,  \mu^{\ups} \Bigg\} 
\cdot f\big( \bs{\la}^{(p)}, \bs{\la}^{(h)} \big) \Big( 1 + \e{O}\big( \tfrac{ \de }{ L }\big) \Big) \;. 
\enq
Above, I have introduced the measure
\beq
\mf{D}^{ n_{h}^{\ups} }_{\e{red}}  \, \mu^{\ups}   \; = \;\prod_{a=1}^{ n_{h}^{\ups}  } \Big\{    L \wh{\xi}_{1}^{ \; \prime}\big(\mu_a^{\ups} \mid \mf{Y} \big)    \Big\} 
	\cdot \tfrac{ \dd^{ n_{h}^{\ups} } \mu }{ n_{h}^{\ups}! (2\pi)^{ n_{h}^{\ups} } }  
\enq
and likewise with $h\leftrightarrow p$ and $\la\leftrightarrow \mu$. Note the absence of the $-$ sign in front of $ L \wh{\xi}_{1}^{ \; \prime}$. 
It has been absorbed into the change of orientation in the contours $\wh{\msc{C}}_{p/h}^{\ups}$, as compared to $\mf{C}^{(\ups)}_{p/h;\infty}$. 

The measures involving $\nu^{(1)}_a$ and $\mu_a$ are defined analogously. 
Assuming the summability of the remainder, one can apply this result to the series representing $ \wh{\mc{C}}^{\,(\ga)}_2(m,t)$, 
hence recasting it as  
\bem
 \wh{\mc{C}}^{\,(\ga)}_2(m,t) \,  =  \,  (-1)^{m \op{s}_{\ga}}  \sul{ \bs{n} \in \wh{\mf{S}}   }{}   \sul{ n_p^{\ups}-n_h^{\ups}=\ell_{\ups} }{}
 \pl{ r \in  \mf{N}_{\e{st}} }{} \;\Bigg\{  \Int{ \big(\msc{C}_r\big)^{n_r} }{}  \mf{D}^{n_r}_{\e{red}} \nu^{(r)} \Bigg\} 
\cdot  \Int{  \wh{\msc{C}}_{ n_p ; n_h }^{\; (\de)}  }{} \hspace{-2mm}  \mf{D}^{ n_{p} }_{\e{red}}  \nu^{(1)} \cdot \mf{D}^{ n_{h} }_{\e{red}}  \,  \mu   \\
 \times  \pl{\ups= \pm }{} \bigg\{ \Int{ \big( \wh{\msc{C}}_p^{\ups} \big)^{ n_p^{\ups} } }{}    \mf{D}^{n_{p}^{\ups} }_{\e{red}} \la^{\ups}   
 \cdot  \Int{  \big( \wh{\msc{C}}_h^{\ups} \big)^{n_h^{\ups}}  }{}    \mf{D}^{n_{h}^{\ups}  }_{\e{red}}\,  \mu^{\ups}  \bigg\}
\cdot \wh{\msc{F}}^{(\ga)}\big( \mf{R} \big) \cdot \ex{\i m  \mc{U}(\mf{R},\op{v}) }  \Big( 1 + \e{O}\big( \tfrac{ \de }{ L }\big) \Big)  \;. 
\label{ecriture serie pour C hat 2 apres partitionement integrales}
\end{multline}
The summation in the series defining $ \wh{\mc{C}}^{\,(\ga)}_2(m,t)$ runs through all the positive 
integers $n_h, \{n_r\}_{r \in \mf{N}}$ and the integers $ \ell_{\ups} =n_p^{\ups}-n_h^{\ups} $ gathered in $\bs{n} = \big( n_h, \{n_r\}_{r \in \mf{N} }, \ell_{\ups} \big)$ which runs through the set 
\beq
\wh{\mf{S}} \; = \; \bigg\{ \big( n_h, \{n_r\}_{r \in \mf{N} }, \ell_{\ups} \big) \; : \;  n_h  \, + \, \op{s}_{\ga} \; = \; \sul{\ups \in \{\pm\} }{} \ell_{\ups} \, + \, \sul{ r\in \mf{N}  }{} r n_r 
\;\;  , \;  \; n_h^+ + n_h + n_h^- \, \in \,  \intn{ 1 }{\varkappa_{L} } \bigg\} \;. 
\enq
Also, one sums up over all positive integers $n_{p/h}^{\ups}$ satisfying to  $n_p^{\ups}-n_h^{\ups}=\ell_{\ups}$. 

Observe that owing to the properties of the contours, the collection of rapidities $\mf{R}$ appearing in each multidimensional integral of the series \eqref{ecriture serie pour C hat 2 apres partitionement integrales}
partitions exactly as in \eqref{decomposition de R mode massifs et massless}. Hence, one can replace $\wh{\msc{F}}^{(\ga)}\big( \mf{R} \big) $ in  \eqref{ecriture serie pour C hat 2 apres partitionement integrales}
by its large-L asymptotics taking into account the more refined information on the \textit{locii} of the rapidities of the particles and holes and 
whose explicit form is given in \eqref{ecriture DA FF apres partition modes massifs et masse nulle}. 
These asymptotics contain $\e{O}(\de \ln \de)$ corrections which correspond to dropping the dependence of the form factor on certain "massless" rapidities $\la^{\ups}$ and $\mu^{\ups}$. 
 The factorisation \eqref{ecriture DA FF apres partition modes massifs et masse nulle} of the asymptotics  of $\wh{\msc{F}}^{(\ga)}\big( \mf{R} \big) $
allows one to take a partial thermodynamic limit of  $\wh{\mc{C}}^{\,(\ga)}_2(m,t)$, namely send $L\tend +\infty$ in the part of the series \eqref{ecriture serie pour C hat 2 apres partitionement integrales}
that is relative to the "massive" modes parametrised by $\mf{Y}$. 
In order to write down the result, it is enough to observe that, owing to the form taken by the counting function \eqref{equations donnant fct cptge r}, 
the joint particle-hole $n_p+n_h$ real dimensional manifold of integration $ \wh{\msc{C}}_{ n_p ; n_h }^{\; (\de)} $ 
factorises into a simple Cartesian product in the thermodynamic limit:
\beq
\lim_{L\tend + \infty} \Big\{ \wh{\msc{C}}_{ n_p ; n_h }^{\; (\de)} \Big\} \; = \; \Big( \msc{C}^{(\de)}_{p} \Big)^{n_p} \times  \Big( \msc{C}^{(\de)}_{h} \Big)^{n_h} 
\enq
where the contours $\msc{C}^{(\de)}_{p/h}$ are as depicted in Figure \ref{Figure contour Gamma p et h reduits delta deformes a la limite thermo}. 
\begin{figure}[ht]
\begin{center}

\begin{pspicture}(12,9)

\psline[linestyle=dashed, dash=3pt 2pt](0,7.25)(11.8,7.25)
\psline[linewidth=2pt]{->}(11.8,7.25)(11.9,7.25)

 \rput(11.8,6.9){$\R $}

\psdots(3.5,8.2)(3.5,7.25)(3.5,6.3)(8.5,8.2)(8.5,7.25)(8.5,7.25)(8.5,6.3)

 \psline[linecolor=blue](0,8.55)(11.8,8.55)
 \psline[linecolor=blue,linewidth=2pt]{->}(1.5,7.25)(1.6,7.25)
 \psline[linecolor=blue,linewidth=2pt]{->}(10.5,7.25)(10.6,7.25)
 \psline[linecolor=blue,linewidth=2pt]{->}(6,8.55)(5.9,8.55)
 \psline[linecolor=orange,linewidth=2pt]{->}(6,7.25)(6.1,7.25)
\rput(11.5,8.2){$\R+\i\tfrac{\pi}{2}$}

 \psline[linecolor=blue](0,7.25)(2.25,7.25)
 \pscurve[linecolor=blue](2.25,7.25)(2.5,6.9)(3,6.4)(3.5,6.3)
 
 \psline{<->}(2.25,7.4)(3.5,7.4)
 \rput(2.9,7.6){$\e{O}(\de)$}

 \pscurve[linecolor=orange](3.5,8.2)(4,8.1)(4.5,7.6)(4.75,7.25)
 \psline[linecolor=orange](4.75,7.25)(7.25,7.25)
 \pscurve[linecolor=orange](7.25,7.25)(7.5,6.9)(8,6.4)(8.5,6.3)

 \pscurve[linecolor=blue](8.5,8.2)(9,8.1)(9.5,7.6)(9.75,7.25)
 \psline[linecolor=blue](9.75,7.25)(11.7,7.25)

\rput(3.4,7){$ -q  $}
\rput(8.7,7){$ q  $}

\rput(3.2,8.2){$\la_{-;\ua}$}
\rput(3.9,6.2){$\la_{-;\da}$}

\rput(8.5,8){$\la_{+;\ua}$}
\rput(8.5,6.1){$\la_{+;\da}$}


\psline(0,5.9)(12,5.9)


\psline[linestyle=dashed, dash=3pt 2pt](0,4.25)(11.8,4.25)
\psline[linewidth=2pt]{->}(11.8,4.25)(11.9,4.25)

 \rput(11.8,3.9){$\R $}

\psdots(3.5,5.2)(3.5,4.25)(3.5,4.25)(3.5,3.3)(8.5,5.2)(8.5,4.25)(8.5,4.25)(8.5,3.3)

 \psline[linecolor=blue](0,5.55)(11.8,5.55)
 \psline[linecolor=blue,linewidth=2pt]{->}(1.5,4.25)(1.6,4.25)
 \psline[linecolor=blue,linewidth=2pt]{->}(10.5,4.25)(10.6,4.25)
 \psline[linecolor=blue,linewidth=2pt]{->}(6,5.55)(5.9,5.55)
 \psline[linecolor=orange,linewidth=2pt]{->}(6,4.25)(6.1,4.25)
\rput(11.5,5.2){$\R+\i\tfrac{\pi}{2}$}

 \psline[linecolor=blue](0,4.25)(2.25,4.25)
 \pscurve[linecolor=blue](2.25,4.25)(2.5,4.6)(3,5.1)(3.5,5.2)
 
 \psline{<->}(2.25,4.1)(3.5,4.1)
 \rput(2.9,3.9){$\e{O}(\de)$}

 \pscurve[linecolor=orange](3.5,3.3)(4,3.4)(4.5,3.9)(4.75,4.25)
 \psline[linecolor=orange](4.75,4.25)(7.25,4.25)
 \pscurve[linecolor=orange](7.25,4.25)(7.5,4.6)(8,5.1)(8.5,5.2)

 \pscurve[linecolor=blue](8.5,3.3)(9,3.4)(9.5,3.9)(9.75,4.25)
 \psline[linecolor=blue](9.75,4.25)(11.7,4.25)

\rput(3.2,4.4){$ -q  $}
\rput(8.3,4.4){$ q  $}

\rput(3.9,5.1){$\la_{-;\ua}$}
\rput(3.5,3.1){$\la_{-;\da}$}

\rput(8.9,5.1){$\la_{+;\ua}$}
\rput(8.5,3.1){$\la_{+;\da}$}

\psline(0,2.9)(12,2.9)


\psline[linestyle=dashed, dash=3pt 2pt](0,1.25)(11.8,1.25)
\psline[linewidth=2pt]{->}(11.8,1.25)(11.9,1.25)

 \rput(11.8,0.9){$\R $}

\psdots(3.5,2.2)(3.5,1.25)(3.5,1.25)(3.5,0.3)(8.5,2.2)(8.5,1.25)(8.5,1.25)(8.5,0.3)

  \psline[linecolor=blue](0,2.55)(11.8,2.55)
 \psline[linecolor=blue,linewidth=2pt]{->}(1.5,1.25)(1.6,1.25)
 \psline[linecolor=blue,linewidth=2pt]{->}(10.5,1.25)(10.6,1.25)
 \psline[linecolor=blue,linewidth=2pt]{->}(6,2.55)(5.9,2.55)
 \psline[linecolor=orange,linewidth=2pt]{->}(6,1.25)(6.1,1.25)
\rput(11.5,2.2){$\R+\i\tfrac{\pi}{2}$}

 \psline[linecolor=blue](0,1.25)(2.25,1.25)
 \pscurve[linecolor=blue](2.25,1.25)(2.5,1.6)(3,2.1)(3.5,2.2)
 
 \psline{<->}(2.25,1.1)(3.5,1.1)
 \rput(2.9,0.9){$\e{O}(\de)$}

 \pscurve[linecolor=orange](3.5,0.3)(4,0.4)(4.5,0.9)(4.75,1.25)
 \psline[linecolor=orange](4.75,1.25)(7.25,1.25)
 \pscurve[linecolor=orange](7.25,1.25)(7.5,0.9)(8,0.4)(8.5,0.3)

 \pscurve[linecolor=blue](8.5,2.2)(9,2.1)(9.5,1.6)(9.75,1.25)
 \psline[linecolor=blue](9.75,1.25)(11.7,1.25)

\rput(3.2,1.4){$ -q  $}
\rput(8.7,1){$ q  $}

\rput(3.9,2.1){$\la_{-;\ua}$}
\rput(3.5,0.1){$\la_{-;\da}$}

\rput(8.5,2){$\la_{+;\ua}$}
\rput(8.5,0.1){$\la_{+;\da}$}

\end{pspicture}
\caption{  Particle $\msc{C}_{p}^{(\de)}$ -in blue- and hole $\msc{C}_{h}^{(\de)}$ -in orange- contours in the vicinity of $\R$ and in the thermodynamic limit. The contours are plotted for the three regimes of the velocity 
$\op{v}=\tf{m}{t}$ appearing from bottom to top $|\op{v}|>\op{v}_F$, $\op{v}_F>\op{v}>0$ and $0>\op{v}>-\op{v}_F$. The contour $\msc{C}_{p}^{(\de)}$ and $\msc{C}_{p}^{(\de)}$
start at the points $\la_{\pm;\ua/\da}=p_1^{-1}\big( \pm \tfrac{D}{2} + \i \eps_{\ua/\da} \de \big)=\pm q + \e{O}(\de)$, where $\eps_{\ua}=+1$ and $\eps_{\da}=-1$, 
and then, over a distance of the order of $\de$ they joint with the real axis. 
\label{Figure contour Gamma p et h reduits delta deformes a la limite thermo} }
\end{center}

\end{figure}

Thus, since 
\beq
\lim_{L\tend + \infty} \Big\{ \wh{\mc{C}}^{\,(\ga)}_2(m,t) \Big\} \, = \, \big< \sg_1^{\ga^{\prime}}(t) \sg_{m+1}^{\ga}  \big> \; ,
\enq
one obtains the below representation for the thermodynamic limit of the two-point function 
\bem
\big< \sg_1^{\ga^{\prime}}(t) \sg_{m+1}^{\ga} \big> \; = \;   (-1)^{m \op{s}_{\ga} }   \sul{ \bs{n} \in \mf{S}   }{}   
 \pl{ r \in  \mf{N} }{} \;\Bigg\{  \Int{  \big(\msc{C}_r^{(\de)} \big)^{n_r} }{}  \f{ \dd^{n_r}\nu^{(r)} }{ n_r! \cdot (2\pi)^{n_r}  } \pl{a=1}{n_r}\Big[ \ex{\i m u_r(\nu_a^{(r)}, \op{v})} \Big]   \Bigg\} 
\cdot  \Int{ \big( \msc{C}_{h}^{(\de)} \big)^{n_h}   }{}  \f{ \dd^{n_h}\mu  }{ n_h! \cdot (2\pi)^{n_h}  }  \pl{a=1}{n_h}\Big[ \ex{-\i m u_1(\mu_a, \op{v})} \Big] \\
 \times   \mc{F}^{(\ga)}\big( \mf{Y} \big) \cdot 
 \lim_{L\tend + \infty}  \bs{:} \bigg\{  \pl{\ups= \pm }{} \wh{\mc{B}}_{\ell_{\ups}}^{\, (\ups)}\Big[ \mf{F}_{\ups}\big(\mf{Y} \mid * \big) \mid \mf{Y} \Big] \bigg\}  \bs{:} \big(1+\e{O}(\de \ln \de) \big)  \;. 
\label{ecriture limit thermo fct 2 pts avec fnelle Bups}
\end{multline}
Here, I remind that $\mf{N}=\mf{N}_{\e{st}} \cup  \{ 1 \}$ and that the functions $u_r(\mu,\op{v})$ have been defined in \eqref{definition fct ur et vitesse v}. 
Also, one should understand that 
\beq
\msc{C}_{1}^{(\de)}\equiv \msc{C}^{(\de)}_{p} \qquad \e{and} \qquad \msc{C}_r^{(\de)} \,= \, \msc{C}_r \quad \e{for} \quad r \in \mf{N}_{\e{st}} 
\enq
Finally, the summation runs through all positive integers $n_h$ and $\{n_r\}_{r \in \mf{N} }$ and relative integers $\ell_{\ups}$, gathered in $\bs{n}=\big( n_h, \{n_r\}_{r \in \mf{N} }, \ell_{\ups} \big)$
which runs through
\beq
\mf{S} \; = \; \bigg\{ \big( n_h, \{n_r\}_{r \in \mf{N} }, \ell_{\ups} \big) \; : \;  n_h  \, + \, \op{s}_{\ga} \; = \; \sul{\ups \in \{\pm\} }{} \ell_{\ups} \, + \, \sul{ r\in \mf{N}  }{} r n_r  \bigg\} \;. 
\label{definition domaine de sommation mathfrak S}
\enq
Finally, \eqref{ecriture limit thermo fct 2 pts avec fnelle Bups} involves the functionals  $\wh{\mc{B}}_{ \ell_{\ups}}^{\, (\ups)}$. 
These act on the tower of discrete form factors $\mf{F}_{\ups}$ associated with all the possible choices of the "collapsing" rapidities $\la^{\ups}, \mu^{\ups}$
that are compatible with a given choice of $\mf{Y}$. In taking the thermodynamic limit, it is assumed that one can 
first evaluate the action of the functionals and only then compute the effect of the $\bs{:} * \bs{:}$ ordering. 
It remains to provide an explicit definition of the functional $\wh{\mc{B}}_{ \ell }^{\, (\ups)}$. Given $\ell\in \mathbb{N}$, 
it acts on a collection of functions $f_{n_p,n_h}\big( \{ \la_a  \}_1^{ n_p }   ;  \{ \mu_a  \}_1^{ n_h }    \big)$
of two sets of variables $\{\la_a  \}_1^{ n_p } $ and  $\{ \mu_a  \}_1^{ n_h }$ with $n_p, n_h \in \mathbb{N}$ such that the numbers $n_p$ of $\la$ and $n_h$ of $\mu$ variables are
constrained by the relation $n_p-n_h=\ell$. The action of $\wh{\mc{B}}_{ \ell }^{\, (\ups)}$ takes the explicit form:
\bem
\wh{\mc{B}}_{\ell  }^{\, (\ups)} \big[ f_{*,*} \mid \mf{Y} \big] \; = \;   \sul{  n_p   -n_h    = \ell  }{} \Int{ \wh{\msc{C}}_p^{\ups} }{} \f{ \dd^{n_p }\la }{ n_p! (2\pi)^{n_p} }
\pl{ a=1 }{ n_p } \Big[  L\, \wh{\xi}^{\, \prime}_{1}\big( \la_a \mid \mf{Y}\big) \,  \ex{\i m  u_1(\la_a, \op{v}) }   \Big]   \\
\times \Int{ \wh{\msc{C}}_h^{\ups} }{} \f{ \dd^{n_h}\mu }{ n_h! (2\pi)^{n_h} }
\pl{ a=1 }{ n_h } \Big[  L\, \wh{\xi}^{\, \prime}_{1}\big( \mu_a \mid \mf{Y}\big)  \, \ex{-\i m  u_1(\mu_a, \op{v})}   \Big]
\cdot f_{n_p,n_h}\big( \{ \la_a \}_1^{ n_p }   ;  \{ \mu_a  \}_1^{ n_h }    \big) \;. 
\label{definition fnelle B ell de ups}
\end{multline}
Note that $\wh{\mc{B}}_{\ell_{\ups}}^{\, (\ups)}\Big[ \mf{F}_{\ups}\big(\mf{Y} \mid * \big) \mid \mf{Y} \Big] $ acts on the $\Ups_{\ups}^{(p)}\cup \Ups_{\ups}^{(h)}$ variables gathered
in $\mf{Z}^{\ups}_{\mf{Y}}$ as in  \eqref{definition des variables Z ups}. 

The action of $\wh{\mc{B}}_{\ell_{\ups}  }^{\, (\ups)} $ on $\mf{F}_{\ups}$, after the $\bs{:}*\bs{:}$ operator ordering and up to some $\de$ and $m,t$  dependent corrections,
is computed in Appendix \ref{Appendix Restricted sums}, \textit{c.f.} \eqref{ecriture resultat action operateur B ups}. 
  This result  entails the below expression 
for the thermodynamic limit of the form factor expression of the two-point function:
\bem
\big< \sg_1^{\ga^{\prime}}(t) \sg_{m+1}^{\ga} \big> \; = \;   (-1)^{m \op{s}_{\ga} }   \sul{  \bs{n} \in \mf{S}   }{}   
 \pl{ r \in  \mf{N} }{} \;\Bigg\{  \Int{ \big( \msc{C}_r^{(\de)} \big)^{n_r} }{}  \f{ \dd^{n_r}\nu^{(r)} }{ n_r! \cdot (2\pi)^{n_r}  }   \Bigg\} 
\cdot  \Int{ \big( \msc{C}_{h}^{(\de)} \big)^{n_h}   }{}  \f{ \dd^{n_h}\mu  }{ n_h! \cdot (2\pi)^{n_h}  } \cdot 
\f{  \mc{F}^{(\ga)}\big( \mf{Y} \big) \cdot  \ex{ \i m \msc{U}(\mf{Y}, \op{v})}    }{ \pl{\ups= \pm }{}  \big[  -  \i m_{\ups}  \big]^{   \vth_{\ups}^2(\mf{Y}) }  }      \\
 \times    \bigg( 1+\e{O}\Big( \de \ln \de +     \sum_{ \ups = \pm } \Big\{ \de^2 |m_{\ups}| +  \de  \big| \ln | m_{\ups} |  \big| + \ex{-|m_{\ups}|\de} \Big\} \Big)  \bigg)  \;. 
\label{ecriture limit thermo fct 2 pts forme prefinale avec discontinuites}
\end{multline}
Here, 
\beq
m_{\ups}=\ups m - \op{v}_F t
\enq
and it is assumed that $\op{v}=\tf{m}{t}\not= \pm \op{v}_F$. 
The function $\msc{U}(\mf{Y}, \op{v})$ has been introduced in \eqref{definition de energi impuslion reduite condensation massless}. $\mf{S}$ is as in \eqref{definition domaine de sommation mathfrak S}. 
Finally,  $\de>0$ is a control parameter that can be taken as small as necessary, but finite nonetheless. Indeed, owing to the singular behaviour of $\mc{F}^{(\ga)}\big( \mf{Y} \big) $
described by \eqref{ecriture comportement local FF smooth a rapidite coincidantes} when rapidities $\nu^{(1)}$ of the particles  or rapidities $\mu$ of the holes 
approach the endpoints of the Fermi zone $\pm q$, each individual particle-hole integral diverges in the $\de\tend 0^+$ limit. Furthermore, the \textit{a priori} control on 
the remainder blows up in the $\de \tend 0$ limit. 
Taken as a whole, the series has of course a well-defined $\de \tend 0^+$ limit, but taking this limit on the level of \eqref{ecriture limit thermo fct 2 pts forme prefinale avec discontinuites}
would demand additional resummations, along the line of ideas developed in \cite{KozReducedDensityMatrixAsymptNLSE,KozTerNatteSeriesNLSECurrentCurrent}. This 
is beyond the scope of the present analysis and, anyway, doing so would spoil the very structure of the form factor series  \eqref{ecriture limit thermo fct 2 pts forme prefinale avec discontinuites}
which makes it so useful for studying asymptotic regimes of correlation functions.

\subsection{The dressed momentum picture}
\label{SousSectionFcts2PtsDansRepImpulsion}

There is a final transformation of \eqref{ecriture limit thermo fct 2 pts forme prefinale avec discontinuites} that is necessary so as to put it in a form 
appropriate for further applications. Indeed, owing to the jump conditions \eqref{ecriture conditions saut FF}-\eqref{ecriture variables pour saut FF}
satisfied by the form factor density  $\mc{F}^{(\ga)}\big( \mf{Y} \big)$, those \eqref{ecriture conditions saut pour exposant et impulsion energie} 
satisfied by the exponent $\vth_{\ups}(\mf{Y})$ \eqref{definition exponsant critique ell shifte}
and by the combination of energy and momentum $\msc{U}\big( \mf{Y}, \op{v} \big)$ \eqref{definition de energi impuslion reduite condensation massless}, 
the integrand in each multiple integral is discontinuous in respect to $\nu_a^{(1)}$ when this variable passes from  $-\infty+\i \tf{\pi}{2}$ to  $-\infty$. 
The form of the jump conditions does however allow one to re-organise the series in such a form that the continuity between the mentioned points is achieved. 
Having a continuous, and in fact smooth, integrand along the integration contour is an important ingredient for the asymptotic analysis of the singular structure of 
dynamic response functions in the vicinity of the excitation thresholds. 

One starts by partitioning the contour $\msc{C}_1^{(\de)}=\msc{C}_p^{(\de)}$ as $\msc{C}_{1}^{(\de)} \, = \, \msc{C}_{1;L}\cup \msc{C}_{1;R}$ where 
\beq
\msc{C}_{1;L} \; = \; \msc{C}_1^{(\de)} \cap \Big\{ \Re(\la)\leq -q \; , \; |\Im(\la)| < \eps \Big\} \qquad \e{and} \qquad  \msc{C}_{1;R}=\msc{C}_1^{(\de)}\setminus \msc{C}_{1;L}\;. 
\enq
Here $\eps$ is taken small enough. Upon splitting the integration domain $\msc{C}_1^{(\de)}$ as above and then using the symmetry of the integrand, one recasts  \eqref{ecriture limit thermo fct 2 pts forme prefinale avec discontinuites}
as 
\bem
\big< \sg_1^{\ga^{\prime}}(t) \sg_{m+1}^{\ga} \big> \; = \;   (-1)^{m \op{s}_{\ga} }   \sul{  \bs{n} \in \mf{S}_{L,R}   }{}   
 \pl{ r \in  \mf{N}_{\e{st}} }{} \;\Bigg\{  \Int{ \big(\msc{C}_r^{(\de)}\big)^{n_r} }{}  \f{ \dd^{n_r}\nu^{(r)} }{ n_r! \cdot (2\pi)^{n_r}  }  \Bigg\} 
\cdot  \Int{ \big(\msc{C}_{h}^{(\de)}\big)^{n_h}   }{}  \f{ \dd^{n_h}\mu  }{ n_h! \cdot (2\pi)^{n_h}  }  \cdot  \pl{A\in \{L,R\}}{} \Bigg\{  \Int{ \msc{C}_{1;A} }{}  \f{ \dd^{n_{1;A} } \nu^{(1;A)} }{ n_{1;A}! \cdot (2\pi)^{n_{1;A}}  }    \Bigg\}  \\
 \times \f{  \mc{F}^{(\ga)}\big( \mf{Y}_{L,R} \big) \cdot  \ex{ \i m \msc{U}(\mf{Y}_{L,R}, \op{v})}    }{ \pl{\ups= \pm }{}  \big[  -  \i m_{\ups}  \big]^{   \vth_{\ups}^2(\mf{Y}_{L,R}) }  }   \cdot 
 \bigg( 1+\e{O}\Big( \de \ln \de +     \sum_{ \ups = \pm } \Big\{ \de^2 |m_{\ups}| +  \de  \big| \ln | m_{\ups} |  \big| + \ex{-|m_{\ups}|\de} \Big\} \Big)  \bigg)  \;. 
\end{multline}
There, the integration variables $\mf{Y}_{L,R}$ and the set $\mf{S}_{L,R}$ over which the various integers are being summed up, take the form 
\beqa
\mf{S}_{L,R} & = & \bigg\{ \bs{n}\in \big( n_h, n_{1;L}, n_{1;R}, \{n_r\}_{r \in \mf{N}_{\e{st}} }, \ell_{\ups} \big) \; : \;  n_h  \, + \, \op{s}_{\ga} \; = \; \sul{\ups \in \{\pm\} }{} \ell_{\ups} \, + \, n_{1;L} \, + \, n_{1;R}\, + \, \sul{ r\in \mf{N}_{\e{st}}  }{} r n_r   \bigg\}  \\
\mf{Y}_{L,R} & = & \bigg\{ \{ \mu_a \}_{1}^{n_h} ;  \{\nu_a^{(1;L)}\}_1^{n_{1;L}}\cup  \{\nu_a^{(1;R)}\}_1^{n_{1;R}} \cup \Big\{ \big\{ \nu_{a}^{(r)}\big\}_{a=1}^{n_r} \Big\}_{r\in \mf{N}_{\e{st}}} ; \{ \ell_{\ups} \} \bigg\} \;. 
\eeqa
Next one shifts the $\ell_{\ups}$ summation variables as $\ell_{\ups} \hookrightarrow \ell_{\ups} + \ups n_{1;L} u_1^{+}$ where I remind that $u_{1}^{+}=-\e{sgn}\big( \pi - 2 \zeta \big)$
as introduced in  \eqref{definition parametre u r sigma}.  
I stress that this change of variables is compatible with the constraint on the summation integers appearing in $\mf{S}_{L,R}$ in that it does not alter them. It then remains to observe that 
\beq
n_{1;L} \;  =  \; \sul{ A = \{L,R\} }{} \sul{ a=1  }{ n_{1;A} } \bs{1}_{ \msc{C}_{1;L} }\big( \nu_{a}^{(1;A)}\big)
\enq
where $\bs{1}_{A}$ is the indicator function of the set $A$, and then carry out backwards the contour decomposition so as to get 
\bem
\big< \sg_1^{\ga^{\prime}}(t) \sg_{m+1}^{\ga} \big> \; = \;   (-1)^{m \op{s}_{\ga} }   \sul{ \bs{n}\in \mf{S}   }{}   
 \pl{ r \in  \mf{N}  }{} \;\Bigg\{  \Int{ \big( \msc{C}_r^{(\de)} \big)^{n_r} }{}  \f{ \dd^{n_r}\nu^{(r)} }{ n_r! \cdot (2\pi)^{n_r}  }  \Bigg\} 
\cdot  \Int{ \big(\msc{C}_{h}^{(\de)}\big)^{n_h}   }{}  \f{ \dd^{n_h}\mu  }{ n_h! \cdot (2\pi)^{n_h}  }      \\
 \times \f{  \mc{F}^{(\ga)}\big( \mf{Y} \big) \cdot  \ex{ \i m \msc{U}(\mf{Y}, \op{v})}    }{ \pl{\ups= \pm }{}  \big[  -  \i m_{\ups}  \big]^{   \vth_{\ups}^2(\mf{Y}) }  }   \cdot 
 \bigg( 1+\e{O}\Big( \de \ln \de +     \sum_{ \ups = \pm } \Big\{ \de^2 |m_{\ups}| +  \de  \big| \ln | m_{\ups} |  \big| + \ex{-|m_{\ups}|\de} \Big\} \Big)  \bigg) \;. 
\end{multline}
In this series of multiple integrals expansion, the integration variables take the form
\beq
\mf{Y} \; = \;  \bigg\{ \{ \mu_a \}_{1}^{n_h} ;     \Big\{ \big\{ \nu_{a}^{(r)}\big\}_{a=1}^{n_r} \Big\}_{r\in \mf{N} } ; \Big\{ \ell_{\ups} + \ups \sul{a=1}{n_1} \bs{1}_{\msc{C}_{1;L}}\big(\nu_a^{(1)} \big) u_1^+ \Big\} \bigg\} 
\enq
and the summation is as given in \eqref{definition domaine de sommation mathfrak S}.
The jump conditions \eqref{ecriture conditions saut FF} and \eqref{ecriture conditions saut pour exposant et impulsion energie} satisfied by  the various building blocks of the integrand then ensure that the latter is a smooth function in respect to variables 
$\nu_{a}^{(1)}$ moving along $\msc{C}_1^{(\de)}$. 

The additional shifts in respect to the integers $\ell_{\ups}$ which are present in $\msc{U}(\mf{Y},\op{v})$ can be reabsorbed by
re-defining the dressed momentum of the particles as: 
\beq
\wh{p}_1(\la) \, = \, p_1(\la) \, + \, 2p_{F} u_1^{+} \bs{1}_{\msc{C}_{1;L}}(\la) \; + \; 2\pi \bs{1}_{ \msc{C}_{1;L} \cup \big\{ \R+\i\tfrac{\pi}{2}\big\} }(\la) \;. 
\enq
Note that adding a shift by $2\pi$ does not alter the expression since the function $p_1$ only appears in the combination  $\ex{\i m p_{1}(\nu_a^{(1)})}$. 
The matter is that due to the presence of the additional shifts, the function $\wh{p}_1$ is continuous along $\msc{C}_1^{(\de)}$ and, in fact, a diffeomorphism onto  
$\wh{p}_1(\msc{C}_1^{(\de)} )$.

\noindent It then remains to implement the changes of variables
\beq
\e{for} \;  r\in \mf{N}_{\e{st}} \;\;, \;\;   k_a^{(r)} \; = \; p_r\big( \nu_a^{(r)} \big)    \quad \e{and} \quad 
\left\{ \ba{ccc}  k_a^{(1)} & = & \wh{p}_1 \big( \nu_a^{(1)}  \big) \vspace{2mm} \\ 
		      t_a & = & p_1(\mu_a)    \ea  \right.   \;. 
\enq
This transforms the oriented integration contours into the oriented curves
\beq
\msc{I}_h \, = \, p_1\big(\,  \msc{C}_{h}^{(\de)} \, \big) \;\;  , \quad  \msc{I}_1 \, = \, \wh{p}_1\big( \, \msc{C}_{1}^{(\de)} \,  \big) \quad \e{and} \quad 
 \msc{I}_r \, = \, p_r\big( \, \msc{C}_{r} \,  \big) \;. 
\label{ecriture des contours integration dans les variables impulsion}
\enq
 The resulting series takes the form 
\bem
\big< \sg_1^{\ga^{\prime}}(t) \sg_{m+1}^{\ga} \big> \; = \;   (-1)^{m \op{s}_{\ga} }   \sul{  \bs{n}\in \mf{S}   }{}   
 \pl{ r \in  \mf{N} }{} \;\Bigg\{  \Int{ \big( \msc{I}_r \big)^{n_r} }{}  \f{ \dd^{n_r}k^{(r)} }{ n_r! \cdot (2\pi)^{n_r}  } \pl{a=1}{n_r}\Big[   \ex{\i m \mf{u}_r(k_a^{(r)}, \op{v})}   \Big]   \Bigg\} 
\cdot  \Int{   \big( \msc{I}_h \big)^{n_h}   }{}  \f{ \dd^{n_h}t   }{ n_h! \cdot (2\pi)^{n_h}  }  \pl{a=1}{n_h}\Big[ \ex{-\i m \mf{u}_1(t_a, \op{v})} \Big] \\
 \times    \ov{\mc{F}}^{(\ga)}\big( \mf{K} \big) \cdot 
\pl{\ups= \pm }{} \bigg\{  \f{  \ex{ \i m \ups \ell_{\ups} p_{F} }   }{  \big[  -  \i m_{\ups}  \big]^{  \De_{\ups}(\mf{K})   }  }  \bigg\} \cdot 
 \bigg( 1+\e{O}\Big( \de \ln \de +     \sum_{ \ups = \pm } \Big\{ \de^2 |m_{\ups}| +  \de  \big| \ln | m_{\ups} |  \big| + \ex{-|m_{\ups}|\de} \Big\} \Big)  \bigg)  \;. 
\label{ecriture limit thermo fct 2 pts forme finale}
\end{multline}
There, I have introduced 
\beq
\mf{K} \; = \; 
 \bigg\{ \{ t_a  \}_{1}^{n_h} ;  \Big\{ \big\{ k_a^{(r)}  \big\}_{a=1}^{n_r} \Big\}_{r\in \mf{N} }
 ; \Big\{ \ell_{\ups}  \Big\} \bigg\} \;. 
\enq
Furthermore, the functions arising in the integrand take the form 
\beq
 \ov{\mc{F}}^{(\ga)}\big( \mf{K} \big) \; = \; \mc{F}^{(\ga)}\big( \mf{Y}_{p^{-1}\cdot \mf{K} } \big)  \pl{ r \in  \mf{N}_{\e{st}} }{} \pl{a=1}{n_r}  \Bigg\{ \f{ 1 }{  p_r^{\prime}\circ p_r^{-1}\big(k_a^{(r)} \big) } \Bigg\}
\pl{a=1}{n_h} \Bigg\{ \f{1}{ p_1^{\prime}\circ p_1^{-1}\big(t_a \big) } \Bigg\}  \cdot \pl{a=1}{n_1}  \Bigg\{ \f{ 1 }{ p_1^{\prime}\circ \wh{p}_1^{\, -1}\big(k_a^{(1)} \big)  }  \Bigg\}
\label{definition bar FF density}
\enq
$\De_{\ups}(\mf{K})  \, = \, \vth_{\ups}^2\big(\mf{Y}_{p^{-1}\cdot \mf{K} } \big)   $ and 
\beq
 \mf{Y}_{p^{-1}\cdot \mf{K} } \; = \; 
 \bigg\{ \{  p_1^{-1}\big(t_a \big) \}_{1}^{n_h} ;  \{  \wh{p}_1^{\, -1}\big(k_a^{(1)} \big)  \}_1^{ n_{1} }\cup   \Big\{ \big\{ p_r^{-1}\big(k_a^{(r)} \big) \big\}_{a=1}^{n_r} \Big\}_{r\in \mf{N}_{\e{st}} }
 ; \Big\{ \ell_{\ups} + \ups u_1^{+}\sul{a=1}{n_1} \bs{1}_{\msc{C}_{1;L}}\big( \wh{p}_1^{\, -1}\big(k_a^{(1)} \big) \big) \Big\} \bigg\} \;. 
\enq

Finally, I agree upon 
\beq
\mf{u}_r\big(k,\op{v} \big) \, = \, k -\f{\mf{e}_r(k)}{\op{v} } \qquad \e{with} \qquad \mf{e}_{1}  \, = \, \veps_{1}\circ \wh{p}_1^{\,-1} \quad \e{and} \quad  \mf{e}_{r}  \, = \, \veps_{r}\circ p_r^{-1} \;, \; r \in \mf{N}_{\e{st}} \;.  
\label{definition energie dans representation impulsion}
\enq

I stress that the Jacobian in \eqref{definition bar FF density} appears without the absolue value since the orientation of the contours in \eqref{ecriture des contours integration dans les variables impulsion}
is preserved. In particular, within such a construction, the intervals $\msc{I}_{h}, \msc{I}_{r}$ are skimmed through from the smallest to the largest element.

\subsection{The dynamic response functions}
\label{SousSection fonctions de reponse dynamiques}

Observe that when a particle $k_a^{(1)}$ or a hole $t_a$ momentum approaches one of the endpoints of its respective domain of integration, then the associated 
 oscillatory phases take complex values hence generating an exponential decay in $m, t \tend +\infty$. 
On the one hand, the presence of complex valued oscillatory phases is not that convenient for taking the space and time Fourier transform. 
On the other hand, as will be confirmed below,  such integrations only contribute as higher order corrections in $m_{\ups}$. 
Thus prior to computing the dynamic response functions, it appears convenient to recast the form factor series in such a way 
that the integrations outside of real intervals are included into corrections.

\begin{figure}[ht]
\begin{center}

\begin{pspicture}(12,9)

\psline[linestyle=dashed, dash=3pt 2pt](0,7.25)(11.8,7.25)
\psline[linewidth=2pt]{->}(11.8,7.25)(11.9,7.25) 

\rput(11.8,6.9){$\R $}

\psdots(0.5,8.2)(0.5,7.25)(1.75,7.25)(4.25,7.25)(5.5,6.3)(5.5,7.25)(5.5,8.2)(6.75,7.25)(9.75,7.25)(11,7.25)(11,6.3)

\psline[linecolor=blue,linewidth=2pt]{->}(8.5,7.25)(8.6,7.25)
\psline[linecolor=orange,linewidth=2pt]{->}(3,7.25)(3.1,7.25)


\pscurve[linecolor=orange](0.5,8.2)(1,8.1)(1.5,7.6)(1.75,7.25)
\psline[linecolor=orange](1.75,7.25)(4.25,7.25)
\pscurve[linecolor=orange](4.25,7.25)(4.5,6.9)(5,6.4)(5.5,6.3)
 
\rput(1.75,7){$\eps \! - \! p_F$}
\rput(4.25,7.5){$p_F\! - \! \eps$}  
\rput(6.75,7){$p_F\! + \! \eps$}  
\rput(9.75,7.5){$p_{\e{mx}}\! - \! \eps$}  
\rput(11,7){$p_{\e{mx}}$}  

\pscurve[linecolor=blue](5.5,8.2)(6,8.1)(6.5,7.6)(6.75,7.25)
\psline[linecolor=blue](6.75,7.25)(9.75,7.25)
\pscurve[linecolor=blue](9.75,7.25)(10,6.9)(10.5,6.4)(11,6.3)

\rput(0.5,7){$ -p_F  $}
\rput(5.4,7){$ p_F  $}

\rput(11.4,6.3){$p_{-;\ua}$}
\rput(0.1,8.1){$p_{-;\da}$}

\rput(5.5,8){$p_{+;\ua}$}
\rput(5.9,6.3){$p_{+;\da}$}

\rput(1.5,8.1){$\mf{I}_{h}^{(L)}$}
\rput(3,6.9){$\msc{J}_h^{(\eps)}$}
\rput(4.6,6.2){$\mf{I}_{h}^{(R)}$}

\rput(6.5,8.1){$\mf{I}_{1}^{(L)}$}
\rput(8,6.9){$\msc{J}_1^{(\eps)}$}
\rput(10.1,6.2){$\mf{I}_{1}^{(R)}$}


\psline(0,5.5)(12,5.5)


\psline[linestyle=dashed, dash=3pt 2pt](0,4.25)(11.8,4.25)
\psline[linewidth=2pt]{->}(11.8,4.25)(11.9,4.25) 

\rput(11.8,3.9){$\R $}

\psdots(0.5,3.3)(0.5,4.25)(1.75,4.25)(4.25,4.25)(5.5,3.3)(5.5,4.25)(5.5,5.2)(6.75,4.25)(9.75,4.25)(11,4.25)(11,5.2)

\psline[linecolor=blue,linewidth=2pt]{->}(8.5,4.25)(8.6,4.25)
\psline[linecolor=orange,linewidth=2pt]{->}(3,4.25)(3.1,4.25)


\pscurve[linecolor=orange](0.5,3.3)(1,3.4)(1.5,3.9)(1.75,4.25)
\psline[linecolor=orange](1.75,4.25)(4.25,4.25)
\pscurve[linecolor=orange](4.25,4.25)(4.5,4.6)(5,5.1)(5.5,5.2)
 
\rput(1.75,4.5){$\eps \! - \! p_F$}
\rput(4.25,4){$p_F\! - \! \eps$}  
\rput(6.75,4.5){$p_F\! + \! \eps$}  
\rput(9.75,4){$p_{\e{mx}}\! - \! \eps$}  
\rput(11,4){$p_{\e{mx}}$}  

\pscurve[linecolor=blue](5.5,3.3)(6,3.4)(6.5,3.9)(6.75,4.25)
\psline[linecolor=blue](6.75,4.25)(9.75,4.25)
\pscurve[linecolor=blue](9.75,4.25)(10,4.6)(10.5,5.1)(11,5.2)

\rput(0.5,4){$ -p_F  $}
\rput(5.4,4){$ p_F  $}

\rput(11.4,5.1){$p_{-;\ua}$}
\rput(0.1,3.3){$p_{-;\da}$}

\rput(5.9,5.2){$p_{+;\ua}$}
\rput(5.1,3.3){$p_{+;\da}$}

\rput(1.3,3.2){$\mf{I}_{h}^{(L)}$}
\rput(3,3.9){$\msc{J}_h^{(\eps)}$}
\rput(4.6,5.1){$\mf{I}_{h}^{(R)}$}

\rput(6.5,3.2){$\mf{I}_{1}^{(L)}$}
\rput(8,3.9){$\msc{J}_1^{(\eps)}$}
\rput(10.1,5.1){$\mf{I}_{1}^{(R)}$}


\psline(0,2.5)(12,2.5)

6.3

\psline[linestyle=dashed, dash=3pt 2pt](0,1.25)(11.8,1.25)
\psline[linewidth=2pt]{->}(11.8,1.25)(11.9,1.25) 

\rput(11.8,0.9){$\R $}

\psdots(0.5,0.3)(0.5,1.25)(1.75,1.25)(4.25,1.25)(5.5,0.3)(5.5,1.25)(5.5,2.2)(6.75,1.25)(9.75,1.25)(11,1.25)(11,2.2)

\psline[linecolor=blue,linewidth=2pt]{->}(8.5,1.25)(8.6,1.25)
\psline[linecolor=orange,linewidth=2pt]{->}(3,1.25)(3.1,1.25)


\pscurve[linecolor=orange](0.5,0.3)(1,0.4)(1.5,0.9)(1.75,1.25)
\psline[linecolor=orange](1.75,1.25)(4.25,1.25)
\pscurve[linecolor=orange](4.25,1.25)(4.5,0.9)(5,0.4)(5.5,0.3)
 
\rput(1.75,1.5){$\eps \! - \! p_F$}
\rput(4.25,1.5){$p_F\! - \! \eps$}  
\rput(6.75,1){$p_F\! + \! \eps$}  
\rput(9.75,1){$p_{\e{mx}}\! - \! \eps$}  
\rput(11,1){$p_{\e{mx}}$}  

\pscurve[linecolor=blue](5.5,2.2)(6,2.1)(6.5,1.6)(6.75,1.25)
\psline[linecolor=blue](6.75,1.25)(9.75,1.25)
\pscurve[linecolor=blue](9.75,1.25)(10,1.6)(10.5,2.1)(11,2.2)

\rput(0.5,1){$ -p_F  $}
\rput(5.4,1){$ p_F  $}

\rput(11.4,2.1){$p_{-;\ua}$}
\rput(0.1,0.3){$p_{-;\da}$}

\rput(5.5,2){$p_{+;\ua}$}
\rput(5.9,0.3){$p_{+;\da}$}

\rput(1.3,0.2){$\mf{I}_{h}^{(L)}$}
\rput(3,0.9){$\msc{J}_h^{(\eps)}$}
\rput(4.6,0.2){$\mf{I}_{h}^{(R)}$}

\rput(6.5,2.1){$\mf{I}_{1}^{(L)}$}
\rput(8,0.9){$\msc{J}_1^{(\eps)}$}
\rput(10.1,2.1){$\mf{I}_{1}^{(R)}$}

\end{pspicture}
\caption{  Particle $ \mf{I}_{1}^{(L)}\cup \msc{J}_1^{(\eps)} \cup \mf{I}_{1}^{(R)}$ -in blue- and hole $ \mf{I}_{h}^{(L)}\cup \msc{J}_h^{(\eps)} \cup \mf{I}_{h}^{(R)}$ -in orange- deformed contours
in the momentum representation. The contours are plotted for the three regimes of the velocity 
$\op{v}=\tf{m}{t}$ appearing from bottom to top $|\op{v}|>\op{v}_F$, $\op{v}_F>\op{v}>0$ and $0>\op{v}>-\op{v}_F$. The contour $\msc{C}_{p}^{(\de)}$ and $\msc{C}_{p}^{(\de)}$
start at the points $p_{\pm;\ua/\da}$ such that $\Im(p_{\pm;\ua/\da}) =  \i \eps_{\ua/\da} \de $, where $\eps_{\ua}=+1$ and $\eps_{\da}=-1$, 
and then, over a distance of the order of $\eps$ they joint with the real axis. Finally, one has $p_{\e{mx}}=2\pi \!-\!p_F\!-\!2p_F\e{sgn}(\pi-2\zeta)$. 
\label{Figure contour Gamma p et h reduits delta deformes a la limite thermo rep impulsion} }
\end{center}

\end{figure}

For such a purpose, by using the holomorphicity of the integrands, one slightly deforms the contours $\msc{I}_{h}$ and $\msc{I}_1$ according to Fig. \ref{Figure contour Gamma p et h reduits delta deformes a la limite thermo rep impulsion}
what ends up with the replacements
\beq
\msc{I}_{h} \; \hookrightarrow \; \mf{I}_{h}^{(L)}\cup \msc{J}_h^{(\eps)} \cup \mf{I}_{h}^{(R)} \quad , \quad \msc{I}_{1} \; \hookrightarrow \;  \mf{I}_{1}^{(L)}\cup \msc{J}_1^{(\eps)} \cup \mf{I}_{1}^{(R)}  \;. 
\enq
Above, $\eps>0$ can be taken as small as necessary, 
\beq
\msc{J}_h^{(\eps)}\,=\, \intoo{-p_F+\eps}{ p_F- \eps} \quad , \quad \msc{J}_1^{(\eps)}\,=\, \intoo{p_F+\eps}{ 2\pi -p_F - \eps -2p_F \e{sgn}(\pi-2\zeta)  } 
\label{introduction des intervalles Jh et J1 epsilon deformees}
\enq
while the curves $\mf{I}_{h}^{(L/R)}$, resp. $\mf{I}_{1}^{(L/R)}$, joint the left/right endpoints of $\msc{J}_{h}^{(\eps)}$, resp. $\msc{J}_{1}^{(\eps)}$, 
to the nearby endpoints of the original intervals $\msc{I}_h$ and $\msc{I}_1$, this by solely passing in the complex half-plane where this endpoint belongs to. 
As a consequence, it follows that 
\beq
|\ex{-\i m \mf{u}_1(t,\op{v})} | < 1  \quad  \e{for}  \quad t \in \e{Int}\big(\mf{I}_{h}^{(L/R)}\big) \quad 
\e{and},\, \e{likewise}, \quad  |\ex{ \i m \mf{u}_1(k,\op{v})} | < 1  \quad \e{for} \quad  k \in \e{Int}\big(\mf{I}_{1}^{(L/R)}\big) \; . 
\enq
By following an analogous procedure to the one described in the previous section, one reorganises the form factor expansion given in  \eqref{ecriture limit thermo fct 2 pts forme finale}
as 
\bem
\big< \sg_1^{\ga^{\prime}}(t) \sg_{m+1}^{\ga} \big> \; = \;   (-1)^{m \op{s}_{\ga} }   \sul{  \bs{n} \in \mf{S}_{\e{part}}   }{}   
 \pl{ r \in  \mf{N}_{\e{st}} }{} \;\Bigg\{  \Int{ \big(\msc{I}_r\big)^{n_r} }{}  \f{ \dd^{n_r}k^{(r)} }{ n_r! \cdot (2\pi)^{n_r}  } \pl{a=1}{n_r}\Big[   \ex{\i m \mf{u}_r(k_a^{(r)}, \op{v})}   \Big]   \Bigg\} \\
\times  \Int{ \big(\msc{J}_1^{(\eps)}\big)^{n_1} }{}  \f{ \dd^{n_1}k^{(1)} }{ n_1! \cdot (2\pi)^{n_1}  } \pl{a=1}{n_1}\Big[   \ex{\i m \mf{u}_1(k_a^{(1)}, \op{v})}   \Big] 
\cdot   \Int{  \big(\msc{J}_h^{(\eps)}\big)^{n_h}    }{}  \f{ \dd^{n_h}t   }{ n_h! \cdot (2\pi)^{n_h}  }  \pl{a=1}{n_h}\Big[ \ex{-\i m \mf{u}_1(t_a, \op{v})} \Big] \\
 \times  \mc{W}_{ n_h^{(*)}, n_1^{(*)} } \Bigg[ \,   \ov{\mc{F}}^{(\ga)}\big( \mf{K}_{\e{part}} \big) \cdot 
\pl{\ups= \pm }{} \bigg\{  \f{  \ex{ \i m \ups \ell_{\ups} p_{F} }   }{  \big[  -  \i m_{\ups}  \big]^{  \De_{\ups}( \mf{K}_{\e{part}} )   }  }  \bigg\}  \cdot 
 \bigg( 1+\e{O}\Big( \de \ln \de +     \sum_{ \ups = \pm } \Big\{ \de^2 |m_{\ups}| +  \de  \big| \ln | m_{\ups} |  \big| + \ex{-|m_{\ups}|\de} \Big\} \Big)  \bigg) \Bigg]  \;. 
\label{ecriture serie FF momentum partitionne sur integrale complexe et reguliere}
\end{multline}
where
\beqa
\mf{S}_{\e{part}} & = & \bigg\{ \bs{n} \in \big( \{ n_h^{(L)},n_h,n_h^{(R)} \} , \{ n_1^{(L)},n_1,n_1^{(R)} \} ,  \{n_r\}_{r \in \mf{N}_{\e{st}} }, \ell_{\ups} \big) \; : \;  
  n_h  \, + \, \op{s}_{\ga}   \; = \; \sul{\ups \in \{\pm\} }{} \ell_{\ups}    +    \sul{ r\in \mf{N}   }{} r n_r    + \hspace{-2mm} \sul{ A = \{L,R\} }{} \hspace{-2mm} \big[n_1^{(A)}\, -n_h^{(A)}\big]  \bigg\}  \nonumber \\
\mf{K}_{\e{part}} & = & \bigg\{ \Big\{ \{ t_a^{(L)} \}_{1}^{n_h^{(L)}} \cup  \{ t_a \}_{1}^{n_h} \cup \{ t_a^{(R)} \}_{1}^{n_h^{(R)}} \Big\};
\Big\{ \{k_a^{(1;L)}\}_1^{n_{1;L}}\cup \{k_a^{(1)}\}_1^{n_{1}} \cup  \{ k_a^{(1;R)}\}_1^{n_{1;R}} \Big\} \cup 
\Big\{ \big\{ k_{a}^{(r)}\big\}_{a=1}^{n_r} \Big\}_{r\in \mf{N}_{\e{st}}} ; \{ \ell_{\ups} \} \bigg\} \;. 
\nonumber
\eeqa
$\mc{W}_{ n_h^{(*)}, n_1^{(*)} } $ is an integral operator acting on the variables $\{t_a^{(*)}\}$ and $\{k_a^{(1;*)}\}$ present in $ \mf{K}_{\e{part}} $
as
\bem
\mc{W}_{ n_h^{(*)}, n_1^{(*)} }\Big[ f \Big] \; = \; \pl{ A \in \{L,R\} }{} \;\Bigg\{  \Int{ \mf{J}_1^{(A)} }{}  \f{ \dd^{ n_1^{(A)} } k^{(1;A)} }{ n_{1}^{(A)} \cdot (2\pi)^{ n_{1}^{(A)} }  } 
\pl{a=1}{n_{1}^{(A)}}\Big[   \ex{\i m \mf{u}_r(k_a^{(1;A)}, \op{v})}   \Big]   \Bigg\}  \\
\times   \pl{ A \in \{L,R\} }{} \;\Bigg\{   \Int{ \mf{J}_h^{(A)} }{}  \f{ \dd^{ n_h^{(A)} } t^{(A)} }{ n_{h}^{(A)} \cdot (2\pi)^{ n_{h}^{(A)} }  }  
\pl{ a=1 }{ n_{h}^{(A)} }\Big[ \ex{-\i m \mf{u}_1(t_a^{(A)}, \op{v})} \Big]   \Bigg\} \;
f\Big(   \{ t_a^{(L)} \}_{1}^{n_h^{(L)}} , \{ t_a^{(R)} \}_{1}^{n_h^{(R)}} ,  \{k_a^{(1;L)}\}_1^{n_{1;L}} ,  \{ k_a^{(1;R)}\}_1^{n_{1;R}}    \Big) \;. 
\end{multline}

The functions occurring in the argument of $\mc{W}_{ n_h^{(*)}, n_1^{(*)} }$ in \eqref{ecriture serie FF momentum partitionne sur integrale complexe et reguliere}
exhibit power-law singularities when the particle, resp. hole, momenta approach $p_F$ or $2\pi-p_F-2p_F\e{sgn}(\pi-2\zeta)$, resp. $\pm p_F$,
as can be inferred from the local behaviour of form factor densities in rapidity space 
\eqref{ecriture comportement local FF smooth a rapidite coincidantes}-\eqref{ecriture fonction singuliere D}.
These information and a straightforward application of Watson lemma in the vicinity of the points $\pm(p_F-\eps)$, $p_{F}+\eps$, $2\pi-p_F-2p_F\e{sgn}(\pi-2\zeta)-\eps$
along with the exponential decay of the exponents of oscillatory phases outside of these endpoints
on $\mf{J}_{h/1}^{(L, R)}$ ensure that,  whenever $n_h^{(A)}\not=0$ or $n_1^{(A)}\not=0$
for some $A\in \{L,R\}$, the second line of \eqref{ecriture serie FF momentum partitionne sur integrale complexe et reguliere} behaves at most as $\e{O}\Big( \sul{ \ups = \pm }{} |m_{\ups}|^{-1} \Big) $. 
This behaviour can then be included into the corrections, hence leading to the below form of the massless form factor expansion for two-point functions
\bem
\big< \sg_1^{\ga^{\prime}}(t) \sg_{m+1}^{\ga} \big> \; = \;   (-1)^{m \op{s}_{\ga} }   \sul{  \mf{S}   }{}   
 \pl{ r \in  \mf{N} }{} \;\Bigg\{  \Int{ \big( \msc{J}_r^{(\eps)} \big)^{n_r} }{}  \f{ \dd^{n_r}k^{(r)} }{ n_r! \cdot (2\pi)^{n_r}  } \pl{a=1}{n_r}\Big[   \ex{\i m \mf{u}_r(k_a^{(r)}, \op{v})}   \Big]   \Bigg\} 
\cdot  \Int{ \big( \msc{J}_h^{(\eps)} \big)^{n_h}   }{}  \f{ \dd^{n_h}t   }{ n_h! \cdot (2\pi)^{n_h}  }  \pl{a=1}{n_h}\Big[ \ex{-\i m \mf{u}_1(t_a, \op{v})} \Big] \\
 \times    \ov{\mc{F}}^{(\ga)}\big( \mf{K} \big) \cdot 
\pl{\ups= \pm }{} \bigg\{  \f{  \ex{ \i m \ups \ell_{\ups} p_{F} }   }{  \big[  -  \i m_{\ups}  \big]^{  \De_{\ups}(\mf{K})   }  }  \bigg\} \cdot 
 \bigg( 1+\e{O}\Big( \de \ln \de +     \sum_{ \ups = \pm } \Big\{ \de^2 |m_{\ups}| +  \de  \big| \ln | m_{\ups} |  \big| + \ex{-|m_{\ups}|\de}+ \f{1}{|m_{\ups}|} \Big\} \Big)  \bigg)  \;. 
\label{ecriture form amenagee masslessFF expansion}
\end{multline}
Here, $\msc{J}_1^{(\eps)}$ and $\msc{J}_h^{(\eps)}$ are as introduced earlier on while $\msc{J}_r^{(\eps)}=\msc{I}_r$ for $r \in \mf{N}_{\e{st}}$. 

The representation \eqref{ecriture form amenagee masslessFF expansion} is almost in good form so as to allow for the 
computation of the time and space Fourier transform of $\big< \sg_1^{\ga^{\prime}} \sg_{m+1}^{\ga}(t) \big>$. One only needs to 
put the remainder terms into a more uniform form. 
Recall that the parameter $\de>0$  can be set to the desired value. In particular, one can take
\beq
\de \, = \,  \f{ C }{ [|m_+|+|m_-|+1]^{1-\tf{\tau}{2}} }
\enq
with $C>0$ and small enough so as to ensure a "minimal" smallness of $\de$ and where $0<\tau<1$ is arbitrary.  Substituting this in the remainder and using that
 one has $\ex{-|m_{\ups}|\de}=\e{O}\big( [|m_{\ups}|\de]^{-\f{2}{\tau}}  \big)$ in the regime of interest, one gets that 
\beq
\e{O}\Big( \de \ln \de +     \sum_{ \ups = \pm } \Big\{ \de^2 |m_{\ups}| +  \de  \big| \ln | m_{\ups} |  \big| + \ex{-|m_{\ups}|\de}+ \f{1}{|m_{\ups}|} \Big\} \Big)  
\; = \; \e{O}\Big( \sum_{ \ups = \pm } \f{1}{|m_{\ups}|^{1-\tau}}  \Big)\;. 
\enq
Once that the remainders have been replaced as above, the dynamic response functions given in \eqref{definition dynamic response function}  can be computed by using the results gathered in Appendix \ref{Appendix discreto continuous FT},
Proposition \ref{Proposition Calcul des TFs}. \textit{Per se}, the application of that result demands a slightly better control on the asymptotic expansion of the remainder than the one given just above. 
However, it is relatively clear that by pushing further the techniques of asymptotic expansions developed in this work one would indeed get a remainder that indeed takes the desired form, as given in \eqref{forme DA W}.
Then, the dynamic response function takes the form 
\beq
\msc{S}^{(\ga)}(k,\om) \; = \;     \sul{   \bs{n} \in  \mf{S}   }{}    \msc{S}^{(\ga)}_{\bs{n}}(k,\om) 
\label{ecriture dynamic response factor}
\enq
where the summation runs through $\bs{n}=(n_h,n_1, n_{r_1},\dots, n_{ r_{ |\mf{N}_{\e{st}}|} },\ell_{+},\ell_{-})$ with $\mf{N}_{\e{st}}=\{r_1,\dots, r_{ |\mf{N}_{\e{st}}|}  \}$. The summand and $\msc{S}^{(\ga)}_{\bs{n}}(k,\om) $
corresponds to the contribution to the dynamic response function of all the excited states that are characterised by  fixed Umklapp integers $\ell_{\ups}$ and having $n_h$ holes in the bulk of the Fermi zone, 
 $ n_1 $ particles lying uniformly away from the endpoints of the Fermi zone and having $n_{r}$ $r$-string excitations, with $r\in \mf{N}_{\e{st}}$. It takes the form 
\bem
\msc{S}^{(\ga)}_{\bs{n}}(k,\om)  \, = \, 
 \Int{  \big( \msc{J}_h^{(\eps)} \big)^{n_h}    }{}  \dd^{n_h}t  \cdot \pl{ r \in  \mf{N} }{} \; \bigg\{  \Int{ \big( \msc{J}_r^{(\eps)} \big)^{n_r} }{}   \dd^{n_r}k^{(r)}   \bigg\}  \cdot \;  \wt{\mc{F}}^{(\ga)}\big( \mf{K} \big)
  \sul{s\in \mathbb{Z} }{} \pl{\ups= \pm }{} \bigg\{  \Xi\Big(\,  \wh{\mf{z}}_{\ups}\big( \mf{K}; s\big)  \Big) \cdot \Big[ \, \wh{\mf{z}}_{\ups}\big( \mf{K}; s\big)  \Big]^{ \De_{\ups}(\mf{K}) -1 }  \bigg\}   \\
 \times   
\bigg( 1+\e{O}\Big(   \sum_{ \ups = \pm } \big|\, \wh{\mf{z}}_{\ups}\big( \mf{K}; s\big) \big|^{1-\tau}   \Big)  \bigg)  \;. 
\label{ecriture contrib excitation donnee facteur structure}
\end{multline}
The integrand of \eqref{ecriture contrib excitation donnee facteur structure} is built up from two contributions. The first one is smooth in $\mf{K}$  and corresponds to a dressing of the form factor density 
\beq
\wt{\mc{F}}^{(\ga)}\big( \mf{K} \big) \;=\; \f{ (2\pi)^2 \cdot \ov{\mc{F}}^{(\ga)}\big( \mf{K} \big)  \cdot    \big[ 2\op{v}_F \big]^{-\De_{+}  ( \mf{K} ) -\De_{-} ( \mf{K} ) +1 } }
				    { n_h!   (2\pi)^{n_h} \cdot  \pl{ r \in  \mf{N} }{}  \big\{  n_r! \, (2\pi)^{n_r} \big\}   \cdot  \Ga\Big(\De_{+} \big( \mf{K} \big)  \Big)  \cdot  \Ga\Big(\De_{-} \big( \mf{K} \big)   \Big)} 
\enq
with $\ov{\mc{F}}^{(\ga)}\big( \mf{K} \big)$, $\De_{\ups}(\mf{K})$ defined resp. in \eqref{definition bar FF density} and below it. 
The second contribution introduces singularities in the integrand as it is the one responsible for the existence of an edge singular behaviour of the 
spectral functions. This contributions is given in terms of powers of the functions 
\beq
\wh{\mf{z}}_{\ups}\big( \mf{K}; s\big)  \; = \;  \om - \mc{E}(\mf{K}) +  \ups \op{v}_{F} \big[ k - \mc{P}(\mf{K})  + 2\pi s \big]    
\enq
in which  
\beq
\mc{P}(\mf{K}) \, = \, \sul{ r \in  \mf{N} }{}  \sul{a=1}{n_r} k_a^{(r)}  \, + \, p_{F}\sul{\ups=\pm}{}  \ups \ell_{\ups} 
+ \pi  \op{s}_{\ga} - \sul{a=1}{n_h} t_a
\enq
and
\beq
\mc{E}(\mf{K}) \, = \, \sul{ r \in  \mf{N} }{}  \sul{a=1}{n_r} \mf{e}_r\big( k_a^{(r)} \big) \,- \,  \sul{a=1}{n_h} \mf{e}_1(\mu_a) \;. 
\enq
I remind that $\mf{e}_a$ has been defined in \eqref{definition energie dans representation impulsion}. Also, the integrations runs along the intervals 
$\msc{J}_h^{(\eps)}$ and $ \msc{J}_1^{(\eps)}$ given in \eqref{introduction des intervalles Jh et J1 epsilon deformees} while $\msc{J}_r^{(\eps)}\,=\, \msc{I}_{r}$
for $r \in \mf{N}_{\e{st}}$. I stress again that $\eps>0$ and $\tau>0$ are control parameters which are arbitrary but should be taken finite so that the controls on the remainders do not blow up.

Finally, one should observe that the summation over $s$ in \eqref{ecriture contrib excitation donnee facteur structure} simply translates the fact that the spectral function is a $2\pi$ periodic function of $k$, owing to the 
discrete nature of the XXZ chain.

\section{Phenomenological form of  massless form factor expansions for a model belonging to a Luttinger liquid universality class}
\label{Section extension a autres modeles dans class de LL}

I will now argue that the form taken by the form factor expansions in the massless regime of the XXZ chain obtained above captures the full structure of a form factor expansion arising 
in a massless one dimensional model belonging to the Luttinger liquid universality class. As will be shown in forthcoming publications, it is this structure that is responsible for all the universal features shared by
models belonging to this class. Hence having at one's disposal the general phenomenological form of a form factor expansion in a model belonging to the Luttinger liquid universality class 
may lead to an even deeper insight into the universal properties of such models. 

Given a massless one dimensional model, one expects that, when the volume gets large enough, the excited states can be interpreted as being built up from  various species of particle excitations and/or 
from hole excitations located inside the Fermi zone  
of the model. I will assume in the following that the model has one Fermi zone, as in the XXZ chain case, although the case of several Fermi zones can be treated analogously. 
Given that there is one Fermi zone, there will be one species of particles which can generate excitations right above that zone and hence that has a massless spectrum. 
In fact, the massless part of the spectrum corresponds to making excitations located directly on the endpoints of the Fermi zone. 

The various elementary excitations can be parametrised by collections of their rapidities, similarly to \eqref{ecriture rapidites macr decrivant etat}. 
Also, it is natural to expect that these rapidities will encapsule enough data so as to fully parametrise the observables of the model associated with a given excited state. 
For a model in finite volume, the rapidities will be quantised. The quantisation equations will take the generic multi-variable form as in \eqref{equations pour rapidities part trous et cordes}. 
Still, for a generic model, the functions $\wh{\xi}_{a}$ are unknown; thus the quantisation equations can take, in principle, any form and do not carry as such any data. 
However, if one adds the physical requirement that the elementary excitations should propagate as free independent particles in the infinite volume limit, then
the functions realising the quantisation equations should take the form \eqref{equations donnant fct cptge r}. 
The $p_a$'s will then be some unexplicit model dependent functions in \textit{one} variable and the $1/L$ corrections involving $F_a$ will introduce the dependence on the other rapidities. 

\vspace{2mm}

Owing to the multi-particle species interpretation of the model's spectrum of excitations, the relative to the ground state excitation energy and momentum
will have the form given in \eqref{ecriture forme energie relative excitation grand L}, \eqref{ecriture forme leading energie ex}, \eqref{ecriture forme leading impulsion ex}. 
The difference with the XXZ chains is that the momentum and energy associated with the various elementary excitation species are now unknown, model dependent, functions. 
Imagine a situation where a particle and a hole both collapse onto a given endpoint of the Fermi zone. To the leading order in their distance to the Fermi zone, one should not be able to distinguish between
this situation and the one where this particle and hole are both absent in the excited state. Thus, regarding to massless excitations, be it particle or hole ones, only the relative differences $\ell_{\pm}=n_p^{\pm}-n_h^{\pm}$
between the number of particles $n_p^{\pm}$ and holes $n_h^{\pm}$ collapsing on a given Fermi boundary should matter. This reasoning means that 
upon a splitting of the excitation's rapidities into massive and massless modes as in \eqref{decomposition de R mode massifs et massless}-\eqref{definition rapidites massives reduites},
the quantisation functions should enjoy the property \eqref{ecriture propriete reduction fct comptage}. 

Until this stage the reasonings should capture the structure of a very large class, if not all, of massless one-dimensional models. What however fixes the universality class is the structure of the form factors of local 
operators. Indeed it was unravelled in \cite{KozMailletMicroOriginOfc=1CFTUniversality} that a model belongs to the Luttinger liquid universality class 
if and only if the matrix elements of its local operators taken between two massless excited states, \textit{viz}. the excited states whose energies
differs from the one of the ground state by $1/L$ corrections, what corresponds to rapidities collapsing onto the endpoints of the Fermi zone, take a very specific form.
In the special case of ground to excited states, the arguments developed in the work  \cite{KozMailletMicroOriginOfc=1CFTUniversality} means that the form factors squared involving 
solely excitations on the left and/or right Fermi boundary are given by
\beq
  {\bf : } \pl{ \ups \in \{ \pm \} }{} \Big\{ \mf{F}_{\ups}\big( \mf{Y} \mid \mf{Z}^{\ups}_{\mf{Y}} \big)  \Big\}  {\bf :}   \; .
\enq
These building blocks take exactly the same form as for the XXZ chain \eqref{ecriture explicite FF discret}, see  \cite{KozMailletMicroOriginOfc=1CFTUniversality} for more details. 
The function $\vth_{\ups}(\mf{Y})$ arising in these expression is now some non-explicit, model dependent, function. Also, the differential operators $\mf{g}_{\ups}$
which involve the dependence on the massless rapidities collapsing on the endpoints of the Fermi zone will take  a much more involved form than \eqref{definition explicite operateur diff g}.
However, due to the indistinguishability phenomenon discussed earlier when particle and hole excitations collapse on the Fermi boundaries, 
one may reasonably expect that $\mf{g}_{\ups}=\e{O}\big( \tf{\de}{L} \big)$, exactly as for the XXZ spin chain. 
This closes the discussion of the contribution of the massless modes to the form factors. 

It remains to discuss the structure induced by the massive modes. It appears reasonable to assume that such modes lead to a smooth structure of form factors; 
such a behaviour is confirmed by the structure of form factors in massive quantum integrable field theories or those of massive excitations in the XXZ spin-$1/2$ chain. 
This means that if an excited state contains, on top of a swarm of massless excitations, $n_{\e{ex}}$ massive excitations, then the form factor 
should be weighted by an additional power of the volume $(1/L)^{n_{\e{ex}}}$ and be multiplied by a massive excitation form factor density function $\mc{F}^{(\ga)}(\mf{Y})$.
One can reasonably assume $\mc{F}^{(\ga)}(\mf{Y})$ to be a smooth function of the 
massive excitations rapidities.  The latter will be some non-explicit function whose details depend on the model.  
In fact, in order for this prefactor to have a true interpretation in terms of a form factor density,  
this contribution should also be weighted by the density of each massive mode which corresponds, to the leading order in $L$, 
to the derivative of the quantisation function in respect to its principal variable, exactly as in \eqref{ecriture DA FF apres partition modes massifs et masse nulle}. 
I nonetheless do stress that the explicit presence of this factor is not that important since it can 
be incorporated in the definition of the form factor density $\mc{F}^{(\ga)}(\mf{Y})$.

Once that all these ingredients relative to the microscopic structure of the Hilbert space are set, it is evident that one can re-do the steps outlined in the core of the paper. 
Indeed, the analysis was not relying at all on the specific properties of the functions that where dealt with. 
It is true that some of the steps of the analysis built on the use of complex variables techniques. 
One way to deal with the issue is to assume that, in a general model, the observables -excitation energies, momenta, form factor densities,...- will be analytic functions of the massive modes' rapidities, 
at least in some small vicinity of the curves where the quantised rapidities condense. Otherwise, in the smooth case, one will have to modify certain arguments by using 
local polynomial approximations allowing one to bypass the lack of analyticity, but globally the final result should still hold.  
More precisely, with such a structure at hand, equation \eqref{ecriture limit thermo fct 2 pts forme prefinale avec discontinuites}
appears to grasp the correct functional form of a form factor expansion in a massless model belonging to the Luttinger liquid universality class. 
As it was explained in Section \ref{SousSectionFcts2PtsDansRepImpulsion}, some more properties of the model -the structure of cuts of the form factors in the complex plane- are necessary so as to pass onto the momentum picture. 
It is hard to imagine how to postulate these for a generic model, especially that these will quite probably depend on the fine details of a model.
Still, it seems not so crazy to expect that the form factor denisties would be smooth functions of the various excitation momenta
in the dressed momentum representation, \textit{viz}. when one parametises each excitation in terms of its associated momentum. 
Hence, a representation in the spirit of \eqref{ecriture limit thermo fct 2 pts forme finale} should also hold, with the sole exception that one ought to substitute the appropriate to the model 
and operator expression -usually unknown- for the form factor density of the massive modes. 
This entails the functional form of the dynamic response functions since the passage from that representation to the one given in \eqref{ecriture contrib excitation donnee facteur structure}
is rather direct.

\section{Conclusion}

The present work developed techniques which allow one, starting from the large-volume behaviour of form factors of local operators in the massless regime of the XXZ chain, to write down the 
thermodynamic limit of a form factor expansion of the zero-temperature dynamical two-point functions in this model. In particular, in the thermodynamic limit, 
it carries a constructive and explicit renormalisation of the form factor series which stems from the dressing of the interactions by the swarm of massless models. The method developed in this work appears quite general and uses a rather weak input on the details of the model other than
the typical features of its universality class, the Luttinger liquid. The latter property allows one to advocate a certain universal form taken by the form factor expansions in one-dimensional models, 
not necessary integrable, belonging to this universality class. 

The obtained massless form factor series depends on a control parameter $\de$ which provides one with a scale allowing one to separate between the massive and massless modes of the model. 
Although leading to certain unexplicit terms, this dependence is already enough so as to extract all the physically interesting properties of the two-point functions such as the
long-distance and large-time asymptotic behaviour of dynamic two-point functions or the singular structure of the dynamic response functions. The techniques for achieving this goal and the corresponding results will 
appear in forthcoming publications. 

One should note that there would be no principal obstacle to obtain similar representations for the XXX chain or for the massless regime of the XXZ corresponding to $ \cos(\zeta)>1$
and $h_{\e{c;1}}>h>h_{\e{c;2}}$ where $h_{\e{c;1}}$ is the upper, resp. $h_{\e{c;2}}$ is the lower, critical magnetic field.
Also, with a little more work, one could generalise the results of \cite{KozProofOfAsymptoticsofFormFactorsXXZBoundStates} so as to obtain the large-volume behaviour of 
the XXZ chain's form factors corresponding to local operators taken between two excited states above the ground state. The adaptation  of the techniques developed in the present work 
along with the formulae that were obtained in \cite{KozKitMailTerMultiRestrictedSums} would then allow to conform the present technique so as to construct 
massless form factor series representation for multi-point dynamical correlation functions of the XXZ chain.

\section*{Acknowledgment}

K.K.K. acknowledges support from a CNRS and ENS de Lyon. The author is indebted to M. Brockmann, J.-S. Caux, F. Göhmann, J.M. Maillet, G. Niccoli for stimulating discussions
on various aspects of the project.

\appendix

\section*{Appendix}

\section{Observables in the infinite XXZ chain}
\label{Appendix Observables dans XXZ}

\subsection{Solutions to the linear integral equations}
\label{Appendix Lin Int Eqns Defs et al}

The observables describing the thermodynamic limit of the spin-$1/2$ XXZ chain are characterised by means of a collection of functions 
solving linear integral equations. These equations are driven by the integral kernel 
\beq
K(\la\mid \eta )  \, = \,   \f{ \sin(2\eta)   }{ 2\pi \sinh(\la + \i \eta)  \sinh(\la  - \i \eta)  }  \;. 
\label{ecriture fonction K de lambda et eta}
\enq

To introduce all of the functions of interest to this work, one starts by defining the $Q$-dependent dressed energy which allows one to construct the Fermi zone of the model. 
It is defined as the solution to the linear integral equation
\beq
\veps(\la\mid Q) \, + \, \Int{-Q}{Q} K\big(\la-\mu\mid \zeta \big) \, \veps(\la\mid Q)  \cdot \dd \mu \; = \;  h - 4 \pi J \sin(\zeta) K\big( \la \mid \tfrac{1}{2}\zeta \big)  \;. 
\label{definition energie habille et energie nue}
\enq
The endpoint of the Fermi zone is defined as the unique \cite{KozDugaveGohmannThermoFunctionsZeroTXXZMassless} positive solution $q$ to $\veps(q\mid q)=0$. 
Then, the function $\veps_1(\la)\equiv \veps(\la\mid q)$ corresponds to the dressed energy of the particle-hole excitations of the model. 
The functions 
\beq
\veps_r(\la)\; = \; r h - 4 \pi J \sin(\zeta) K\big( \la \mid \tfrac{r}{2}\zeta \big)   \, -\, \Int{-q}{q} K_{r}\big(\la-\mu \big) \veps_1(\mu)  \cdot \dd \mu 
\label{definition r energi habille}
\enq
with $r\geq 2$ and 
\beq
K_{r}(\la) \,  = \,  K\Big(\la \mid \tfrac{1}{2} \zeta(r+1) \Big) \, + \,  K\Big(\la \mid \tfrac{1}{2} \zeta(r-1) \Big)
\enq
correspond to the dressed energies of the $r$-bound state excitations.
For any $0<\zeta<\tf{\pi}{2}$ and under some additional constraints for $\tf{\pi}{2}< \zeta < \pi$, one can show \cite{KozProofOfStringSolutionsBetheeqnsXXZ}
that for any $r \in \mf{N}_{\e{st}}$, there exists $c_r>0$ such that $\veps_{r}(\la+\i\de_{r}  \tf{\pi}{2}) \geq  c_r>0$ for any $\la\in \R$. 
Still, one expects -on the basis of a numerical investigation of the solutions to \eqref{definition r energi habille}- that this lower bound holds throughout the whole massless regime $0<\zeta<\pi$, 
irrespectively of the mentioned additional conditions on $\zeta$. Also, I recall that $\de_{r}$ is such that the central rapidity of the $r$-string evolves, in the thermodynamic limit, on $\R+\i \de_{r} \tf{\pi}{2}$. 
Finally, when $r=1$, both options $\de_{1}\in \{0,1\}$ are possible.

In order to introduce the dressed momenta of the $r$-bound states and of the particle-hole excitations, I first need to 
define the $r$-bound state bare phases $\theta_r$ :
\beq
\theta_r(\la) \, = \,  \theta\Big(\la \mid \tfrac{1}{2} \zeta(r+1) \Big) \, + \,  \theta\Big(\la \mid \tfrac{1}{2} \zeta(r-1) \Big)  \quad \e{for} \;\;  r \geq 2
\quad \e{and} \qquad
\theta_1(\la) \, = \, \theta(\la\mid \zeta) 
\enq
with 
\beq
\theta(\la\mid \eta) \, = \, 2\pi \Int{ \Ga_{\la}  }{}  K(\mu-0^+\mid \eta ) \cdot \dd \mu  \;. 
\enq
The contour of integration appearing above corresponds to the union of two segments $ \Ga_{\la} \; = \; \intff{ 0 }{ \i \Im(\la) }\cup \intff{\i \Im(\la) }{ \la } $ and the $-0^+$ prescription indicates that the poles 
of the integrand at $\pm \i \eta +\i \pi \mathbb{Z}$ should be avoided from the left.

Then, the function  
\bem
p_r(\la)\; = \; \theta\big(\la\mid \tfrac{r}{2}\zeta \big)  \, -\, \Int{-q}{q} \theta_{r}\big(\la-\mu \big) p^{\prime}_1(\mu)  \cdot \f{ \dd \mu }{2\pi} \\
\, + \, \pi \ell_r(\zeta) + p_{F}m_r(\zeta)
-2p_{F} \sul{ \sg=\pm }{} \big(1\, - \, \de_{\sg,-}\de_{r,1} \big) \e{sgn}\Big( 1- \tfrac{2}{\pi} \cdot \wh{\tfrac{r+\sg}{2}\zeta} \Big) \cdot \bs{1}_{ \mc{A}_{r,\sg} } (\la)\;, 
\label{definition r moment habille}
\end{multline}
extended by $\i \pi$-periodicity to $\Cx$, corresponds to the dressed momentum of the $r$-bound states. Above, I have introduced
\beq
 \ell_r(\zeta)=1-r+2 \lfloor \tfrac{r \zeta}{2\pi} \rfloor  \qquad \e{and} \qquad 
   m_r(\zeta)=2-r - \de_{r,1} + 2 \sul{ \ups = \pm }{} \lfloor \zeta \tfrac{r + \ups   }{2\pi} \rfloor  \;. 
\enq
Furthermore, $\bs{1}_{A}$ stands for the indicator function of the set $A$, and 
\beq
\wh{\eta} \, = \, \eta  - \pi \lfloor \tfrac{ \eta }{ \pi } \rfloor \qquad \e{while} \qquad  \mc{A}_{r,\sg} \, = \,
 \Big\{ \la \in \Cx \; : \; \tfrac{\pi}{2}\geq |\Im(\la) | \geq \e{min}\big(   \wh{\tfrac{r+\sg}{2}\zeta} ,  \pi- \wh{\tfrac{r+\sg}{2}\zeta} \big)  \Big\}  \;. 
\enq
In order to obtain $p_r$, one should first solve the linear integro-differential equation for $p_1$ and then 
use $p_1$ to define $p_r$ by \eqref{definition r moment habille}.  $p_1$ corresponds to the dressed momentum of the particle or hole excitations and  $p_F=p_1(q)$ corresponds to the Fermi momentum.

One can show \cite{KozProofOfStringSolutionsBetheeqnsXXZ}, under similar conditions on $\zeta$ as for the dressed energy, that, for any $\la \in \R$
%
%
\beq
\big| p^{\prime}_{r}\big(\la  +\i \de_{r} \tfrac{\pi}{2} \big) \big| \; > \; 0 \quad \e{when} \quad r \in \mf{N}_{\e{st}} 
\qquad \e{and} \qquad 
\e{min}\Big( p^{\prime}_{1}\big(\la\big) \, , \, - p^{\prime}_{1}\big(\la  + \i \tfrac{\pi}{2} \big)   \Big) \; > \; 0   \;. 
\label{ecriture equation positivite pr prime}
\enq
Again, a numerical investigation indicates that \eqref{ecriture equation positivite pr prime} does, in fact, hold irrespectively of the value of $\zeta$. 

Finally, the $r$-bound dressed phase is defined as the solution to 
\beq
\phi_{r}(\la,\mu) \, = \, \f{ 1  }{ 2 \pi }  \theta_{r}\big( \la -\mu  \big) \, - \, \Int{-q}{q} K(\la-\nu)\,  \phi_{r}(\nu, \mu ) \cdot \dd \nu  \; + \; \f{m_{r}(\zeta)}{2}
\label{definition dressed phase}
\enq
and the dressed charge solves
\beq
Z(\la)\, + \,  \Int{-q}{q} K(\la-\mu)\,  Z(\mu ) \cdot \dd \mu   \, = \,  1 \;. 
\label{definition dressed charge}
\enq
The dressed charge is related to the dressed phase by the identities \cite{KorepinSlavnovNonlinearIdentityScattPhase}:
\beq
\phi_1(\la,q) \, - \,   \phi_1(\la,-q) \, + \, 1 \; = \; Z(\la) \quad \e{and} \quad 1+\phi_1(q,q) - \phi_1(-q,q) \, = \, \f{1}{ Z(q) }  \;. 
\label{ecriture identites entre phase et charge habilles} 
\enq
It is easy to see that the difference of boundary values of the dressed phase 
\beq
\De \phi_{r}^{\sg}(\la,\mu) \; = \; \phi_{r}\Big( \la , \mu + \i \mf{f}_{ \tfrac{ r+\sg }{ 2 } \zeta} +\i 0^+ \Big) \, - \, \phi_{r}\Big( \la , \mu + \i \mf{f}_{ \tfrac{ r+\sg }{ 2 } \zeta }-\i 0^+ \Big)   \;,
\enq
with $\mf{f}_{\eta}$ defined below of \eqref{ecriture des domaines de saut pour FF density}, 
satisfies to the linear integral equation
\beq
\De \phi_{r}^{\sg}(\la,\mu) +\Int{-q}{q}  K(\la-\nu)\,  \De \phi_{r}^{\sg}(\nu, \mu ) \cdot \dd \nu \; = \; u_{r}^{\sg} \bs{1}_{\intoo{\mu}{+\infty}}(\la) \qquad \la, \mu \in \R  \;.  
\enq
Here, $u_r^{\sg}$ is as given in \eqref{definition parametre u r sigma}. In particular, for $\mu < -q$, it holds that 
\beq
\De \phi_{r}^{\sg}(\la,\mu) \; = \; u_{r}^{\sg} \cdot  Z(\la) \qquad \e{for} \qquad \la \in \intff{-q}{q} \;. 
\enq

\subsection{The bound states}
\label{Appendix Sous-section cordes}

It has been proven in \cite{KozProofOfStringSolutionsBetheeqnsXXZ} that for $\tf{\pi}{2}<\zeta< \pi$ an $r$ bound state  centred on $\R+\i s \tf{\pi}{2}$, $s\in \{0,1\}$, exists if and only if the 
below constraints are all simultaneously satisfied:
\beq
(-1)^s \sin\big[ k \zeta \big] \cdot \sin\big[ (r-k)\zeta \big] \, > \, 0 \qquad \e{for} \; k=1,\dots, r-1. 
\enq
These are precisely the conditions argued earlier by Suzuki-Takahashi \cite{TakahashiSuzukiFiniteTXXZandStrings}
and, subsequently, in \cite{FowlerZotosConditionsOfStringExistenceXYZAndSineGordon,HidaConditionExistenceStringsXXZ,KorepinAnalysisofBoundStateConditionMassiveThirring}.

However, for $0< \zeta < \tf{\pi}{2}$, the work \cite{KozProofOfStringSolutionsBetheeqnsXXZ} proved that an $r$ bound state centred on $\R - \i \kappa_{r} \tf{\pi}{2}$, exists if and only if the 
below constraints are all simultaneously satisfied:
\beq
(-1)^{ \kappa_{r} } \cdot \sin\bigg(  \f{\pi \zeta}{\pi-\zeta} (k-p) \bigg) \cdot \sin\bigg(  \f{\pi \zeta}{\pi-\zeta} (r-k+p-\kappa_{r}-1) \bigg) \, > \, 0 
\label{ecriture des condition existence corde zero zeta pi sur 2}
\enq
for $k\, = \, 1,\dots, r-2$ $and$ $k \in \intn{ w_{p}+1 }{ w_{p+1}-1 }$. 
Above, $\kappa_{r}=\lfloor (r-1)\tf{\zeta}{\pi} \rfloor$  and 
\beq
w_{p} \, = \, \Big\lfloor  \Big( p-\tfrac{\kappa_{r}}{2}+(r-1)\tfrac{\zeta}{2\pi}  \Big) \f{ \pi }{ \zeta } \Big\rfloor \;. 
\enq

\section{The operator $\wh{\mc{B}}_{\ell_{\ups}  }^{\, (\ups)}$ and restricted sums}
\label{Appendix Restricted sums}

The present appendix is devoted to the computation of the action of the functional  $\wh{\mc{B}}_{\ell_{\ups}}^{\, (\ups)}$ on the discrete form factor 
$\mf{F}_{\ups}(\mf{Y}\mid \mf{Z}_{\ups})$, this up to some $\de$, $m$ and $L$ dependent corrections. The main result of this computation is the representation 
\beq
\bs{:} \lim_{L\tend +\infty}  \pl{\ups= \pm }{}  \wh{\mc{B}}_{\ell_{\ups}  }^{\, (\ups)} \Big[ \mf{F}_{\ups}\big(\mf{Y} \mid * \big) \mid \mf{Y} \Big] \bs{:} \; = \;  
\pl{\ups=\pm }{} \Bigg\{  \f{  \ex{ \i m \ups \ell_{\ups} p_{F} }   }{   \big[ -\i m_{\ups} \big]^{   \vth_{\ups}(\mf{Y})^2 }  }    \Bigg\} \cdot 
\bigg( 1+\e{O}\Big(      \sum_{ \ups = \pm }   \Big\{ \de  \big| \ln |m_{\ups}| \big| + \de^2 |m_{\ups}| +  \ex{-\de |m_{\ups}|}  \Big\} \Big)  \bigg)  
\label{ecriture resultat action operateur B ups}
\enq
where $m_{\ups}=\ups m -\op{v}_F t$.

In the handlings that follow, I will assume that the remainders
are such that it is allowed to carry out the various interchanges of symbols. One first focuses on $ \wh{\mc{B}}_{\ell_{\ups}  }^{\, (\ups)} \Big[ \mf{F}_{\ups}\big(\mf{Y} \mid * \big) \mid \mf{Y} \Big] $.
Then, the first step consists in changing the variables in \eqref{definition fnelle B ell de ups} as 
\beq
\tfrac{1}{2\pi} \, \wh{\xi}_{1}\big( \mu_a   \mid \mf{Y} \big) \, = \, t_a \qquad \e{and} \qquad \tfrac{1}{2\pi} \, \wh{\xi}_{1}\big( \la_a   \mid \mf{Y} \big) \, = \, k_a  \;. 
\enq
Upon inserting the explicit expression for the discrete form factor $\mf{F}_{\ups}$ given in \eqref{ecriture explicite FF discret}, this yields
\bem
\wh{\mc{B}}_{\ell_{\ups}  }^{\, (\ups)} \Big[ \mf{F}_{\ups}\big(\mf{Y} \mid * \big) \mid \mf{Y} \Big] \; = \; 
\f{ G^2\big(1- \ups   \vth_{\ups}(\mf{Y}) - \ups \nu_{\ups} - \ell_{\ups} \big)     }{ G^2\big(1- \ups   \vth_{\ups}(\mf{Y}) - \ups \nu_{\ups} \big) }  \cdot \bigg( \f{ 2 \pi }{ L } \bigg)^{  ( \vth_{\ups}(\mf{Y})  + \nu_{\ups} )^2 }  \\
\times \sul{ n_p   -n_h   = \ell_{\ups}  }{} \Int{ I_p^{(\ups)} }{} \f{ \dd^{n_p }k  }{ n_p!   }
\pl{ a=1 }{ n_p } \Big[  L\,  \ex{\i m  \wt{u}_1(k_a, \op{v}) + \wt{\bs{\mf{g}}}_{\ups}(k_a) }   \Big] 
\cdot \Int{ I_h^{(\ups)} }{} \f{ \dd^{n_h}t }{ n_h! } \pl{ a=1 }{ n_h } \Big[  L\,\ex{-\i m  \wt{u}_1(t_a, \op{v}) - \wt{\bs{\mf{g}}}_{\ups}(t_a) }    \Big] \cdot 
\mc{R} \Big(   \mf{Z} \mid - \ups   \vth_{\ups}(\mf{Y}) - \ups \nu_{\ups} - \ell_{\ups}    \Big)  \;.
\label{serie pour Bhat apres 1er chgmt vars}
\end{multline}
Here, the function $\mc{R}$ is as defined in \eqref{definition densite R discrete} and depends on the collection of integration variables
\beq
\mf{Z}\; = \; \bigg\{ \Big\{  \ups \big(L k_a - N_{\ups} \big)  \Big\}_1^{n_p} \; ; \;   \Big\{  \ups \big(N_{\ups} -  L t_a  \big)  -1 \Big\}_1^{n_h} \bigg\}\;. 
\enq
Also, the integration in \eqref{serie pour Bhat apres 1er chgmt vars} runs over the intervals 
\beq
I_p^{(\ups)} \; = \; \ups \Big[ \tfrac{ N_{\ups}-\ups/4 }{L}  \, ; \,  \tfrac{ N_{\ups}-\ups/4 }{L} + \i \tfrac{\de \mf{s}_{u}^{\ups}}{2\pi} \Big]
\qquad \e{and} \qquad  
 I_h^{(\ups)} \; = \;  -\ups \Big[   \tfrac{ N_{\ups}-3\ups/4 }{L} \, ; \, \tfrac{ N_{\ups}-3\ups/4 }{L} - \i \tfrac{ \de \mf{s}_{u}^{\ups} }{ 2\pi } \Big]  
\enq
where I remind that $\mf{s}_{u}^{\ups}\, = \, \e{sgn}\big( u_1^{\prime}(\ups q , \bf{v} ) \big)$. Finally, I have set 
\beq
 \wt{u}_1(s, \op{v}) \; = \;  u_1\Big(  \,  \wh{\xi}^{\, -1}_1( 2\pi  s\mid\mf{Y}) , \op{v}\Big)  
 \qquad \e{and} \qquad 
 \wt{\bs{\mf{g}}}_{\ups}(s) \; = \;  \bs{\mf{g}}_{\ups}\Big(   \, \wh{\xi}^{\, -1}_1(2\pi  s\mid\mf{Y}) \Big)   \;. 
\enq

At this stage, it remains to shift the integration variables by $\tfrac{1}{L} N_{\ups}$ and expand the functions appearing in the exponents around $0$. One has
\beq
 \wt{ \bs{\mf{g}} }_{ \ups }\Big( \tfrac{ N_{\ups} }{L} + s \Big) \; = \; \e{O}\Big( \de \sul{ \ups^{\prime} = \pm }{} \Dp{ \nu_{\ups^{\prime}} }  \Big)
\enq
Above, $\e{O}\Big( \de  \sum_{ \ups^{\prime} = \pm }  \Dp{ \nu_{\ups^{\prime}} } \Big)$ refers to the presence of some function of the differential 
operators $  \Dp{ \nu_{\pm} }  $ such that each derivative is always preceded by a prefactor which is, at most, of the magnitude $\e{O}(\de)$. 
Also, one has the expansion 
\bem
 m \, \wt{u}_1\Big( \tfrac{ N_{\ups} }{L} + s, \op{v} \Big) \; = \;  m \, \wt{u}_1\Big( \tfrac{ N_{\ups} }{L}, \op{v} \Big)  \, + \, m \, s\cdot  \wt{u}_1^{\, \prime}\Big( \tfrac{ N_{\ups} }{L} , \op{v} \Big)  
\; + \; \e{O}\big( \de^2 ( |m|+|t|) \big)  \\  
\; = \; \ups m p_F   \, + \, m \, s\cdot  \wt{u}_1^{\, \prime}\Big( \tfrac{ N_{\ups} }{L} , \op{v} \Big)  
\; + \; \e{O}\Big( (\de^2 + \tfrac{1}{L})\sul{\ups=\pm}{} |m_{\ups}| \Big)  \;.
\end{multline}
Here, one obtains the second line by noting that 
\beq
m \wt{u}_1\Big( \tfrac{ N_{\ups} }{L}, \op{v} \Big)=  \ups m p_F + \e{O}\bigg( \f{ |m|+|t| }{ L }  \bigg)
\enq
and by  using that $2m=m_+-m_-$ and $ - 2v_F t = m_++m_-$.

All these simplifications yield
\bem
\wh{\mc{B}}_{\ell_{\ups}  }^{\, (\ups)} \Big[ \mf{F}_{\ups}\big(\mf{Y} \mid * \big) \mid \mf{Y} \Big] \; = \; \ex{\i \ups  m  \ell_{\ups} p_{F} }
\f{ G^2\big(1- \ups   \vth_{\ups}(\mf{Y}) - \ups \nu_{\ups} - \ell_{\ups} \big)     }{ G^2\big(1- \ups   \vth_{\ups}(\mf{Y}) - \ups \nu_{\ups} \big) }\cdot 
 \bigg( \f{ 2 \pi }{ L } \bigg)^{  ( \vth_{\ups}(\mf{Y})  + \nu_{\ups} )^2 }  \\
\hspace{2cm} \times  \sul{ n_p   -n_h   = \ell_{\ups}  }{}  \f{ L^{n_p+n_h} }{ n_p! \, n_h! } \Int{ -\tfrac{\ups}{4L} }{  -\tfrac{\ups}{4L} + \i \tfrac{ \de \mf{s}_{u}^{\ups} }{ 2\pi }  } \hspace{-4mm}  \dd^{n_p }k  
  \hspace{-2mm}  \Int{ -\tfrac{3 \ups}{4L} }{  - \tfrac{3 \ups}{4L} - \i\tfrac{ \de \mf{s}_{u}^{\ups} }{ 2\pi }  }  \hspace{-4mm} \dd^{n_h}t \;  
\exp\bigg\{ \i m  \, \wt{u}_1^{\, \prime}\Big( \tfrac{ N_{\ups} }{L} , \op{v} \Big) \big[ \sul{ a=1 }{ n_p } k_a \, - \, \sul{ a=1 }{ n_h } t_a  \big]  \bigg\}   
\cdot (\ups)^{n_p}\cdot (-\ups)^{n_h}\\
\times  \mc{R} \Big(   \mf{Z} \mid - \ups   \vth_{\ups}(\mf{Y}) - \ups \nu_{\ups} - \ell_{\ups}    \Big)  \cdot
  \bigg( 1+\e{O}\Big(  \de^2 \sul{\ups=\pm}{} |m_{\ups}|  + \tfrac{1}{L} \sul{\ups=\pm}{} |m_{\ups}|  + \de  \sul{ \ups^{\prime} = \pm }{} \Dp{ \nu_{\ups^{\prime}} } \Big)  \bigg) \;.
\label{representation legerement transformee de B hat}
\end{multline}
In \eqref{representation legerement transformee de B hat}, the argument $\mf{Z} $ of $\mc{R}$ reads
\beq
\mf{Z}  \; = \;  \bigg\{ \Big\{  \ups L k_a  \Big\}_1^{n_p} \; ; \;   \Big\{  - \ups L  t_a -1  \Big\}_1^{n_h} \bigg\} \;. 
\enq
Even though the remainders involving differential operators appear in the \textit{rhs} of \eqref{representation legerement transformee de B hat}, one should bear in mind
that they will act on the \textit{lhs} of the series once that the $\bs{:} * \bs{:}$ order is imposed.

Recall that, in fact, one is only interested in computing the thermodynamic limit of the action of the functionals $\wh{\mc{B}}_{\ell_{\ups}  }^{\, (\ups)}$. Thus, one can add
additional terms to the integrals that will presumably not change the value of the thermodynamic limit, much in the spirit of the replacements
that have been discussed in Section \ref{Sousection serie auxiliare C2}. Indeed, the below replacement in $k$ or $t$-based one-dimensional integrals only
produces $\e{O}(L^{-1})$ corrections 
\beq
\ups L  \hspace{-4mm} \Int{ -\tfrac{\ups}{4L} }{  -\tfrac{\ups}{4L} + \i \tfrac{ \de \mf{s}_{u}^{\ups} }{ 2\pi }  } \hspace{-4mm}  \dd k  \; \hookrightarrow 
\Oint{ \Ga^{(\ups)}_p }{   }   \dd k  \f{  L \mf{s}_{u}^{\ups} }{ \ex{ 2\i \pi  L k \mf{s}_{u}^{\ups}  } \, - \, 1  } \Big(1+\e{O}\big(L^{-1}\big) \Big)
\; + \; L \mf{s}_{u}^{\ups} \Int{ \Ga_p^{(\ups, \eta_{\ups}(\op{v}))}  }{} \dd k
\label{ecriture remplacement contours Ip vers Gamma p loc}
\enq
and
\beq
-\ups L  \hspace{-4mm} \Int{ -\tfrac{3\ups}{4L} }{  -\tfrac{3\ups}{4L} - \i \tfrac{ \de \mf{s}_{u}^{\ups} }{ 2\pi } } \hspace{-4mm}  \dd t  \; \hookrightarrow 
\Oint{ \Ga^{(\ups)}_h }{   }   \dd t  \f{  L  \mf{s}_{u}^{\ups} }{ 1\, - \, \ex{ - 2\i \pi  L t  \mf{s}_{u}^{\ups}  }  } \Big(1+\e{O}\big(L^{-1}\big) \Big)
\; - \; L \mf{s}_{u}^{\ups} \Int{ \Ga_h^{(\ups, \ov{\eta}_{\ups}(\op{v}))}  }{} \dd t  \;. 
\label{ecriture remplacement contours Ih vers Gamma h loc}
\enq
The contours $\Ga^{(\ups)}_{p/h}$ arising in \eqref{ecriture remplacement contours Ip vers Gamma p loc}-\eqref{ecriture remplacement contours Ih vers Gamma h loc} are depicted in Figure \ref{Figure contour Gamma pm p et h pour fnelle Bell} and 
\beq
\eta_{\ups}(\op{v}) \;  = \; \left\{  \ba{ccc}  \ua &  \e{if} &  \mf{s}_{u}^{\ups} = +  \hspace{2mm} \\ 
					      \da &  \e{if} &  \mf{s}_{u}^{\ups} = - \ea  \right.  \qquad \e{and} \qquad 
\ov{\eta}_{\ups}(\op{v}) \;  = \; \left\{  \ba{ccc}  \da &  \e{if} &  \mf{s}_{u}^{\ups} = +  \hspace{2mm} \\ 
					      \ua &  \e{if} &  \mf{s}_{u}^{\ups} = - \ea  \right.  \;. 
\enq
\begin{figure}[ht]
\begin{center}

\begin{pspicture}(6.5,4.5)


\psline[linestyle=dashed, dash=3pt 2pt](0,3.5)(5.5,3.5)
\psline[linewidth=2pt]{->}(5.5,3.5)(5.6,3.5)

 \rput(6,3.5){$\R$}

 \psline{<->}(4.7,3.5)(4.7,4.25)
 \rput(4.9,3.9){$ \tfrac{ \de }{ 2\pi } $}
 
 \psdots(2.5,3.5)(3.5,3.5)

\rput(4,3.15){$\tfrac{3}{4L}$}
\rput(2,3.15){$\tfrac{1}{4L}$}

\psline(3.5,2.75)(3.5,4.25)
\psline(3.5,2.75)(5.5,2.75)
\psline(3.5,4.25)(5.5,4.25)
\psline[linewidth=2pt]{<-}(4.5,2.75)(4.4,2.75)
\psline[linewidth=2pt]{<-}(4.4,4.25)(4.5,4.25)

\psline(2.5,2.75)(2.5,4.25)
\psline(0,2.75)(2.5,2.75)
\psline(0,4.25)(2.5,4.25)
\psline[linewidth=2pt]{<-}(1.5,2.75)(1.4,2.75)
\psline[linewidth=2pt]{<-}(1.4,4.25)(1.5,4.25)

\rput(4.8,3.15){$ \Ga^{(-)}_{h} $}
\rput(1.3,3.2){$ \Ga^{(-)}_{p} $}

\rput(-0.1,4.35){$ \Ga^{(-,\ua)}_{p} $}
\rput(-0.1,2.85){$ \Ga^{(-,\da)}_{p} $}

\rput(6,4.35){$ \Ga^{(-,\ua)}_{h} $}
\rput(6,2.85){$ \Ga^{(-,\da)}_{h} $}

\psline(-0.5,2.25)(6.5,2.25)


\psline[linestyle=dashed, dash=3pt 2pt](0,1)(5.5,1)
\psline[linewidth=2pt]{->}(5.5,1)(5.6,1)

 \rput(6,1){$\R$}

 \psline{<->}(4.7,1)(4.7,1.75)
 \rput(4.9,1.4){$ \tfrac{ \de }{ 2\pi } $}
 
 \psdots(2.5,1)(3.5,1)

\rput(4,0.65){$-\tfrac{1}{4L}$}
\rput(2,0.65){$-\tfrac{3}{4L}$}

\psline(3.5,0.25)(3.5,1.75)
\psline(3.5,0.25)(5.5,0.25)
\psline(3.5,1.75)(5.5,1.75)
\psline[linewidth=2pt]{->}(4.4,0.25)(4.5,0.25)
\psline[linewidth=2pt]{->}(4.5,1.75)(4.4,1.75)

\psline(2.5,0.25)(2.5,1.75)
\psline(0,0.25)(2.5,0.25)
\psline(0,1.75)(2.5,1.75)
\psline[linewidth=2pt]{->}(1.4,0.25)(1.5,0.25)
\psline[linewidth=2pt]{->}(1.5,1.75)(1.4,1.75)

\rput(4.8,0.6){$ \Ga^{(+)}_{p} $}
\rput(1.1,0.7){$ \Ga^{(+)}_{h} $}

\rput(-0.1,1.8){$ \Ga^{(+,\ua)}_{h} $}
\rput(-0.1,0.3){$ \Ga^{(+,\da)}_{h} $}

\rput(6,1.75){$ \Ga^{(+,\ua)}_{p} $}
\rput(6,0.25){$ \Ga^{(+,\da)}_{p} $}

\end{pspicture}
\caption{ Contours $\Ga^{(+)}_{p/h}$ -depicted in the bottom graph- and $\Ga^{(-)}_{p/h}$ -depicted in the top graph-. One has $ \Ga^{(\ups,\ua)}_{p/h}= \Ga^{(\ups)}_{p/h} \cap \Big\{ \R + \i \tf{ \de }{ 2\pi } \Big\}$ and 
$ \Ga^{(\ups,\da)}_{p/h}= \Ga^{(\ups)}_{p/h}  \cap \Big\{ \R - \i \tf{ \de }{ 2\pi } \Big\}$.
\label{Figure contour Gamma pm p et h pour fnelle Bell} }
\end{center}

\end{figure}

\noindent It appears convenient to introduce the integration measures
\beq
 \mc{D}_{\ups }^{n_p} k \; = \; \pl{a=1}{n_p} \bigg\{ \f{  L  \mf{s}_{u}^{\ups} }{ \ex{ 2\i \pi  L k_a  \mf{s}_{u}^{\ups}  } \, - \, 1  } \bigg\} \cdot \f{\dd^{n_p} k}{n_p!} 
\qquad \e{and} \qquad 
 \mc{D}_{\ups }^{n_h} t \; = \;   \pl{a=1}{n_h} \bigg\{ \f{  L  \mf{s}_{u}^{\ups} }{ 1\, - \, \ex{ - 2\i \pi  L t_a  \mf{s}_{u}^{\ups}  }  } \bigg\} \cdot \f{ \dd^{n_h} t }{ n_h!} \;. 
\enq
The contour substitutions  \eqref{ecriture remplacement contours Ip vers Gamma p loc} and \eqref{ecriture remplacement contours Ih vers Gamma h loc} introduce a new partitioning
of the integration variables in \eqref{representation legerement transformee de B hat} what recasts the functional as 
\bem
\wh{\mc{B}}_{\ell_{\ups}  }^{\, (\ups)} \Big[ \mf{F}_{\ups}\big(\mf{Y} \mid * \big) \mid \mf{Y} \Big] \; = \; \ex{\i \ups  m   \ell_{\ups} p_{F} }
\f{ G^2\big(1- \ups   \vth_{\ups}(\mf{Y}) - \ups \nu_{\ups} - \ell_{\ups} \big)     }{ G^2\big(1- \ups   \vth_{\ups}(\mf{Y}) - \ups \nu_{\ups} \big) }\cdot 
 \bigg( \f{ 2 \pi }{ L } \bigg)^{  ( \vth_{\ups}(\mf{Y})  + \nu_{\ups} )^2 }  \\
\times \sul{ \substack{ \mf{n}_{\ups}+\mf{l}_{\ups} \\ = \ell_{\ups} } }{} \sul{ \substack{ n_x   -n_y \\   = \mf{n}_{\ups} }  }{}   \sul{ \substack{ n_p   -n_h  \\  = \mf{l}_{\ups} } }{} \;
\Oint{ \Ga_{p}^{(\ups)} }{   } \mc{D}_{\ups }^{n_p} k  \; \Oint{ \Ga_{h}^{(\ups)} }{   } \mc{D}_{\ups }^{n_h} t 
 \hspace{-2mm} \Int{  \Ga_p^{(\ups, \eta_{\ups}(\op{v}))} }{}  \hspace{-3mm}  \dd^{n_x} x  \f{ (\mf{s}_{u}^{\ups}  L )^{n_x} }{ n_x! }
 \hspace{-2mm}  \Int{  \Ga_h^{(\ups, \ov{\eta}_{\ups}(\op{v}))} }{}  \hspace{-3mm}  \dd^{n_y} y \f{ (-  \mf{s}_{u}^{\ups}  L )^{n_y}  }{ n_y! } \\
\times \exp\bigg\{ \i m  \, \wt{u}_1^{\, \prime}\big( \tfrac{ N_{\ups} }{L} , \op{v} \big) \big[ \sul{ a=1 }{ n_p  } k_a \,+ \,  \sul{ a=1 }{ n_x  } x_a \, - \, \sul{ a=1 }{ n_h } t_a \, -  \, \sul{ a=1 }{ n_y }  y_a \big]  \bigg\}   \\
\times  \mc{R} \Big(   \mf{Z}_{x,y} \cup \mf{Z}_{k,t}  \mid - \ups   \vth_{\ups}(\mf{Y}) - \ups \nu_{\ups} - \ell_{\ups}    \Big)  \cdot
  \bigg( 1+\e{O}\Big(  \de^2 \sul{\ups=\pm}{} |m_{\ups}|  + \tfrac{1}{L} \sul{\ups=\pm}{} |m_{\ups}|  + \de  \sul{ \ups^{\prime} = \pm }{} \Dp{ \nu_{\ups^{\prime}} } \Big)   \bigg) \;.
\end{multline}
Here, I have introduced two sets of variables
\beq
\mf{Z}_{x,y}=\Big\{ \big\{ \ups L x_a \big\}_1^{n_x} \, ; \, \big\{ -\ups L y_a -1\big\}_1^{n_y}  \Big\} \qquad \e{and} \qquad 
\mf{Z}_{k,t}=\Big\{ \big\{ \ups L k_a \big\}_1^{n_p} \, ; \, \big\{ -\ups L t_a -1 \big\}_1^{n_h}  \Big\} 
\enq
and the union of sets  $ \mf{Z}_{x,y} \cup \mf{Z}_{k,t}$ means joining together the $k$ and $x$ variables as well as the $t$ and $y$ variables. 
The contour integrals over $ \Ga^{(\ups)}_{p}$ and $ \Ga^{(\ups)}_{h}$ can be evaluated by taking the residues at 
\beq
k_a \, = \, \f{\ups}{L}p_a \; , \qquad t_a \, = \, - \f{\ups}{L}(h_a+1) \quad \e{with} \quad p_a, h_a \in \mathbb{N} 
\enq
and by using the symmetry of the integrand. Also, by using the analyticity of the integrand, one can deform the integration domains as
\beq
 \Ga_p^{(\ups, \eta_{\ups}(\op{v}))} \; \hookrightarrow \; -\ups \mf{s}_{u}^{\ups}  \Big[ \i \mf{s}_{u}^{\ups} \tfrac{\de}{2\pi} \, ; \, \i \mf{s}_{u}^{\ups} \infty \Big[
\qquad \e{and} \qquad 
 \Ga_h^{(\ups, \ov{\eta}_{\ups}(\op{v}))} \; \hookrightarrow \; -\ups \mf{s}_{u}^{\ups}  \Big[ -\i \mf{s}_{u}^{\ups} \tfrac{\de}{2\pi} \, ; \,- \i \mf{s}_{u}^{\ups} \infty \Big[
\enq
where the $ -\ups \mf{s}_{u}^{\ups} $ pre-factor corresponds to the orientation of the interval. 
Thus, eventually, one gets 
\bem
\wh{\mc{B}}_{\ell_{\ups}  }^{\, (\ups)} \Big[ \mf{F}_{\ups}\big(\mf{Y} \mid * \big) \mid \mf{Y} \Big] \; = \; \ex{\i \ups  m  \ell_{\ups} p_{F} }
\f{ G^2\big(1- \ups   \vth_{\ups}(\mf{Y}) - \ups \nu_{\ups} - \ell_{\ups} \big)     }{ G^2\big(1- \ups   \vth_{\ups}(\mf{Y}) - \ups \nu_{\ups} \big) }\cdot 
 \bigg( \f{ 2 \pi }{ L } \bigg)^{  ( \vth_{\ups}(\mf{Y})  + \nu_{\ups} )^2 }  \\
\times \sul{ \substack{ \mf{n}_{\ups}+\mf{l}_{\ups} \\ = \ell_{\ups} } }{} \sul{ \substack{ n_x   -n_y \\   = \mf{n}_{\ups} }  }{}  
  \Int{  \tfrac{\de}{2\pi} }{ +\infty }    \dd^{n_x} x  \f{ (- \i \ups \mf{s}_{u}^{\ups}  L )^{n_x} }{ n_x! }
   \Int{ \tfrac{\de}{2\pi} }{ + \infty }    \dd^{n_y} y \f{ (- \i \ups \mf{s}_{u}^{\ups}  L )^{n_y}  }{ n_y! }  
 \exp\bigg\{ - m  \, \big|\,  \wt{u}_1^{\, \prime}\big( \tfrac{ N_{\ups} }{L} , \op{v} \big) \big| \cdot \Big[   \sul{ a=1 }{ n_x  } x_a \, + \,  \sul{ a=1 }{ n_y }  y_a \Big]  \bigg\}  \\
\sul{ \substack{ \mc{Z}_{p,h} \, :  \\ \, n_p-n_h \,  = \, \mf{l}_{\ups} }  }{}\;  
\ex{ \i \tfrac{m}{L} \ups \, \wt{u}_1^{\, \prime}\big( \tfrac{ N_{\ups} }{L} , \op{v} \big)\, \mc{J}(\mc{Z}_{p,h})  } 
\cdot  \mc{R} \Big(   \wt{\mf{Z}}_{x,y} \cup \mc{Z}_{p,h}  \mid - \ups   \vth_{\ups}(\mf{Y}) - \ups \nu_{\ups} - \ell_{\ups}    \Big)  \\
\times
  \bigg( 1+\e{O}\Big(  \de^2 \sul{\ups=\pm}{} |m_{\ups}|  + \tfrac{1}{L} \sul{\ups=\pm}{} |m_{\ups}|  + \de  \sul{ \ups^{\prime} = \pm }{} \Dp{ \nu_{\ups^{\prime}} } \Big)   \bigg) \;.
\label{ecriture action B ups comme somme restreinte}
\end{multline}
Above, it is undercurrent that
\beq
\mc{Z}_{p,h}  \, = \, \Big\{ \{p_a\}_1^{n_p} ; \{h_a \}_1^{n_h} \Big\} \quad \e{and}   \quad  
\wt{\mf{Z}}_{x,y} \, =\,  \Big\{ \big\{ \ups \i L \mf{s}_{u}^{\ups}  x_a \big\}_1^{n_x} \, ; \, \big\{ \ups L \i  \mf{s}_{u}^{\ups}  y_a -1 \big\}_1^{n_y}  \Big\} 
\enq
as well as 
\beq
\mc{J}(\mc{Z}_{p,h}) \; = \;  \sul{a=1}{n_p}   p_a  + \sul{a=1}{n_h}  (h_a+1) \;. 
\enq
Finally, the summation in \eqref{ecriture action B ups comme somme restreinte}
runs through all the possible choices of the set $\mc{Z}_{p,h}$, \textit{viz}. of $p_a, h_a \in \mathbb{N} $ such that $n_p-n_h = \mf{l}_{\ups}$ and $ p_1<\dots < p_{n_p}$, resp. $h_1<\dots < h_{n_h}$. 
Straightforward estimates and the lower bound $x_a, y_a \geq \tf{\de}{2\pi}$ ensures that one has the large-$L$ behaviour 
\beq
 \mc{R} \Big(   \wt{\mf{Z}}_{x,y} \cup \mc{Z}_{p,h}  \mid \nu   \Big) \, = \, \Big( \f{L}{2\pi}\Big)^{ \mf{n}_{\ups}^{2}+2 \mf{n}_{\ups}(\nu+\mf{l}_{\ups}) }
\cdot \f{   \mc{D}_{\ups} \big(    \mc{Z}_{x,y}  \mid \nu +\mf{l}_{\ups}  \big)  }{   \big(-\i \ups \mf{s}_{u}^{\ups} L \big)^{n_x+n_y} } \cdot  \mc{R} \Big(    \mc{Z}_{p,h}  \mid \nu   \Big)
\cdot \f{ \pl{a=1}{n_p}  \vp_{x,y}^{(\ups)} \Big( \tfrac{p_a}{L} \Big) }{  \pl{a=1}{n_h}  \vp_{x,y}^{(\ups)} \Big( -\tfrac{h_a+1}{L} \Big)   }
\cdot \bigg( 1 \, + \, \e{O}\Big( \f{n_x+n_y}{\de L } \Big) \bigg) \;. 
\enq
Here $\mc{Z}_{x,y}  \, = \, \Big\{ \{x_a\}_1^{n_x} ; \{y_a \}_1^{n_y} \Big\}$,
\beq
 \mc{D}_{\ups} \big(    \mc{Z}_{x,y}  \mid \nu   \big) \; = \; (-1)^{ \mf{n}_{\ups} } \cdot \big( 2\i \pi \ups \mf{s}_{u}^{\ups} \big)^{ \mf{n}_{\ups}^{2}+2 \mf{n}_{\ups}\nu }
\cdot \Big( \f{\sin(\pi \nu) }{ \pi } \Big)^{2n_y} \cdot \f{ \pl{a=1}{n_x} x_a^{2\nu} }{ \pl{a=1}{n_y} y_a^{2\nu} }
\cdot \f{ \pl{a<b}{n_x}(x_a-x_b)^2 \cdot \pl{a<b}{n_y}(y_a-y_b)^2  }{  \pl{a=1}{n_x} \pl{b=1}{n_y} (x_a+y_b)^2 }
\enq
and
\beq
 \vp_{x,y}^{(\ups)}(w) \, = \, \f{ \pl{a=1}{n_x}  \Big( 1 \, - \, \f{ w }{ \i \ups \mf{s}_{u}^{\ups} x_a } \Big)^2   }{ \pl{a=1}{n_y}  \Big( 1 \, + \, \f{ w }{ \i \ups \mf{s}_{u}^{\ups} y_a } \Big)^2    }\;. 
\enq
One is now in position to take a partial $L\tend +\infty$ limit of the series. By using that $\wt{u}_1^{\, \prime}\big( \tfrac{ N_{\ups} }{L} , \op{v} \big) = 2\pi \ups m_{\ups}$, with $m_{\ups}=\ups m -\op{v}_{F} t$
one obtains 
\bem
\wh{\mc{B}}_{\ell_{\ups}  }^{\, (\ups)} \Big[ \mf{F}_{\ups}\big(\mf{Y} \mid * \big) \mid \mf{Y} \Big] \; = \; \ex{\i m \ups \ell_{\ups} p_{F} }
\sul{ \substack{ \mf{n}_{\ups}+\mf{l}_{\ups} \\ = \ell_{\ups} } }{} \sul{ \substack{ n_x   -n_y \\   = \mf{n}_{\ups} }  }{}  
  \Int{  \tfrac{\de}{2\pi} }{}     \f{  \dd^{n_x} x }{ n_x! }
   \Int{ \tfrac{\de}{2\pi} }{}     \f{ \dd^{n_y} y  }{ n_y! }  
 \exp\bigg\{ - 2\pi |m_{\ups}| \cdot \Big[   \sul{ a=1 }{ n_x  } x_a \, + \,  \sul{ a=1 }{ n_y }  y_a \Big]  \bigg\}  \\
\times  \mc{D}_{\ups} \big(    \mc{Z}_{x,y}  \mid  - \ups   \vth_{\ups}(\mf{Y}) - \ups \nu_{\ups} - \mf{n}_{\ups}    \big)  
\cdot  \mc{G}_{\ups}\big( \mc{Z}_{x,y} \big) 
 \cdot
  \bigg( 1+\e{O}\Big(  \de^2 \sul{\ups=\pm}{} |m_{\ups}|   + \de  \sul{ \ups^{\prime} = \pm }{} \Dp{ \nu_{\ups^{\prime}} } \Big)  \bigg) \;.
\label{ecriture action B ups comme series regulatrice massive plus somme restreinte modifiee}
\end{multline}
Above, I have introduced 
\bem
\mc{G}_{\ups}\big( \mc{Z}_{x,y} \big) \; = \;  \f{ G^2\big(1- \ups   \vth_{\ups}(\mf{Y}) - \ups \nu_{\ups} - \ell_{\ups} \big)     }{ G^2\big(1- \ups   \vth_{\ups}(\mf{Y}) - \ups \nu_{\ups} \big) }\cdot  
\lim_{L\tend + \infty} \Bigg\{
 \bigg( \f{ 2 \pi }{ L } \bigg)^{  ( \ups \vth_{\ups}(\mf{Y})  + \ups \nu_{\ups} + \mf{n}_{\ups} )^2 } \\
\times \sul{ \substack{ \mc{Z}_{p,h} \, :  \\ \, n_p-n_h \,  = \, \mf{l}_{\ups} }  }{} 
\ex{ \i  \ups \tfrac{m}{L}   \, \wt{u}_1^{\, \prime}\big( \tfrac{ N_{\ups} }{L} , \op{v} \big)\, \mc{J}(\mc{Z}_{p,h})  }   \cdot  \mc{R} \Big(  \mc{Z}_{p,h}  \mid - \ups   \vth_{\ups}(\mf{Y}) - \ups \nu_{\ups} -\mf{n}_{\ups}- \mf{l}_{\ups}    \Big)
 \cdot \f{ \pl{a=1}{n_p}  \vp_{x,y}^{(\ups)} \Big( \tfrac{p_a}{L} \Big) }{  \pl{a=1}{n_h}  \vp_{x,y}^{(\ups)} \Big( -\tfrac{h_a+1}{L} \Big)   } \Bigg\} \;. 
\label{definition fonction G ups somme restreinte deformee}
\end{multline}

One can convince oneself that the term in \eqref{ecriture action B ups comme series regulatrice massive plus somme restreinte modifiee} corresponding to $n_x$ $x$-integrations and 
$n_y$ $y$-integrations can be bounded by $\ex{-|m_{\ups}| \de (n_x+n_y)}$ provided that $\de |m_{\ups}|>c>0$ for some $c>0$

  Therefore, the only term in \eqref{ecriture action B ups comme series regulatrice massive plus somme restreinte modifiee} not giving rise to an exponential behaviour in $\de |m_{\ups}|$
 corresponds to taking $\mf{n}_{\ups}=0$ and $n_x=n_y=0$. For that particular choice, one has that $\vp_{x,y}^{(\ups)} \big(w \big)=1$ and then the 
 series in \eqref{definition fonction G ups somme restreinte deformee} corresponds to a so-called restricted sum. 
When $ \ell=0$, these  appeared for the first time in the context of harmonic analysis on $GL_{\infty}$ \cite{KerovOlshanskiVershikFirstOccurenceAbstracxtLevelMagicFormula,OlshanskiPointProcessAndInfiniteSymmetricGroup} 
and, for any $\ell \in \mathbb{Z}$,  in the study of the large-$x$ and low-temperature expansion of correlation function in the non-linear Schrödinger model 
\cite{KozMailletSlaLowTLimitNLSE}. It has been shown in \cite{KerovOlshanskiVershikFirstOccurenceAbstracxtLevelMagicFormula} for $\ell=0$ and in  
\cite{KozKitMailSlaTerRestrictedSums} for general $\ell$, see also \cite{KozKitMailTerMultiRestrictedSums,KozMailletMicroOriginOfc=1CFTUniversality} for other proofs, 
that the following summation identity holds
\beq
\f{ G^2(1+\nu-\ell) } { G^2(1+\nu) }   
\Big( \f{2\pi}{  L } \Big)^{\nu^2 } 
 \sul{ \mc{Z}_{p,h} \, : \, n_p-n_h \,  = \, \ell  }{}
 \ex{  \i  \tfrac{  x }{ L }  \mc{J}(\mc{Z}_{p,h})  } \cdot \mc{R}\big(  \mc{Z}_{p,h} \mid \nu -\ell \big)  
\; = \;    \bigg( \f{ \tf{2\pi}{L}  }{ 1- \ex{ \i \tfrac{ x }{ L } } } \bigg)^{ \nu^2 } .
\label{ecriture formule somme restreinte}
\enq
The summation identity allows one to compute readily the $L\tend +\infty$ limit of $\mc{G}_{\ups}\big( \big\{ \{ \emptyset \} ; \{ \emptyset \} \big\} \big)$, hence leading to 
\beq
\wh{\mc{B}}_{\ell_{\ups}  }^{\, (\ups)} \Big[ \mf{F}_{\ups}\big(\mf{Y} \mid * \big) \mid \mf{Y} \Big] \; = \;   \f{ \ex{\i \ups  m \ell_{\ups} p_{F} } }{ [-\i m_{\ups}]^{  (  \vth_{\ups}(\mf{Y})  +  \nu_{\ups} )^2 }  }  
\cdot \bigg( 1+ \e{O}\Big(  \de^2 \sul{\ups=\pm}{} |m_{\ups}|  +\sum_{ \ups  = \pm }  \ex{-   |m_{\ups}|\de}  + \de  \sul{ \ups^{\prime} = \pm }{} \Dp{ \nu_{\ups^{\prime}} } \Big)  \;. 
\enq
It then remains to take the product and compute the effect of the operator ordering $\bs{:}*\bs{:}$. Due to the form of the leading term, its action only produces logarithmic corrections, \textit{viz}. 
the operator ordering amounts to the substitution 
\beq
\e{O}\Big(  \de  \sum_{ \ups = \pm } \Dp{\nu_{\ups}} \Big)  \; \hookrightarrow \; \e{O}\Big( \de  \sum_{ \ups = \pm }  \ln |m_{\ups}| \Big)  \;, 
\enq
what entails the claim.

\section{A discrete/continuous Fourier transform}
\label{Appendix discreto continuous FT}

The purpose of this appendix is to compute the discrete time and space Fourier transform that arises in the context of studying spectral functions: 
\beq
\mc{T}[W](k,\om) \; = \; \sul{m \in \mathbb{Z} }{} \Int{ \R }{}  \f{ \ex{\i(\om t - k m) } W( m_{+} ,m_-)  }{   \pl{\ups = \pm }{}  \Big( \i \big[\op{v}_F(t-\i0^+)-\ups m  \big] \Big)^{\de_{\ups} }  }
 \cdot \dd t  \qquad \e{with} \quad  m_{\ups} = \ups m-\op{v}_F(t-\i0^+) \;. 
\enq
In this representation, $W$ is a function of two variables that is analytic in $\mathbb{H}^+\times \mathbb{H}^+$ and that admits the large variable asymptotic expansion, uniform on $\ov{\mathbb{H}}^+\times \ov{\mathbb{H}}^+$ of the type 
\beq
W(\vp_+,\vp_-) \;= \;  1 + \sul{\ups=\pm }{} V_{\ups} (\vp_{\ups})
\,+\, \sul{a_+=1}{n_+}\sul{a_-=1}{n_-} w_{a_+,a_-} \cdot \pl{\ups=\pm }{}\bigg\{ \f{ P_{a_{\ups}}^{(\ups)}\big( \ln  [-\i \vp_{\ups} ] \big) }{  [ -\i \vp_{\ups} ]^{\a_{ a_{\ups},\ups } } } \bigg\} \; + \; 
\e{O}\Big( \pl{\ups=\pm }{}[ -\i\vp_{\ups} ]^{-\a_{0,\ups} }  \Big)
\label{forme DA W}
\enq
where 
\beq
V_{\ups} (\vp_{\ups})\, = \, \sul{ a=1 }{ s_{\ups} } v^{ (\ups) }_a  \cdot \f{ Q_{ a }^{(\ups)} \big( \ln [-\i \vp_{\ups}] \big) }{  [ -\i \vp_{\ups} ]^{ \ga_{a,\ups} } }   
\;+\; \e{O}\Big(  [-\i \vp_{\ups} ]^{ -\ga_{0,\ups} }  \Big)\;. 
\enq
Here $P_{ a  }^{ (\ups) }$, $Q_{a }^{(\ups)}$ are some polynomials. The constants controlling the power-law decay are such that 
\beq
0 \, < \, \a_{1,\ups}  \, < \, \cdots \,  <  \,  \a_{n_{\ups},\ups}  <  \a_{0,\ups} 
\qquad \e{and} \qquad 
0 \, < \, \ga_{1,\ups}  \, < \, \cdots \,  <  \,  \ga_{s_{\ups},\ups}  <  \a_{0,\ups} \;. 
\enq
Furthermore, the largest coefficients satisfy 
\beq
\ga_{0,\ups} \, + \, \de_{\ups} \, > \, 1 \qquad \e{and} \qquad \a_{0,\ups} \, + \, \de_{\ups} \, > \,  1 \;. 
\label{ecriture conditions L1 integrables pour les exposants de decroissance}
\enq

\begin{prop}
\label{Proposition Calcul des TFs}

Under the above assumptions, and provided that for some $0<\tau<1$, 
\beq
W(\vp_+,\vp_-) \;= \;  1  \, + \,    \e{O}\bigg(  \sul{\ups=\pm }{} \f{1}{ [-\i \vp_{\ups} ]^{1-\tau} } \bigg)
\label{ecriture asymptotiques de W avec une forme moins fine}
\enq
it holds, in the sense of distributions,  and for $\de_{\ups} \geq 0$,
\bem
\mc{T}[W](k,\om) \; = \; \f{ (2\pi)^2 }{ 2 \op{v}_F  } \sul{n \in \mathbb{Z} }{} 
\pl{\ups= \pm}{} \Bigg\{ \Xi\big[\om + \ups \op{v}_F(k+2\pi n) \big] \cdot  \bigg( \f{\om + \ups \op{v}_F(k+2\pi n) }{2 \op{v}_F} \bigg)^{\de_{\ups}-1}  \Bigg\}  \\
\times \bigg\{ \f{1}{ \Ga(\de_+) \Ga(\de_-)  } \; + \;   \sul{\ups=\pm }{} \e{O}\bigg(  \Big[ \om + \ups \op{v}_F(k+2\pi n)  \Big]^{1-\tau}  \bigg) \bigg\} 
\label{ecriture TF discrete et continue du facteur dumping des bords Fermi}
\end{multline}

\end{prop}

\Proof 

To start with one observes that $\mc{T}(k,\om)$ can be recast as 
\bem
\mc{T}[W](k,\om) \; = \; \sul{m \in \mathbb{Z} }{} \Int{ \R^2 }{}    \f{  \ex{\i(\om t - k x) }  \de(x-m)  }{   \pl{\ups = \pm }{}  \Big( \i \big[\op{v}_F(t-\i0^+)-\ups x  \big] \Big)^{\de_{\ups} }  }
W\Big( x-\op{v}_F(t-\i0^+) ,- x-\op{v}_F(t-\i0^+) \Big) \cdot \dd t\, \dd x   \\
\; = \; \sul{ n \in \mathbb{Z} }{} \Int{ \R^2 }{}  \f{  \ex{\i(\om t - (k+2\pi n ) x) }     }{   \pl{\ups = \pm }{}  \Big( \i \big[\op{v}_F(t-\i0^+)-\ups x  \big] \Big)^{\de_{\ups} }  }
W\Big( x-\op{v}_F(t-\i0^+) ,- x-\op{v}_F(t-\i0^+)\Big) \cdot \dd t\, \dd x  \\ 
\; = \;   \sul{ n \in \mathbb{Z} }{} \Int{ \R^2 }{}   \pl{\ups = \pm }{} \Bigg\{  \f{ \ex{- \i \vp_{\ups} \Om_{\ups,n} }      }
{     \Big( -\i \big[\vp_{\ups}  + \i0^+ \big] \Big)^{\de_{\ups} }  } \Bigg\}
W\Big( \vp_+ + \i0^+  , \vp_- +\i0^+ \Big) \cdot \f{ \dd \vp_{+} \dd \vp_{-} }{2 \op{v}_F } \;. 
\end{multline}
Above, I have set 
\beq
 \Om_{\ups,n} \; = \; \f{ \om + \ups \op{v}_F (k+2\pi n ) }{ 2\op{v}_F } \;. 
\enq
In the intermediate steps one uses the Fourier series expansion of the Dirac Comb: $\sul{m \in \mathbb{Z} }{} \de(x-m) \, = \, \sul{n \in \mathbb{Z} }{} \ex{2 \i \pi n x} $ and, subsequently, 
changes the variables to $\vp_{\ups}=-(v_F t -\ups x )$. 
Since $W$ is analytic and bounded on $\mathbb{H}^+\times\mathbb{H}^+$, it follows by deforming the integrals to $+\i\infty$, that, whenever $\Om_{\ups;n}<0$, the integral vanishes. 
Hence, one has that 

\beq
\mc{T}[W](k,\om)  \; = \;   \sul{ n \in \mathbb{Z} }{}   \pl{\ups = \pm }{}\Big\{   \Xi(\Om_{\ups;n}\big) \big[ \Om_{\ups;n} \big]^{\de_{\ups}-1} \Big\}
\Int{ \R^2 }{}   \pl{\ups = \pm }{} \Bigg\{  \f{ \ex{- \i \vp_{\ups}   }      }
{     \Big(-\i \big[\vp_{\ups}  + \i0^+ \big]  \Big)^{\de_{\ups} }  } \Bigg\}
W\Big(  \tfrac{ \vp_+ }{  \Om_{+;n}  } + \i0^+  , \tfrac{ \vp_- }{  \Om_{-;n}  } +\i0^+ \big) \cdot \f{ \dd \vp_{+} \dd \vp_{-}  }{ 2 \op{v}_F } \;. 
\label{ecriture forme legerement modifiee TF} 
\enq
At this stage it remains to deform the integration contours to the lines $\big\{ \R+\i \eps_+\big\}\times \big\{ \R + \i \eps_- \big\}$ for some $\eps_{\pm}>0$ small enough 
and then inject the asymptotic expansion \eqref{forme DA W} into \eqref{ecriture forme legerement modifiee TF}. This yields
\bem
\mc{T}[W](k,\om)  \; = \;  \f{ 1 }{ 2 \op{v}_F } \sul{ n \in \mathbb{Z} }{}   \pl{\ups = \pm }{} \Big\{   \Xi\big( \Om_{\ups;n} \big)  \cdot \big[ \Om_{\ups;n} \big]^{\de_{\ups}-1} \Big\} \\
\times \Bigg\{ J( \de_{+} ) \, J( \de_{-} ) \;  + \;  \sul{\ups = \pm }{} \sul{ a=1 }{ s_{\ups} } v^{ (\ups) }_a  \cdot J( \de_{-\ups} ) \cdot Q_{ a }^{(\ups)} \big( \Dp{ \ga_{a,\ups} } \big)  
\cdot \Big\{   [ \Om_{\ups;n}  ]^{ \ga_{a,\ups} }    \cdot  J( \de_{\ups} +  \ga_{a,\ups} ) \Big\}  \\
 \; + \; \sul{a_+=1}{n_+} \sul{a_-=1}{n_-}  w_{a_+,a_-} \cdot \pl{ \ups=\pm }{}  \bigg(   P_{ a_{\ups} }^{ (\ups) }\big( \Dp{ \a_{a_{\ups},\ups} } \big) \cdot 
 \Big\{   [ \Om_{\ups;n}  ]^{ \a_{a_{\ups},\ups} }     J( \de_{\ups} +  \a_{a_{\ups},\ups} ) \Big\} 
  \bigg) \; + \; \mf{r}\big(\Om_{+;n},\Om_{-;n} \big)  
  \Bigg\}  \;. 
\label{ecriture du DA de T}
\end{multline}
$J(\de)$ are given as one dimensional integrals
\beq
J(\de) \, = \, \Int{ \R }{}  \f{ \ex{ - \i \vp    }  }{    \big( -\i \big[\vp +\i0^+ \big] \big)^{\de }  } \cdot  \dd \vp 
\; = \; \f{2\pi}{ \Ga(\de) }
\enq
where the explicit result comes from  deforming the contour to a loop around $ -\i\R$ and recognising the definition of the Gamma function.  
Finally, the remainder $ \mf{r}\big(\Om_{+;n},\Om_{-;n} \big)  $ is readily seen to be bounded as 
\bem
\big| \mf{r}\big(\Om_{+;n},\Om_{-;n} \big) \big| \; \leq \; C  \pl{\ups = \pm }{} \Big\{   \Xi\big( \Om_{\ups;n} \big)  \cdot \big[ \Om_{\ups;n} \big]^{\de_{\ups}-1} \Big\}  \\
\times \Bigg\{  \sul{\ups=\pm }{}   \Int{\R }{} \f{ \big[ \Om_{\ups;n} \big]^{\ga_{0,\ups}}  }{  |\vp + \i \eps_{\ups} |^{ \de_{\ups}+\ga_{0,\ups} } } \cdot \dd \vp  
\; + \;  \pl{\ups=\pm }{}   \Int{\R }{} \f{  \big[ \Om_{\ups;n} \big]^{\a_{0,\ups}} }{  |\vp + \i \eps_{\ups} |^{ \de_{\ups}+\a_{0,\ups} } } \cdot \dd \vp  \Bigg\}  \;. 
\end{multline}
The integrals converge due to the conditions \eqref{ecriture conditions L1 integrables pour les exposants de decroissance} and the bounds \eqref{ecriture asymptotiques de W avec une forme moins fine} 
then allow one to provide an estimate on all the power laws appearing in the expansions 
\eqref{ecriture du DA de T}. \qed


\begin{thebibliography}{100}

\bibitem{AizenbergYuzhakovIntRepAndMultidimensionalResidues}
I.A. Aizenberg and A.P. Yuzhakov, \emph{{Integral representations and residues
  in multidimensional complex analysis}}, Graduate Texts in Mathematics,
  vol.~58, American Mathematical Society, 1978.

\bibitem{ArikawaKabrachMullerWieleXXAsymptoticsofFFs}
M.~Arikawa, M.~Kabrach, G.~M\"{u}ller, and K.~Wiele, \emph{{"Spinon excitations
  in the XX chain: spectra, transition rates, observability."}}, J. Phys. A:
  Math. Gen \textbf{\bf{39}} (2006), 10623--10640.

\bibitem{BabujianFoersterKarowskiSomeReviewOfBootstrap}
H.~M. Babujian, A.~Foerster, and M.~Karowski, \emph{{"The Form Factor Program:
  a Review and New Results - the Nested SU(N) Off-Shell Bethe Ansatz."}},
  SIGMA, Proc. of the O'Raifeartaigh Symposium on Non-Perturbative and Symmetry
  Methods in Field Theory (June 2006, Budapest, Hungary) \textbf{\bf{2}}
  (2006), 082.

\bibitem{BarouchMcCoy2ptFctsInXYPlusLongDistAB}
E.~Barouch and B.M. McCoy, \emph{{"Statistical mechanics of XY-model .2.
  Spin-correlation functions."}}, Phys. Rev. A \textbf{\bf 3} (1971), 786--804.

\bibitem{BeckBonnerMullerThomasSpectralFctsXXXGeneralFeatures}
H.~Beck, J.C. Bonner, G.~M\"{u}ller, and H.~Thomas, \emph{{"Quantum spin
  dynamics of the antiferromagnetic linear chain in zero and nonzero magnetic
  field."}}, Phys. Rev. B \textbf{\bf 24} (1981), 1429--1467.

\bibitem{BetheSolutionToXXX}
H.~Bethe, \emph{{"Z\"{u}r Theorie der Metalle: Eigenwerte und Eigenfunktionen
  der linearen Atomkette."}}, Zeitschrift f\"{u}r Physik \textbf{\bf 71}
  (1931), 205--226.

\bibitem{BogoliubiovIzerginKorepinBookCorrFctAndABA}
N.M. Bogoliubov, A.G. Izergin, and V.E. Korepin, \emph{{"Quantum inverse
  scattering method, correlation functions and algebraic Bethe Ansatz."}},
  Cambridge monographs on mathematical physics, 1993.

\bibitem{CardyConformalDimensionsFromLowLSpectrum}
J.L. Cardy, \emph{{"Operator content of two-dimensional conformally invariant
  theories."}}, Nucl. Phys. B \textbf{\bf 270} (1986), 186--204.

\bibitem{ColomoIzerginKorepinTognettiTempCorrFctXX}
F.~Colomo, A.G. Izergin, V.E. Korepin, and V.~Tognetti, \emph{{"Temperature
  correlation functions in the XX0 Heisenberg chain."}}, Teor. Math. Phys.
  \textbf{\bf{94}} (1993), 19--38.

\bibitem{DeVegaWoynarowichFiniteSizeCorrections6VertexNLIEmethod}
H.J. de~Vega and F.~Woynarowich, \emph{{"Method for calculating finite size
  corrections in Bethe Ansatz systems- Heisenberg chains and 6-vertex
  model."}}, Nucl. Phys. B \textbf{\bf{251}} (1985), 439--456.

\bibitem{DelfinoMussardoSimonettiMasslessFFFromMassiveQFTIntoMethod}
G.~Delfino, G.~Mussardo, and P.~Simonetti, \emph{{"Correlation Functions Along
  a Massless Flow."}}, Phys. Rev. D \textbf{\bf{51}} (1995), 6620--6624.

\bibitem{DescloizeauxGaudinExcitationsXXZ+Gap}
J.~des Cloizeaux and M.~Gaudin, \emph{{"Anisotropic linear magnetic chain."}},
  J. Math. Phys. \textbf{\bf{7}} (1966), 1384--1400.

\bibitem{DescloizeauxPearsonExcitationsXXX}
J.~des Cloizeaux and J.J. Pearson, \emph{{"Spin-wave spectrum of the
  antiferromagnetic linear chain."}}, Phys. Rev. \textbf{\bf{128}} (1962),
  2131--2135.

\bibitem{DestriLowensteinFirstIntroHKBAEAndArgumentForStringIsWrong}
C.~Destri and J.H. Lowenstein, \emph{{"Analysis of the Bethe Ansatz equations
  of the chiral invariant Gross-Neveu model."}}, Nucl. Phys. B
  \textbf{\bf{205}} (1982), 369--385.

\bibitem{KozDugaveGohmannThermoFunctionsZeroTXXZMassless}
M.~Dugave, F.~G\"{o}hmann, and K.K. Kozlowski, \emph{{"Functions characterizing
  the ground state of the XXZ spin-1/2 chain in the thermodynamic limit."}},
  SIGMA \textbf{10} (2014), 043, 18 pages.

\bibitem{FaddeevTakhtadzhanSpinOfExcitationsInXXX}
L.D. Faddeev and L.A. Takhtadzhan, \emph{{"What is the spin of a spin wave?"}},
  Phys. Lett. A \textbf{\bf 85} (1981), 375--377.

\bibitem{FowlerZotosConditionsOfStringExistenceXYZAndSineGordon}
M.~Fowler and X.~Zotos, \emph{{"Quantum sine-Gordon thermodynamics:the Bethe
  Ansatz method."}}, Phys. Rev. B \textbf{\bf 24} (1981), 2634--2639.

\bibitem{GohmannKlumperSeelFinieTemperatureCorrelationFunctionsXXZ}
F.~G\"{o}hmann, A.~Kl\"{u}mper, and A.~Seel, \emph{{"Integral representations
  for correlation functions of the XXZ chain at finite temperature."}}, J.
  Phys. A: Math. Gen. \textbf{\bf 37} (2004), 7625--7652.

\bibitem{HidaConditionExistenceStringsXXZ}
K.~Hida, \emph{{"Rigorous derivation of the distribution of the eigenstates of
  the quantum Heisenberg-Ising chain with XY-like anisotropy."}}, Phys.Lett. A
  \textbf{\bf{84}} (1981), 338--340.

\bibitem{IshimuraShibaLinIntEqnsForFiniteHForXXZ}
N.~Ishimura and H.~Shiba, \emph{{"Effect of the magnetic field on the des
  Cloizeaux-Pearson spin wave spectrum."}}, Prog. Theor. Phys. \textbf{\bf{57}}
  (1977), 1862--1873.

\bibitem{ItsIzerginKorepinSlavnovDifferentialeqnsforCorrelationfunctions}
A.R. Its, A.G. Izergin, V.E. Korepin, and N.A. Slavnov, \emph{{"Differential
  equations for quantum correlation functions."}}, Int. J. Mod. Physics
  \textbf{\bf{B4}} (1990), 1003--1037.

\bibitem{ItsSlavnovNLSTimeAndSpaceCorrDualFields}
A.R. Its and N.A. Slavnov, \emph{{"On the Riemann-Hilbert approach to the
  asymptotic analysis of the correlation functions of the Quantum Nonlinear
  Schr\"{o}dinger equation. Non-free fermion case."}}, Theor. Math. Phys.
  \textbf{\bf{119}:2} (1990), 541--593.

\bibitem{JimboMiwaElementaryBlocksXXZperiodicMassless}
M.~Jimbo and T.~Miwa, \emph{{"QKZ equation with $\mid q \mid$ =1 and
  correlation functions of the XXZ model in the gapless regime."}}, J. Phys. A
  \textbf{\bf 29} (1996), 2923--2958.

\bibitem{JimboMiwaMoriSatoSineKernelPVForBoseGaz}
M.~Jimbo, T.~Miwa, Y.~Mori, and M.~Sato, \emph{{"Density matrix of an
  impenetrable Bose gas and the fifth Painlev\'{e} transcendent."}}, Physica D
  \textbf{\bf{1}} (1980), 80--158.

\bibitem{KarowskiWeiszFormFactorsFromSymetryAndSMatrices}
M.~Karowski and P.~Weisz, \emph{{"Exact form factors in (1 + 1)-dimensional
  field theoretic models with soliton behaviour."}}, Nucl. Phys. B \textbf{\bf
  139} (1978), 455--476.

\bibitem{KafmanOnsagerFirstIntroDetRepCorrIsing2D}
B.~Kaufman and L.~Onsager, \emph{{"Crystal statistics. III. Short-range order
  in a binary Ising lattice."}}, Phys. Rev. \textbf{\bf 7} (1949), 1244--1252.

\bibitem{KerovOlshanskiVershikFirstOccurenceAbstracxtLevelMagicFormula}
S.~Kerov, G.~Olshanski, and A.~Vershik, \emph{{"Harmonic analysis on the
  infinite symmetric group. A deformation of the regular representation."}},
  Comptes Rend. Acad. Sci. Paris, S\'{e}r I \textbf{\bf 316} (1993), 773--778.

\bibitem{KirillovSmirnovFirstCompleteSetBootstrapAxiomsForQIFT}
A.N. Kirillov and F.A. Smirnov, \emph{{"A representation of the current algebra
  connected with the SU (2)-invariant Thirring model."}}, Zap. Nauchn. Sem.
  Leningrad. Otdel. Mat. Inst. Steklov. \textbf{\bf 198} (1987), 506--510.

\bibitem{KozKitMailSlaTerXXZsgZsgZAsymptotics}
N.~Kitanine, K.K. Kozlowski, J.-M. Maillet, N.A. Slavnov, and V.~Terras,
  \emph{{"Algebraic Bethe Ansatz approach to the asymptotics behavior of
  correlation functions."}}, J. Stat. Mech: Th. and Exp. \textbf{04} (2009),
  P04003.

\bibitem{KozKitMailSlaTerEffectiveFormFactorsForXXZ}
\bysame, \emph{{"On the thermodynamic limit of form factors in the massless XXZ
  Heisenberg chain."}}, J. Math. Phys. \textbf{50} (2009), 095209.

\bibitem{KozKitMailSlaTerRestrictedSums}
\bysame, \emph{{"A form factor approach to the asymptotic behavior of
  correlation functions in critical models."}}, J. Stat. Mech. : Th. and Exp.
  \textbf{1112} (2011), P12010.

\bibitem{KozKitMailSlaTerThermoLimPartHoleFormFactorsForXXZ}
\bysame, \emph{{"Thermodynamic limit of particle-hole form factors in the
  massless XXZ Heisenberg chain."}}, J. Stat. Mech. : Th. and Exp.
  \textbf{1105} (2011), P05028.

\bibitem{KozKitMailSlaTerRestrictedSumsEdgeAndLongTime}
\bysame, \emph{{"Form factor approach to dynamical correlation functions in
  critical models."}}, J. Stat. Mech. \textbf{1209} (2012), P09001.

\bibitem{KozKitMailTerMultiRestrictedSums}
N.~Kitanine, K.K. Kozlowski, J.-M. Maillet, and V.~Terras,
  \emph{{"Long-distance asymptotic behaviour of multi-point correlation
  functions in massless quantum integrable models."}}, J. Stat. Mech.
  \textbf{1405} (2014), P05011.

\bibitem{KitanineMailletSlavnovTerrasDynamicalCorrelationFunctions}
N.~Kitanine, J.-M. Maillet, N.A. Slavnov, and V.~Terras, \emph{{"Dynamical
  correlation functions of the XXZ spin-$1/2$ chain."}}, Nucl. Phys. B
  \textbf{\bf 729} (2005), 558--580.

\bibitem{KitanineMailletTerrasFormfactorsperiodicXXZ}
N.~Kitanine, J.-M. Maillet, and V.~Terras, \emph{{"Form factors of the XXZ
  Heisenberg spin-$1/2$ finite chain."}}, Nucl. Phys. B \textbf{554} (1999),
  647--678.

\bibitem{KitanineMailletTerrasElementaryBlocksPeriodicXXZ}
\bysame, \emph{{"Correlation functions of the XXZ Heisenberg spin-$1/2$ chain
  in a magnetic field."}}, Nucl. Phys. B \textbf{567} (2000), 554--582.

\bibitem{KlumperBatchelorNLIEApproachFiniteSizeCorSpin1XXZIntroMethod}
A.~Kl\"{u}mper and M.T. Batchelor, \emph{{"An analytic treatment of finite-size
  corrections of the spin-1 antiferromagnetic XXZ chain."}}, J. Phys. A: Math.
  Gen. \textbf{\bf 23} (1990), L189.

\bibitem{KojimaKorepinSlavnovNLSEDeterminatFormFactorAndDualFieldTempeAndTime}
T.~Kojima, V.E. Korepin, and N.A. Slavnov, \emph{{"Determinant representation
  for dynamical correlation functions of the quantum nonlinear Schr\"{o}dinger
  equation."}}, Comm. Math. Phys. \textbf{\bf 188} (1997), 657--689.

\bibitem{KorepinAnalysisofBoundStateConditionMassiveThirring}
V.E. Korepin, \emph{{"Direct calculation of the $S$-matrix in the massive
  Thirring model."}}, Theor. Math. Phys. \textbf{\bf 41} (1979), 169--189.

\bibitem{KorepinSlavnovTimeDepCorrImpBoseGas}
V.E. Korepin and N.A. Slavnov, \emph{{"The time dependent correlation function
  of an impenetrable Bose gas as a Fredholm minor I."}}, Comm. Math.Phys.
  \textbf{\bf 129} (1990), 103--113.

\bibitem{KorepinSlavnovNonlinearIdentityScattPhase}
\bysame, \emph{{"The new identity for the scattering matrix of exactly solvable
  models."}}, Eur. Phys. J. \textbf{\bf B 5} (1998), 555--557.

\bibitem{KozProofOfDensityOfBetheRoots}
K.K. Kozlowski, \emph{{"On condensation properties of Bethe roots associated
  with the XXZ chain."}}, Comm. Math. Phys. \textbf{357}, 3, (2018), 1009-1069.

\bibitem{KozProofOfStringSolutionsBetheeqnsXXZ}
\bysame, \emph{{"On string solutions to the Bethe equations for the XXZ chain:
  a rigorous approach."}}, to appear.

\bibitem{KozTimeDepGSKandNatteSeries}
\bysame, \emph{{Riemann--Hilbert approach to the time-dependent generalized
  sine kernel."}}, Adv. Theor. Math. Phys. \textbf{15} (2011), 1--89.

\bibitem{KozReducedDensityMatrixAsymptNLSE}
\bysame, \emph{{"Large-distance and long-time asymptotic behavior of the
  reduced denisty matrix in the non-linear Schr\"{o}dinger model."}}, Ann.
  Henri-Poincar\'{e} \textbf{16} (2015), 437--534.

\bibitem{KozProofOfAsymptoticsofFormFactorsXXZBoundStates}
\bysame, \emph{{"Form factors of bound states in the XXZ chain."}}, J. Phys. A:
  Math. Theor. Special Issue "Emerging talents" \textbf{50} (2017), 184002.

\bibitem{KozMailletSlaLowTLimitNLSE}
K.K. Kozlowski, J.-M. Maillet, and N.~A. Slavnov, \emph{{"Low-temperature limit
  of the long-distance asymptotics in the non-linear Schr\"{o}dinger model."}},
  J.Stat.Mech. (2011), P03019.

\bibitem{KozMailletMicroOriginOfc=1CFTUniversality}
K.K. Kozlowski and J.M. Maillet, \emph{{"Microscopic approach to a class of 1D
  quantum critical models."}}, J. Phys. A: Math and Theor. Baxter anniversary
  special issue \textbf{48} (2015), 484004.

\bibitem{KozRagoucyAsymptoticsHigherRankModels}
K.K. Kozlowski and E.~Ragoucy, \emph{{"Asymptotic behaviour of two-point
  functions in multi-species models."}}, Nucl. Phys. B (2016).

\bibitem{KozTerNatteSeriesNLSECurrentCurrent}
K.K. Kozlowski and V.~Terras, \emph{{"Long-time and large-distance asymptotic
  behavior of the current-current correlators in the non-linear Schr\"{o}dinger
  model."}}, J. Stat. Mech.: Th. and Exp. (2011), P09013.

\bibitem{LesageSaleurMasslessFFApproachtoFriedelOscillations}
F.~Lesage and H.~Saleur, \emph{{"Form-factors computation of Friedel
  oscillations in Luttinger liquids."}}, J. Phys. A: Math. Gen.
  \textbf{\bf{30}} (1997), L457--L463.

\bibitem{LesageSaleurSkorikMasslessFFApproachtoCurrentCorrelators}
F.~Lesage, H.~Saleur, and S.~Skorik, \emph{{"Form factors approach to current
  correlations in one-dimensional systems with impurities."}}, JNucl. Phys. B
  \textbf{\bf{474}} (1996), 602--640.

\bibitem{LiebMattisSchultzIsingAsFreeFermionModel}
E.H. Lieb, D.C. Mattis, and T.D. Schultz, \emph{{"Two dimensionnal Ising model
  as a soluble problem of many fermions."}}, Rev. Mod. Phys. \textbf{\bf{36}}
  (1964), 856--871.

\bibitem{McCoySomeAsymptoticsForXYCorrelators}
B.M. McCoy, \emph{{"Spin correlation functions in the XY model."}}, Phys. Rev.
  \textbf{\bf{173}} (1968), 531--541.

\bibitem{McCoyPerkShrockSpinTimeAutoCorrAsModSineKernel}
B.M. McCoy, J.H.H. Perk, and R.E. Shrock, \emph{{"Time-dependent correlation
  functions of the transverse Ising chain at the critical magnetic field."}},
  Nucl. Phys. B \textbf{\bf{220}} (1983), 35--47.

\bibitem{MejanSmirnovFormFactorsinPincChiralModelMassless}
P.~Mejean and F.A. Smirnov, \emph{{"Form Factors for Principal Chiral Field
  Model with Wess-Zumino-Novikov-Witten Term."}}, , Int. J. Mod. Phys. A 12,
  3383 (1997) \textbf{\bf{12}} (1997), 3383--3395.

\bibitem{MullerShrockDynamicCorrFnctsTIandXXAsymptTimeAndFourier}
G.~M\"{u}ller and R.E. Shrock, \emph{{"Dynamic correlation functions for
  one-dimensional quantum-spin systems: new results based on a rigorous
  approach."}}, Phys. Rev. B \textbf{\bf{29}} (1984), 288--301.

\bibitem{OlshanskiPointProcessAndInfiniteSymmetricGroup}
G.~Olshanski, \emph{{"Point processes and the infinite symmetric group. Part I:
  The general formalism and the density function."; In: The orbit method in
  geometry and physics: in honor of A. A. Kirillov (C. Duval, L. Guieu, and V.
  Ovsienko, eds.), Birkh\"{a}user, Verlag, Basel}}, Prog. in Math. \textbf{\bf
  213} (2003).

\bibitem{OrbachXXZCBASolution}
R.~Orbach, \emph{{"Linear antiferromagnetic chain with anisotropic
  coupling."}}, Phys. Rev. \textbf{\bf 112} (1958), 309--316.

\bibitem{PerkAuYangTimeDpdtTransverseIsing}
J.H.H. Perk and H.~Au-Yang, \emph{{"New Results for Time-Dependent Correlation
  Functions in the Transverse Ising Chain."}}, J. Stat. Phys. \textbf{\bf 135}
  (2009), 599--619.

\bibitem{PonsotFFDensityinMasslessSSG}
B.~Ponsot, \emph{{"Massless N=1 super-sinh-Gordon: form factors approach."}},
  Phys. Lett. \textbf{\bf 575} (2003), 131--136.

\bibitem{SakaiDynamicalAndTimeDependentCorrelatorsXXZ}
K.~Sakai, \emph{{"Dynamical correlation functions of the XXZ model at finite
  temperature."}}, J. Phys. A: Math Theor \textbf{\bf 40} (2007), 7523--7542.

\bibitem{SlavnovFormFactorsNLSE}
N.A. Slavnov, \emph{{"Non-equal time current correlation function in a
  one-dimensional Bose gas."}}, Theor. Math. Phys. \textbf{\bf 82} (1990),
  273--282.

\bibitem{SlavnovPDE4MultiPtsFreeNLSM}
\bysame, \emph{{"Differential equations for multipoint correlation functions in
  a one-dimensional impenetrable Bose gas."}}, Theor. Math. Phys. \textbf{\bf
  106} (1996), 131--142.

\bibitem{SlavnovComputationDualFieldVaccumExpLongTimeDistTempeRedDensNLSE}
\bysame, \emph{{"Integral equations for the correlation functions of the
  quantum one-dimensional Bose gas."}}, Theor. Math. Phys. \textbf{\bf 121}
  (1999), 1358--1376.

\bibitem{SmirnovReductionsAndClusterPropertyInSineGordonPlusSomeDiscussionsIRLimit}
F.A. Smirnov, \emph{{"Reductions of the sine-Gordon model as a perturbation of
  minimal models of conformal field theory."}}, Nucl. Phys. B \textbf{\bf 337}
  (1990), 156--180.

\bibitem{SmirnovFormFactors}
\bysame, \emph{{"Form factors in completely integrable models of quantum field
  theory."}}, Advanced Series in Mathematical Physics, vol.~14, World
  Scientific, 1992.

\bibitem{TakahashiThermodynamics1DSolvModels}
M.~Takahashi, \emph{{"Thermodynamics of one dimensional solvable models."}},
  Cambridge university press, 1999.

\bibitem{TakahashiSuzukiFiniteTXXZandStrings}
M.~Takahashi and M.~Suzuki, \emph{{"One-dimensional anisotropic Heisenberg
  model at finite temperatures."}}, Prog. Theor. Phys. \textbf{\bf 48} (1972),
  2187--2209.

\bibitem{TracyVaidaTimeDependentTransverseIsingCorrelatorsAsymptotics}
C.A. Tracy and H.G. Vaidya, \emph{{"Transverse time-dependent spin correlation
  functions of the one-dimensional XY model at zero temperature."}}, Physica
  \textbf{\bf A 92} (1978), 1--41.

\bibitem{BarouchTracyMcCOyWuScalingResultsForIsinfPainleveIII}
T.T. Wu, B.M. McCoy, C.A. Tracy, and E.~Barouch, \emph{{"The spin-spin
  correlation function of the two dimensional Ising model: exact results in the
  scaling region."}}, Phys. Rev. B \textbf{\bf 13} (1976), 316.

\end{thebibliography}
\end{document}